%% file: main.tex
\documentclass[nofootinbib,aps,prd,groupedaddress,preprintnumbers,%
  showpacs,showkeys,floatfix,amssymb,amsfonts,superscriptaddress]{revtex4-1}  
\usepackage{hyperref}
\usepackage[usenames]{color}
\usepackage{bbold}
\usepackage{amsmath}
\usepackage{multirow}
\usepackage{graphicx}
\usepackage{xfrac}
\usepackage[utf8]{inputenc}

\usepackage{amsfonts}
\usepackage{amssymb}
\usepackage{extarrows}

\usepackage{slashed}
\usepackage{color}

\newcommand{\GeV}[1]{{\rm GeV}}
\newcommand{\MeV}[1]{{\rm MeV}}
\newcommand{\RIp}{{RI\textquotesingle}}

\newcommand{\Bern}{Institute for Theoretical Physics, Albert Einstein Center for Fundamental Physics,\\University of Bern, Sidlerstrasse 5, CH-3012 Bern, Switzerland}
\newcommand{\hiskp}{HISKP (Theory), Rheinische Friedrich-Wilhelms-Universit\"at Bonn,\\Nussallee 14-16, 53115 Bonn, Germany}
\newcommand{\hpca}{High Performance Computing and Analytics Lab, Rheinische Friedrich-Wilhelms-Universit\"at Bonn,\\ Friedrich-Hirzebruch-Allee 8, 53115 Bonn, Germany}
\newcommand{\CyprusU}{Department of Physics, University of Cyprus, 20537 Nicosia, Cyprus}
\newcommand{\CyprusI}{Computation-based Science and Technology Research Center, The Cyprus Institute,\\20 Konstantinou Kavafi Street, 2121 Nicosia, Cyprus}
\newcommand{\Jena}{University of Jena, Institute for Theoretical Physics, Max-Wien-Platz 1, D-07743 Jena, Germany}
\newcommand{\Parma}{Dipartimento  di  Scienze  Matematiche,  Fisiche  e  Informatiche,  Universit\`a  di  Parma  and  INFN, Gruppo  Collegato  di  Parma,  Parco  Area  delle  Scienze  7/a  (Campus),  43124  Parma,  Italy}
\newcommand{\Romauno}{Dipartimento di Fisica and INFN, Universit\`a di Roma ``La Sapienza",\\Piazzale Aldo Moro 5, I-00185 Roma, Italy}
\newcommand{\Romadue}{Dipartimento di Fisica and INFN, Universit\`a di Roma ``Tor Vergata",\\Via della Ricerca Scientifica 1, I-00133 Roma, Italy}
\newcommand{\Romatre}{Dipartimento di Matematica e Fisica, Universit\`a Roma Tre and INFN, Sezione di Roma Tre,\\Via della Vasca Navale 84, I-00146 Rome, Italy}
\newcommand{\RomatreINFN}{Istituto Nazionale di Fisica Nucleare, Sezione di Roma Tre,\\Via della Vasca Navale 84, I-00146 Rome, Italy}
\newcommand{\temple}{Department of Physics,  Temple University,  Philadelphia, PA 19122 - 1801, USA}
\newcommand{\RomaunoINFN}{Istituto Nazionale di Fisica Nucleare, Sezione di Roma La Sapienza,\\ Piazzale Aldo Moro 5, I-00185 Roma, Italy}
\newcommand{\Edin}{
School of Physics and Astronomy,
The University of Edinburgh, Edinburgh EH9 3FD, UK}
\newcommand{\Odense}{CP$^3$-Origins, University of Southern Denmark, Campusvej 55, 5230 Odense, Denmark}
\newcommand{\Grenoble}{Theory Group, Laboratoire de Physique Subatomique et de Cosmologie, CNRS/IN2P3 38026 Grenoble, France}
\newcommand{\Fermi}{Centro Fermi, Museo Storico della Fisica e Centro Studi e Ricerche ``Enrico Fermi",\\ Piazza del Viminale 1, I-00184 Roma, Italy}
\newcommand{\NIC}{NIC, DESY, Platanenallee 6, D-15738 Zeuthen, Germany}
\newcommand{\Berlin}{Institut f\"ur Physik, Humboldt-Universit\"at zu Berlin, Newtonstrasse 15, 12489 Berlin, Germany}

\begin{document}

\title{Quark masses using twisted mass fermion gauge ensembles}

\author{C.~Alexandrou}\affiliation{\CyprusU}\affiliation{\CyprusI}
\author{S.~Bacchio}\affiliation{\CyprusI}
\author{G.~Bergner}\affiliation{\Jena}
\author{M.~Constantinou}\affiliation{\temple}
\author{M.~Di Carlo}\affiliation{\Edin}\affiliation{\RomaunoINFN}
\author{P.~Dimopoulos}\affiliation{\Parma}
\author{J.~Finkenrath}\affiliation{\CyprusI}
\author{E.~Fiorenza}\affiliation{\Odense}
\author{R.~Frezzotti}\affiliation{\Romadue} 
\author{M.~Garofalo}\affiliation{\hiskp}
\author{K.~Hadjiyiannakou}\affiliation{\CyprusU}\affiliation{\CyprusI}
\author{B.~Kostrzewa}\affiliation{\hpca}
\author{G.~Koutsou}\affiliation{\CyprusI}
\author{K.~Jansen}\affiliation{\NIC}
\author{V.~Lubicz}\affiliation{\Romatre}
\author{M.~Mangin-Brinet}\affiliation{\Grenoble}
\author{F.~Manigrasso}\affiliation{\CyprusU}\affiliation{\Romadue}\affiliation{\Berlin}
\author{G.~Martinelli}\affiliation{\Romauno}
\author{E.~Papadiofantous}\affiliation{\CyprusU}\affiliation{\CyprusI}
\author{F.~Pittler}\affiliation{\CyprusI}
\author{G.C.~Rossi}\affiliation{\Romadue}\affiliation{\Fermi}
\author{F.~Sanfilippo}\affiliation{\RomatreINFN}
\author{S.~Simula}\affiliation{\RomatreINFN}
\author{C.~Tarantino}\affiliation{\Romatre}
\author{A.~Todaro}\affiliation{\CyprusU}\affiliation{\Romadue}\affiliation{\Berlin}
\author{C.~Urbach}\affiliation{\hiskp}
\author{U.~Wenger}\affiliation{\Bern}

\begin{abstract}
\vspace{0.5cm}
\centerline{\includegraphics[scale=0.25]{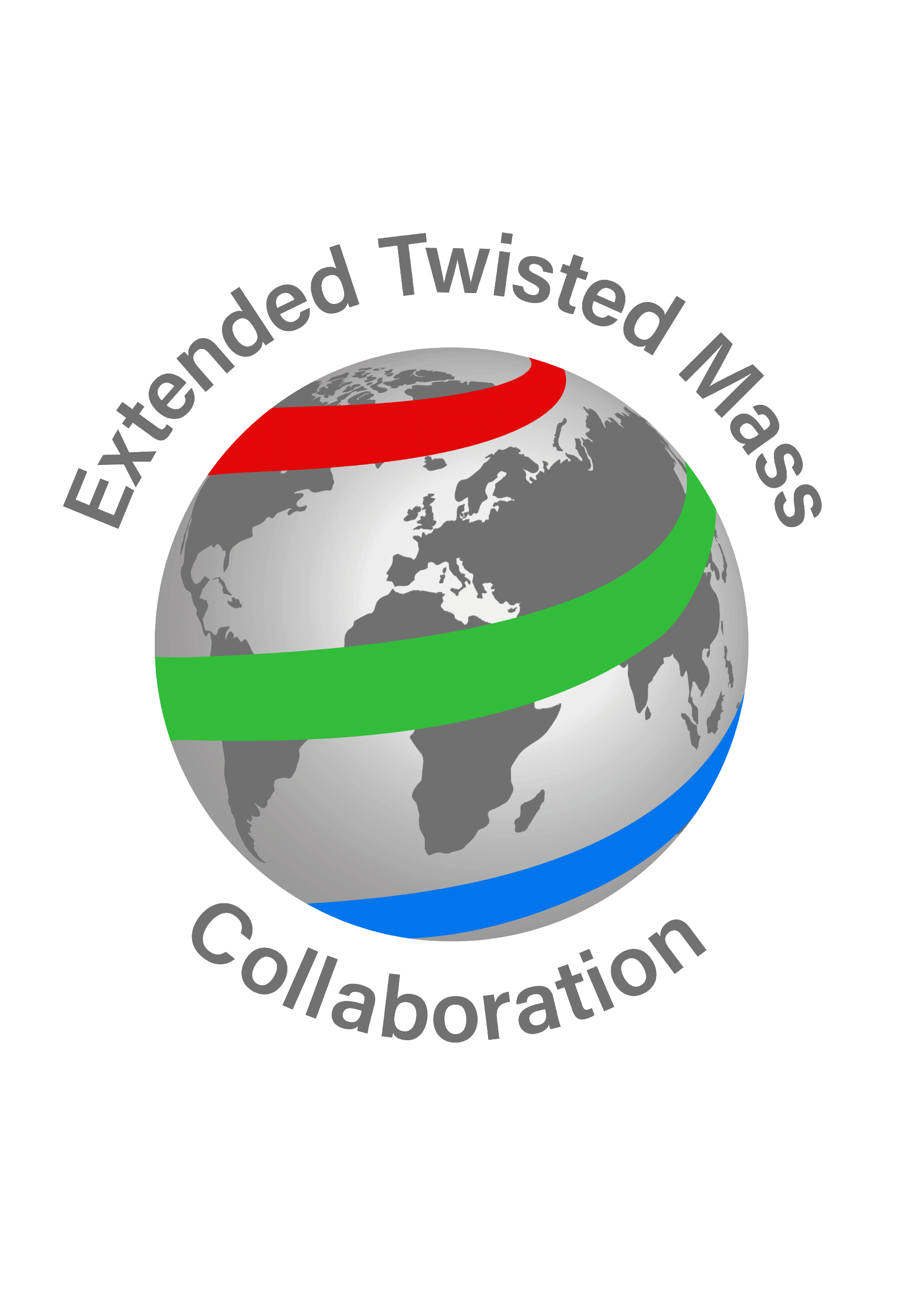}}
\vspace{0.5cm}
We present a calculation of the up, down, strange and charm quark masses performed within the lattice QCD framework. We use the twisted mass fermion action and carry out simulations that include in the sea two light mass-degenerate quarks, as well as the strange and charm quarks. In the analysis we use gauge ensembles simulated at three values of the lattice spacing and with light quarks that correspond to pion masses in the range from 350 MeV to the physical value, while the strange and charm quark masses are tuned approximately to their physical values. We use several quantities to set the scale in order to check for finite lattice spacing effects and in the continuum limit we get compatible results.
 The quark mass renormalization is carried out non-perturbatively using the \RIp-MOM method converted into the $\overline{\rm MS}$ scheme. For the determination of the quark masses we use physical observables from both the meson and the baryon sectors, obtaining $m_{ud} = 3.636(66)(^{+60}_{-57})$~MeV and $m_s = 98.7(2.4)(^{+4.0}_{-3.2})$~MeV in the $\overline{\rm MS}(2\,{\rm GeV})$ scheme and $m_c = 1036(17)(^{+15}_{-8})$~MeV in the $\overline{\rm MS}(3\,{\rm GeV})$ scheme, where the first errors are statistical and the second ones are combinations of systematic errors.
For the  quark mass ratios we get $m_s / m_{ud} = 27.17(32)(^{+56}_{-38})$ and $m_c / m_s = 11.48(12)(^{+25}_{-19})$. 
\end{abstract}

\maketitle

\section{Introduction}
 Quark masses  are essential inputs of the Standard Model (SM) and  play a primary role for the description of a large number of physical processes that can provide insights into the dynamics of the SM as well as in the search of beyond the Standard Model physics. The quark masses together with the strong coupling constant can be regarded as the fundamental parameters of Quantum chromodynamics (QCD), the renormalizable theory of the strong interactions. Therefore, their determination plays a crucial role in the phenomenological description of the plethora of complex phenomena governed by strong nuclear forces taking place in the universe as well as at particle colliders. Lattice QCD provides a non-perturbative approach based on first principles and systematically improvable for determining the quark masses and the strong coupling. In this approach the QCD Lagrangian is defined on a discrete Euclidean 4-dimensional space-time lattice of large but finite extent, which allows for numerical simulation of the theory via Monte Carlo methods. The finite volume and the non-vanishing lattice spacing introduce systematic artefacts, which can be theoretically understood, kept under numerical control and extrapolated away in order to extract the physical quantities of interest.
 
 Theoretical progress in lattice field theory and improvement in numerical algorithms, accompanied with a continuously increasing computational power, are allowing us to perform simulations using physical values of the light-quark masses. However, most of these simulations are still carried out using a single lattice spacing and volume, although this is rapidly changing as more lattice QCD collaborations gain access to larger computational resources and can produce multiple ensembles of gauge configurations generated with physical values of the light quark masses. Such ensembles will be referred to as physical point ensembles.
 In this work we include two physical point ensembles at two different lattice spacings. In order to take the continuum limit we also employ additional ensembles at a coarser lattice resolution with larger than physical pion masses. Globally we thus use ensembles with three values of the lattice spacing and spanning pion masses in the range from about 350 MeV to 135 MeV, which enable us to perform a combined chiral and continuum extrapolations. In order to study systematic effects in the determination of the quark masses, we use two sets of observables to set the scale and to evaluate the quark masses. One set of observables is based on quantities from the meson sector of QCD while the other set relies on baryonic observables. In the former case we use the pion mass and decay constant to set the scale and to determine the average up/down quark mass. The mass of the strange and charm quarks are extracted using the kaon and $D$-meson masses, respectively. 
In the latter case, instead, the masses of the pion and nucleon are employed to set the scale and fix the average up/down quark mass, while the $\Omega^-$ and the $\Lambda_c$ masses determine the strange and charm quark masses, respectively. In this way we obtain a valuable consistency check with respect to the results coming from the mesonic sector. 

For the renormalization of the quark mass we employ a dedicated set of gauge ensembles with four mass-degenerate sea quarks (having mass around half of the strange quark mass). Such a set ensures a good control of the extrapolation to the massless limit.
We perform the computation in an intermediate mass-independent scheme, which is finally converted to the standard $\overline{\rm MS}$ scheme.

The paper is organized as follows: In  Section~\ref{sec:methodology} we describe the gauge ensembles used in this study and explain our methodology. In Sections~\ref{sec:lattice spacing} and \ref{sec:Zp} we present the methods used to set the lattice spacing $a$ and to carry out a non-perturbative computation of the renormalization constant $Z_P$ including  a detailed discussion on the control of hadronic contaminations and other systematic errors. In Sections~\ref{sec:mesons} and \ref{sec:baryons} we describe the extraction of the quark masses and their ratios using inputs from the mesonic and baryonic sectors, respectively. In Section~\ref{sec:conclusions}, we discuss our final results and give our conclusions and outlooks.

\section{Methodology}
\label{sec:methodology}
In twisted-mass lattice QCD~\cite{Frezzotti:2000nk} the discretized Dirac operator in the {\em physical} quark basis is written as 
\begin{equation}
    D(\mu_f)  =  \frac{1}{2} \gamma_\mu  (\nabla_\mu + \nabla^*_\mu  ) - i \gamma _5 \left\{m_{\rm{cr}}(r_f)  - r_f \frac{a}{2} \nabla_\mu \nabla^*_\mu + \frac{c_{SW}(r_f)}{32} \gamma_\mu \gamma_\nu (Q_{\mu\nu}-Q_{\nu\mu}) \right\} +\mu _f ,
\label{eq:dirac}
\end{equation}
where $r_f = \pm 1$, $\nabla_\mu$ and $\nabla^*_\mu$ are nearest-neighbor forward and backward covariant derivatives, $\mu_f$ sets the mass of the quark field $q_f$ of flavor $f$, $c_{SW}$ is the coefficient of the clover-term $Q_{\mu\nu}$~\cite{Sheikholeslami:1985ij} and $m_{cr}$ is the critical value of the ``untwisted'' mass, $m_0$, obtained by requiring the vanishing of the partially conserved {\em axial} current (PCAC) mass, as discussed in Ref.~\cite{Baron:2010bv}.
This condition, referred to as maximal twist, guarantees automatic ${\cal{O}}(a)$-improvement of physical observables~\cite{Frezzotti:2003ni, Frezzotti:2004wz}.
In the twisted-mass fermion formulation at maximal twist, the renormalized quark masses are thus given by
\begin{equation}\label{eq:quark_mass}
    m_f = \frac{\mu_f}{Z_P} = \frac{(a \mu_f)}{a Z_P},
\end{equation}
where $a$ is the lattice spacing and $Z_P$ is the pseudo-scalar renormalization constant. Therefore, the determination of both $Z_P$ and the lattice spacing, combined with inputs from known physical quantities depending on the quark masses, enables us to extract $m_f$. Having gauge ensembles with at least three different lattice spacings at several pion masses allows us to take the continuum limit and by performing a chiral extrapolation we can determine the quark masses at the physical point. 

\subsection{Gauge ensembles}
We use the twisted-mass fermion discretization scheme~\cite{Frezzotti:2000nk,Frezzotti:2003ni} with the inclusion of a clover-term~\cite{Sheikholeslami:1985ij}.  
As already explained, twisted-mass fermions (TMF) provide an attractive formulation for lattice QCD simulations allowing for automatic ${\cal O}(a)$ improvement of physical observables as well as renormalization constants~\cite{Frezzotti:2003ni,Constantinou:2010gr}. This is an important property since quantities of interest have lattice artifacts of ${\cal O}(a^2)$ and are thus closer to the continuum limit. A clover-term is added to the TMF action to suppress $\mathcal{O}(a^2)$ breaking effects between the neutral and charged pions, which eventually leads to the stabilization of simulations with light quark masses close to the physical pion mass. For more details on the TMF formulation see Refs.~\cite{Frezzotti:2005gi,Boucaud:2008xu} and on the simulation and tuning strategies see Refs.~\cite{Abdel-Rehim:2015pwa,Alexandrou:2018egz}.

In this study we analyze ten gauge ensembles simulated at three values of the lattice spacing and at several values of the pion mass, spanning a range from the physical pion mass up to 350~MeV. Some parameters of these ensembles and the values of few key physical quantities are listed in Table~\ref{tab:params}.
More details are given in Ref.~\cite{Alexandrou:2021bfr}. With respect to Ref.~\cite{Alexandrou:2021bfr} the ensemble cC211.20.48 has been added in order to investigate the light quark mass dependence at the finest lattice spacing.

The ensembles are generated with two mass-degenerate light quarks and the strange and charm quarks in the sea ($N_f=2+1+1$ ensembles). The strange and charm {\em sea} quark mass parameters, $a\mu_\sigma$ and $a\mu_\delta$ (see, e.g., Eq.~(8) of Ref.~\cite{Carrasco:2014cwa}) have been adjusted so as to reproduce the phenomenological conditions $m_c/m_s \simeq 11.8$ and $m_{D_s}/f_{D_s} \simeq 7.9$~\cite{Aoki:2019cca}, which are easy to implement with few percent level precision even using simulations with larger than physical pion masses and on lattices of linear size $L \simeq 2.5$ fm, as detailed in Ref.~\cite{Alexandrou:2018egz}. The condition on $m_{D_s}/f_{D_s}$ is sensitive to the charm quark mass while the one on $m_c/m_s$ fixes the strange quark mass. In this way the charm and strange {\em sea} quark mass parameters have been tuned, separately for each lattice resolution (or $\beta$), to bare values that {\em a posteriori} turn out to yield values for the renormalized {\em sea} quark masses that are consistent within statistical errors of few percents with those we determine in this paper, as discussed in the following, at the physical pion mass point, on large volumes and in the continuum limit.

\begin{table}[htb!]
    \centering
\begin{tabular}{|lcccccccr|}
	\hline
	Ensemble       &  $L^3\times T$  & MDUs~ & $~a\mu_\ell$ &   $am_\pi $    &   $af_\pi$    & $~m_\pi L~$ &   $m_N/m_\pi$  & $m_\pi$ [MeV] \\ \hline
	   \multicolumn{9}{c}{$\beta=1.726$, $c_{SW}=1.74$, $a\mu_\sigma = 0.1408$, $a\mu_\delta=0.1521$, $w_0/a = 1.8352~(35)$}    \\ \hline
	cA211.53.24  & $24^3\times48$  & 5026 &   0.00530   & 0.16626 (51)  & 0.07106 (36) &   3.99    &          --      & 346.4 (1.6) \\
	cA211.40.24  & $24^3\times48$  & 5298 &   0.00400   & 0.14477 (70)  & 0.06809 (30) &   3.47    &          --      & 301.6 (2.1) \\
	cA211.30.32  & $32^3\times64$ & 10234  &   0.00300  & 0.12530 (16) & 0.06674 (15) &   4.01    &   4.049 (14) & 261.1 (1.1) \\
	cA211.12.48  & $48^3\times96$  & 2936  &   0.00120   & 0.08022 (18)& 0.06133 (33) &   3.85    &   5.685 (28) & 167.1 (0.8) \\ \hline
	\multicolumn{9}{c}{$\beta=1.778$, $c_{SW}=1.69$, $a\mu_\sigma = 0.1246864$, $a\mu_\delta=0.1315052$, $w_0/a = 2.1299~(16)$ } \\ \hline
	cB211.25.32  & $32^3\times64$  & 3959  &   0.00250   &  0.10475 (45) & 0.05652 (38) &   3.35    &   4.104 (36) & 253.3 (1.4) \\
	cB211.25.48  & $48^3\times96$  & 5246  &   0.00250   & 0.10465 (14) &  0.05726 (12) &   5.02    &   4.124 (17) & 253.0 (1.0) \\
	cB211.14.64  & $64^3\times128$  & 6187 &   0.00140   & 0.07848 (10) &  0.05477 (12) &   5.02    &   5.119 (36) & 189.8 (0.7) \\
	cB211.072.64 & $64^3\times128$  &  3161 &   0.00072   & 0.05659 (8) &  0.05267 (14) &   3.62    &   6.760 (30) & 136.8 (0.6) \\ \hline
	\multicolumn{9}{c}{$\beta=1.836$, $c_{SW}=1.6452$, $a\mu_\sigma = 0.106586$, $a\mu_\delta=0.107146$, $w_0/a = 2.5045~(17)$}  \\ \hline
	cC211.20.48  & $48^3\times96$  & 2000  &   0.00200   & 0.08540 (17) & 0.04892 (13) &   4.13    &   4.244 (25) & 245.73 (98)\\
	cC211.06.80  & $80^3\times160$  & 3207 &   0.00060   & 0.04720 (7) & 0.04504 (10) &   3.78    &   6.916 (19) & 134.3 (0.5) \\ \hline
\end{tabular}
    \caption{Parameters of the $N_f=2+1+1$ ensembles analyzed in this study. In the first column we give the name of the ensemble, in the second the lattice volume, in the third the number of molecular dynamics units simulated per ensemble, in the fourth the twisted-mass parameter, $a\mu_\ell$, for the average up/down (light) quark, in the fifth and in the sixth the pion mass $am_\pi$ and decay constant $af_\pi$ in lattice units from Ref.~\cite{Alexandrou:2021bfr}, in the seventh the pion mass times the lattice spatial length, $m_\pi L$, in the eighth the ratio $m_N / m_\pi$ as determined in Section~\ref{sec:baryons} and, finally, in the last column the pion mass in physical units, using our determination of the gradient-flow scale $w_0$ obtained in Ref.~\cite{Alexandrou:2021bfr} (see later Eq.~\eqref{eq:w0_intro}). We also include for each set of ensembles with the same lattice spacing the coupling constant $\beta$, the clover-term parameter $c_{SW}$, the parameters of the non-degenerate operator $a\mu_\sigma$ and $a\mu_\delta$, related to the renormalized strange and charm sea quark masses~\cite{Frezzotti:2004wz}, and the value of the gradient-flow scale $w_0/a$ determined at the physical pion mass in Ref.~\cite{Alexandrou:2021bfr}.
    }
    \label{tab:params}
\end{table}

\subsection{Osterwalder-Seiler fermions} 

A naive use of the twisted-mass action for non-degenerate strange and charm quarks would lead to an undesired O($a^2$) mixing of the strange and charm flavours in the correlation functions of interest to determine physical quantities~\cite{Frezzotti:2003xj,Baron:2010th}. 
In order to avoid such a mixing in the correlation functions, we adopt a non-unitary lattice setup~\cite{Frezzotti:2004wz} where the twisted-mass action for non-degenerate strange and charm quarks is employed only in the sea sector, while the valence strange and charm quarks that enter the correlation functions are regularized as exactly flavour-diagonal Osterwalder-Seiler fermions~\cite{Osterwalder:1977pc}. Thus, the valence action in the strange and charm sectors ($f = s, ~ c$) reads
 \begin{equation}
    S^f_{OS}(\mu_f)  = a^4 \sum\nolimits_x {\overline{q}_f (x) D(\mu_f) q_f (x)} ~ ,
    \label{eq:OS}
 \end{equation}
where $D(\mu_f)$ is the twisted mass Dirac operator in Eq.~\eqref{eq:dirac} with the same values of $m_0$ and $c_{SW}$ as in the sea sector action used for ensemble generations. As the renormalized strange and charm sea quark masses are matched with few percent relative accuracy to their valence counterparts, no significant unitarity violation is expected in our continuum limit results.

When constructing meson correlation functions, the Wilson parameters of the two valence quarks are always chosen to have opposite values.
This choice guarantees that squared pseudoscalar meson masses, generically indicated by $m_{PS}^2$, differ from their continuum counterparts only by terms of order ${\cal{O}}(a^2 \mu)$~\cite{Frezzotti:2003ni, Frezzotti:2005gi}. 
As we said above, in our lattice setup the (valence) flavour conservation is guaranteed in all correlation functions and automatic ${\cal O}(a)$-improvement is maintained. Of course 
we need to fix the valence strange and charm quark masses, $\mu_s$ and $\mu_c$, by imposing suitable renormalization conditions. For this purpose in the present work we use two different sets of observables. Namely, in one case we use the mass of the physical masses of kaon and $D$ (or $D_s$) mesons and in the other case the masses of the $\Omega$ and $\Lambda_c$ baryons. These two different choices will lead to two different determinations of the strange and charm quark masses, which will enable us to check the consistency of our results when using physical inputs from the mesonic and baryonic sectors.

\section{Scale setting} 
\label{sec:lattice spacing}
As already mentioned, we have ensembles at three different lattice spacings. We will refer to the ensembles in Table~\ref{tab:params} that start with $cA$ in their names as A ensembles, those starting with $cB$ as B ensembles and those with $cC$ as C ensembles (the label $c$ stands for {\em clover}). Each of these groups have the same lattice spacing, with the A ensembles having the largest lattice spacing and C ensembles the smallest one. 
In what follows we will use different quantities to determine the three lattice spacings. This will allow us to check consistency while taking the continuum limit when different inputs are used. In the pion sector, the pion mass and decay constant are used as input. Within this approach one also determines the value of the gradient-flow scale $w_0$. 
We use the iso-symmetric values of the pion  mass and decay constant, given respectively by~\cite{Aoki:2016frl},
\begin{equation}\label{eq:pi_isoqcd}
    m^{isoQCD}_\pi=  135.0(2)\text{~MeV}\quad\text{and}\quad f^{isoQCD}_\pi =  130.4 (2)\text{~MeV} ~ .
\end{equation}
We also compute the value of $w_0/a$ for each ensemble (see Table~\ref{tab:params}) and extrapolate to the physical pion mass and continuum limit. We find $w_0 = 0.17383(63)$ fm~\cite{Alexandrou:2021bfr} and using this value one determines the lattice spacings shown in Table~\ref{tab:scale}. We refer to this determination of the lattice spacings as coming from the ``pion'' sector. Details are given in Ref.~\cite{Alexandrou:2021bfr}. 

Another quantity used for the determination of the lattice spacings is the mass of the nucleon~\cite{Alexandrou:2017xwd,Alexandrou:2014sha}. Details on the extraction of the nucleon mass are given in Section~\ref{sec:baryons}. In order to fit the pion mass dependence of the nucleon mass, we use the well established SU(2) chiral perturbation theory result to one-loop~\cite{Gasser:1987rb,Tiburzi:2008bk}  
\begin{equation}\label{eq:nucl_fit}
m_N(m_\pi) = m^0_{N} - 4 c_1 m_\pi^2 - \frac{3g^2_A}{16\pi f_\pi^2}m_\pi^3 ~ . ~
\end{equation}
The three values of the lattice spacing, which will be denoted by $a_A$, $a_B$ and $a_C$, can be determined from the lattice data for the nucleon and pion masses by rewriting Eq.~(\ref{eq:nucl_fit}) as
\begin{equation}
 (a_im_N) = a_i m_N^0 - 4c_1 \frac{(a_im_\pi)^2}{a_i} - \frac{3g_A^2}{16\pi f_\pi^2} \frac{(a_im_\pi)^3}{a_i^2} ~ , ~
\end{equation}
where $(a_i m_N)$ and $(a_i m_\pi)$ are our lattice QCD results and $i = A, B, C$. 
The three quantities $a_i$ as well as the nucleon mass in the chiral limit, $m^0_N$, are treated as fitting parameters, while the value of $c_1$ is fixed by requiring the reproduction of the physical value of the nucleon mass, $m_N^{isoQCD}$, at the physical pion point (\ref{eq:pi_isoqcd}), namely 
\begin{equation}
\label{eq:c1}
c_1 = \left[ m_N^0 - \frac{3g_A^2}{16\pi f_\pi^2} \left( m_\pi^{isoQCD} \right)^3 - m_N^{isoQCD} \right] / \left[ 4 \left( m_\pi^{isoQCD} \right)^2 \right] ~ . ~
\end{equation}

We restrict ourselves to using ensembles for which the pion mass is below 260 MeV since chiral perturbation theory to higher orders has larger ambiguities. The simulations of the gauge ensembles use mass-degenerate up and down quarks and include no electromagnetic effects. Thus, we use the average value of the proton and neutron mass as our input for fixing the lattice spacings, namely we assume $m_N^{isoQCD} = 0.9389$~GeV in Eq.~(\ref{eq:c1}).
We also use the physical value for the axial charge, $g_A = 1.27641(56)$~\cite{Markisch:2018ndu} and for consistency the physical value of $f_\pi$ from Eq.~(\ref{eq:pi_isoqcd}). The ratio $g_A / f_\pi$ appears in the $m_\pi^3$ term and any residual correction due to strong isospin breaking and electromagnetism is neglected.

\begin{figure}[htb!]
    \centering
    \includegraphics[width=0.9\linewidth]{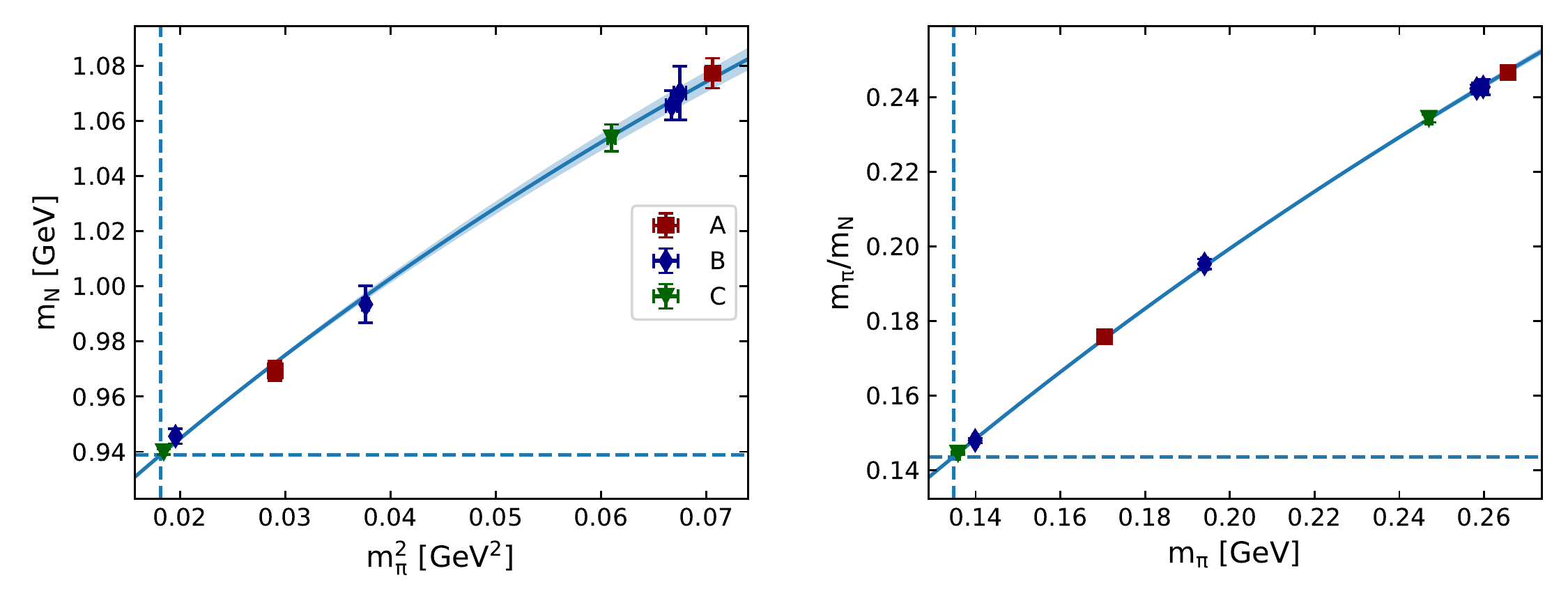}
    \vspace{-0.5cm}
    \caption{Determination of the lattice spacing from the nucleon mass. In the left panel we show the nucleon mass as a function of the pion mass $m_\pi$ squared. In the right panel we show the dimensionless quantity $m_\pi/m_N$ as a function of $m_\pi$, using the lattice spacing extracted from the nucleon mass. The values of $m_\pi/m_N$ are listed in Table~\ref{tab:params}, while $am_N$ and $am_\pi$ are determined in Section~\ref{sec:baryons}. The solid line shows the fit to the lattice QCD data using  Eq.~\eqref{eq:nucl_fit}. The value of $\chi^2/$d.o.f.~is 0.19, where the number of degrees of freedom is five.}
    \label{fig:nucl_pion}
\end{figure}

The result of the fit to the mass of the nucleon $m_N$ is depicted in Fig.~\ref{fig:nucl_pion} and describes very well the data, yielding  $\chi^2/$d.o.f.~= 0.19 where d.o.f.~are the number of degrees of freedom. We find $m^0_N = 0.8737(14)$~GeV, $c_1 = -1.090 (20)$~GeV$^{-1}$ and the values of the lattice spacings shown in the third row of Table~\ref{tab:scale}. In Fig.~\ref{fig:nucl_pion} we also show the ratio $m_N/m_\pi$ and the resulting fit using the parameters extracted from the fit to the nucleon mass. As it can be seen, the data for $m_N/m_\pi$ are well described. 

\begin{figure}[htb!]
\centering
\includegraphics[width=0.5\linewidth]{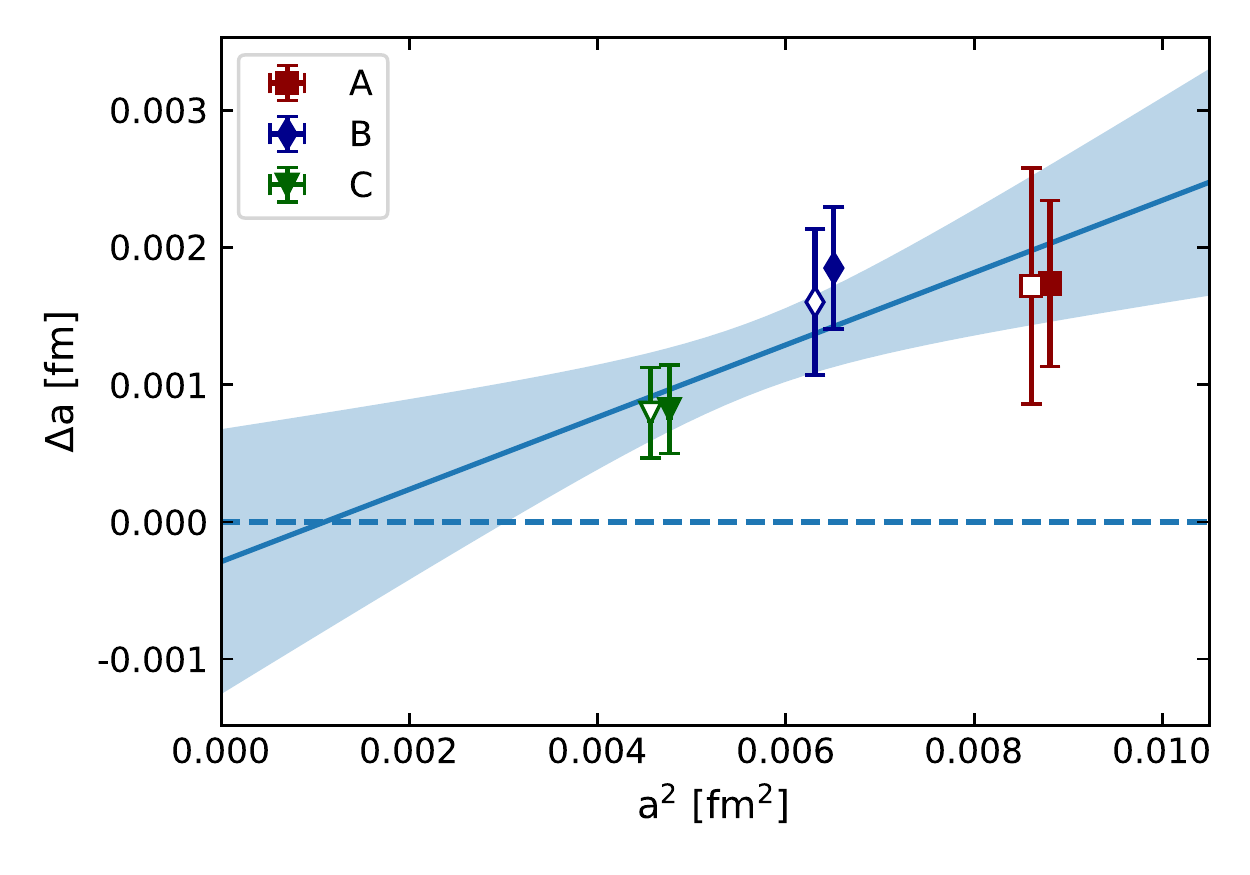}
   \vspace{-0.5cm}
   \caption{The difference $\Delta a$ between the lattice spacings determined from the pion sector and the nucleon mass versus  $a^2$. Full symbols are the lattice spacings determined using all the ensembles for which $m_\pi<260$~MeV. Open symbols, shifted to the left for clarity, are obtained using ensembles for which the pion mass is below 190~MeV. The solid line shows the linear fit in $a^2$ to the results extracted by using ensembles with $m_\pi<260$~MeV (full symbols), which is largely consistent with zero in the continuum limit.}\label{fig:scale}
\end{figure}

The values of the lattice spacing extracted from the pion sector and from the nucleon mass are shown in Table~\ref{tab:scale} and they differ by ${\cal O}(a^2)$ effects. Fitting their difference as a function of $a^2$ is shown in Fig.~\ref{fig:scale}. We observe that in the continuum limit the difference vanishes, as expected for our ${\cal O}(a)$-improved formalism. In what follows we will use both determinations to extract the quark masses. This provides a cross-check for our procedure and for the magnitude of any residual lattice spacing effect.

\begin{table}[htb!]
\centering
\vspace{10pt}
\centering
\begin{tabular}{|l|ccc|}
\hline
Sector & $a_A$ [fm] & $a_B$ [fm] & $a_C$ [fm] \\\hline\hline
Pion &$0.09471(39)$ & $0.08161(30)$ & $0.06942(26)$\\\hline
Nucleon &$0.09295(47)$ & $0.07975(32)$ & $0.06860 (20)$\\\hline\hline
$\Delta a$ &$0.00176(61)$ & $0.00186(44)$ & $0.00082(32)$\\\hline
\end{tabular}
 \caption{The values of the lattice spacing extracted from the pion sector (second row) and using the nucleon mass (third row) for the A-, B- and C-ensembles, denoted by $a_A$, $a_B$ and $a_C$, respectively. In the last row we show the difference $\Delta a$ between the lattice spacings determined using the nucleon mass and from the pion sector.}
    \label{tab:scale}
\end{table}

\section{Computation of $Z_P$} 
\label{sec:Zp}

In order to determine the renormalized quark masses it is crucial to perform an accurate evaluation of the mass renormalization factor $Z_m$, that in the maximally twisted-mass formulation is given by $Z_m = 1/ Z_P$ (see Eq.~(\ref{eq:quark_mass})).
For this reason the details of the procedure we have followed to compute $Z_P$ will be given in this Section.

For the calculation of $Z_P$ we employ the non-perturbative \RIp-MOM renormalization scheme~\cite{Martinelli:1994ty}, which is a mass-independent scheme since the renormalization constants are defined in the massless limit. The choice of this intermediate scheme is convenient in that the scale evolution for the renormalization constants of the operators with non-trivial anomalous dimension is controlled by the renormalized gauge coupling alone. This requires however simulations close enough to the chiral limit, which is not the case of the $N_f=2+1+1$ ensembles of Table~\ref{tab:params}, mainly due to the presence of the heavy sea charm quark. In order to safely take the chiral limit in the computation of the renormalization constants, we have thus separately produced gauge field configurations with four mass-degenerate quarks ($N_f = 4$) at the same value of the coupling $\beta$ and with the same clover term included in the fermionic action as for our A, B and C ensembles of Table~\ref{tab:params}. 
This ensures that in the chiral limit the same massless $N_f = 4$ QCD theory underlies both the ensembles used for computing hadronic observables and setting the energy scale (see Table~\ref{tab:params}) and the ensembles dedicated to the evaluation of the renormalization constants, about which details are given in Table~\ref{tab:Z_ensembles}. The four degenerate quarks are taken with masses from $\sim 8$ to $\sim 16$ times larger than the average up-down quark mass, which simplifies both the simulations and the tuning to maximal twist. The values of the critical mass $m_{cr}$ have been chosen in order to satisfy the maximal twist condition (which is convenient to reduce lattice artifacts) to a very good accuracy level, as it can be deduced from the smallness of the PCAC masses in Table~\ref{tab:Z_ensembles} (actually the tuning is even slightly better than the one corresponding to the $N_f=2+1+1$ ensembles of Table~\ref{tab:params}).

\begin{table}[htb!]
\begin{center}
\begin{tabular}{|ccc|ccc|ccc|}
\hline 
  & $\beta=1.726$ & & & $\beta=1.778$ & & & $\beta=1.836$ &  \\
  & $24^3 \times 48$ & & & $24^3 \times 48$ & & & $32^3 \times 64$ & \\
\hline \hline
$a \mu_{\rm sea}$  &   $a m_{PS}$ & $a m_{PCAC}$ & $a \mu_{\rm sea}$   & $a m_{PS}$ & $a m_{PCAC}$ & $a \mu_{\rm sea}$   & $a m_{PS}$ & $a m_{PCAC}$ \\
\hline
~ 0.0060 ~ & ~ 0.1689(15) ~ & -4.1(1.4)$\times 10^{-4}$ & ~ 0.0075 ~ & ~ 0.1748(15)~ & -2.3(0.8)$\times 10^{-5}$ & ~ 0.0050 ~ & ~ 0.1276(14) ~ & -4.3(3.1)$\times 10^{-5}$ \\ 0.0080 & 0.1905(11) & -4.3(1.1)$\times 10^{-5}$ & 0.0088 & 0.1871(18) & -8.6(8.0)$\times 10^{-5}$  & 0.0065 & 0.1447(14) &+5.9(2.1)$\times 10^{-5}$  \\ 
0.0100 & 0.2155(12) & +1.5(1.3)$\times 10^{-4}$ & 0.0100 & 0.2006(18) & -1.6(0.8)$\times 10^{-4}$ & 0.0080 & 0.1585(14) & +1.6(0.3)$\times 10^{-4}$ \\  
0.0115 & 0.2289(12) & +1.7(1.1)$\times 10^{-4}$ & 0.0115 & 0.2158(11) & +0.2(9.5)$\times 10^{-5}$ & 0.0095 & 0.1744(12) & +2.0(0.3)$\times 10^{-4}$ \\ 
\hline 
\end{tabular}
\caption{Parameters for the $N_f = 4$ ensembles used for the calculation of $Z_P$. By $am_{PS}$ we denote the pseudoscalar (non-singlet) meson mass and by $am_{PCAC}$ the PCAC untwisted mass (see Ref.~\cite{Baron:2010bv}), both given in lattice units. Note that on all the $N_f = 4$ ensembles, taking into account statistical errors, the values of $am_{PCAC}$ are typically 15 to 100 times smaller than the lowest twisted-mass values.} \label{tab:Z_ensembles}
\end{center}
\end{table}

In the \RIp-MOM scheme we obtain the renormalization constant of the (flavour non-singlet) pseudoscalar density operator $O_P = i \bar{q} \gamma_5 q$ with $r_q'=-r_q=-1$ and $\mu_{q'}=\mu_q$ via the following condition~\footnote{The relevant fermionic action density is $\bar{q} D(\mu_q) q + \bar{q'} D(\mu_{q'}) q' $ $= \bar{X} \left[ \gamma_\nu  (\nabla + \nabla^*)_\nu/2 + (m_{cr}(1) -a/2 \nabla^*_\nu \nabla_\nu) + i\gamma_5\tau^3\mu_q + \right.$ $\left. (c_{sw}(1)/32) \gamma_\mu\gamma_\nu (Q_{\mu\nu} - Q_{\nu\mu}) \right] X$, where $X = (\chi_q,\chi_{q'})^T$ is a valence quark field pair in the canonical quark basis for standard Wilson fermions (after which, as usual, the renormalization constants are named) and $q = \exp(i\gamma_5\pi/4) \chi_q$, $q' = \exp(-i\gamma_5\pi/4) \chi_{q'}$, $\bar q = \bar \chi_q  \exp(i\gamma_5\pi/4)$, $\bar q' =\bar \chi_{q'} \exp(-i\gamma_5\pi/4)$. Note in particular that $i \bar q \gamma_5 q' = i \bar\chi_q \gamma_5 \chi_{q'}$.}
\begin{equation}
Z_q^{-1}\,Z_P\,\frac{1}{12} {\rm Tr} \left[\gamma^5 {\cal V}_P(p) \right] \Bigr|_{p^2=\mu_0^2} = 1\,
\label{eq:RIMOM_condition}
\end{equation}
where ${\cal V}_P$ is the pseudoscalar vertex function between quark and antiquark states with momentum $p$ and $Z_q$ is the renormalization constant of the quark field, defined as
\begin{equation}
Z_q = -\frac{i}{12} {\rm Tr} \left[ \frac{ \slashed{p}}{p^2} \,S^{-1}_q(p)\right] \bigg|_{p^2=\mu_0^2} ~ . ~
\label{eq:Zq_RIp}
\end{equation}
Here $S_q$ is the quark propagator at momentum $p$, which is identified with the renormalization scale $\mu_0$.
In this work we adopt the alternative definition of $Z_q$ first proposed in Ref.~\cite{Constantinou:2010gr},
\begin{equation}
Z_q = -\frac{i}{12 N_p} {\sum_\mu}^{\,\large\prime} \,  {\rm Tr} \left[\frac{\gamma_\mu}{\tilde{p}_\mu}\,S_q^{-1}(p)\right]\bigg|_{p^2=\mu_0^2}~, \label{eq:Zq_alt}
\end{equation}
 where the sum $\sum^\prime$ is over the $N_p$ non-vanishing components of the lattice momentum $a \tilde{p}_\mu=\sin(a p_\mu)$. The prescription of Eq.~\eqref{eq:Zq_alt}, unlike the naive \RIp-MOM definition~\eqref{eq:Zq_RIp}, has no lattice artifacts at tree-level and beyond tree-level it exhibits quite small $\mathcal{O}(a^2)$ cutoff effects.

The subtraction of the Goldstone pole in the vertex function $\mathcal{V}_P$ requires a good control of the vertex mass dependence. Therefore, we find it more suitable to adopt a partially quenched (PQ) setup, in which propagators and vertices are computed for multiple values of valence quark masses $\mu_\mathrm{val}$ at fixed sea mass $\mu_\mathrm{sea}$, namely we use
\begin{eqnarray}
a \mu_\mathrm{val}^\mathrm{A} &=& \{0.0060, 0.0080, 0.0100, 0.0115, 0.0130, 0.0150, 0.0170, 0.0190, 0.0210\}~,\nonumber\\
a \mu_\mathrm{val}^\mathrm{B} &=& \{0.0050, 0.0060, 0.0075, 0.0090, 0.0100, 0.0110, 0.0130, 0.0150, 0.0170\}~,\\
a \mu_\mathrm{val}^\mathrm{C} &=& \{0.0040, 0.0050, 0.0065, 0.0080, 0.0095, 0.0110, 0.0125, 0.0140, 0.0155\}~.\nonumber
\end{eqnarray}
The chiral extrapolation of the pseudoscalar vertex is discussed in details in Sec.~\ref{sec:GBpole} below.

 For each ensemble we employ a large number of momenta $a p =2\pi\left(\frac{n_t}{T},\frac{n_x}{L},\frac{n_y}{L},\frac{n_z}{L} \right)$, in the range $(a p)^2 \in [0.24,6.69]$ for the $24^3\times48$  ensembles and $(a p)^2 \in [0.13,5.23]$ for the $32^3\times64$ ensembles. Quark propagators and vertices in momentum space are evaluated using $N_\mathrm{cfg}=200$ gauge configurations. In order to reduce the effect of Lorentz non-invariant cut-off effects, we filter the momenta selecting the ones that are isotropic (``democratic") in the spatial directions, thus satisfying:
\begin{equation}
\frac{\sum_\mu \tilde{p}_\mu^4}{(\sum_\mu \tilde{p}_\mu^2)^2} < 0.28\,.   
\end{equation}
The above constraint ensures that unwanted hypercubic lattice artifacts are suppressed~\cite{Constantinou:2010gr}. We further improve the $Z_P$ estimator by using results from lattice perturbation theory (for more details see, e.g., Refs.~\cite{Constantinou:2014fka,Alexandrou:2015sea}). In summary, we calculate the lattice artifacts at one-loop level and to all orders in the lattice spacing, ${\cal O}(g^2 a^\infty)$. The perturbative corrections to the Green function of the pseudoscalar operator, as well as to the quark propagator, are evaluated for each momentum $a p$ at which the renormalization constants are computed. It should be noted that each value of $a p$ requires a separate calculation of the ${\cal O}(g^2 a^\infty)$ correction, as the perturbative contributions are not analytical and require the numerical evaluation of one-loop integrals. Such contributions also include the leading order terms, ${\cal O}(g^2 a^0)$, that have to be separated from the pure $\mathcal{O}(a^n)$ terms ($n\ne0$). The computation of perturbative lattice artifacts to $\mathcal{O}(g^2a^\infty)$ done in Ref.~\cite{Alexandrou:2015sea} are adapted for the case of the specific definition of $Z_q$ in Eq.~\eqref{eq:Zq_alt}.   Thus, the one-loop perturbative corrections that we use are defined as follows
\begin{eqnarray}
\Delta Z_q^{(1)}(ap) &=& Z^{(1)}_q(\log(ap),ap) - Z^{(1)}_q(\log(ap),0)\,, \\
\Delta \mathcal{V}_P^{(1)}(ap) &=& {\cal V}^{(1)}_P(\log(ap),ap) - {\cal V}^{(1)}_P(\log(ap),0)\,,
\end{eqnarray}
where $Z_q^{(1)}$ and $\mathcal{V}_P^{(1)}$ are the one-loop contributions to the quark field renormalization constant and to the pseudoscalar vertex, respectively.

\noindent Using the above quantities, we extract improved non-perturbative estimates for $Z_q$ and $Z_P$ by modifying the renormalization conditions as:
\begin{eqnarray}
&Z^{\rm impr}_q = Z_q - g^2 \,\Delta Z_q^{(1)}(a\mu_0)\,,&\\
&(Z^{\rm impr}_q)^{-1}\,Z^{\rm impr}_P\,\frac{1}{12} {\rm Tr} \left[\gamma^5 ({\cal V}_P - g^2 \, \Delta \mathcal{V}_P^{(1)}(ap)) \right] \Bigr|_{p^2=\mu_0^2} = 1\, . &   
\label{eq:ZPimpr}
\end{eqnarray}

\subsection{Analysis method and safety checks against hadronic contaminations}
\label{sec:ZP_analysis}

\begin{figure}[htb!]
\centering
\begin{minipage}{0.495\textwidth}
\includegraphics[width=1.\textwidth]{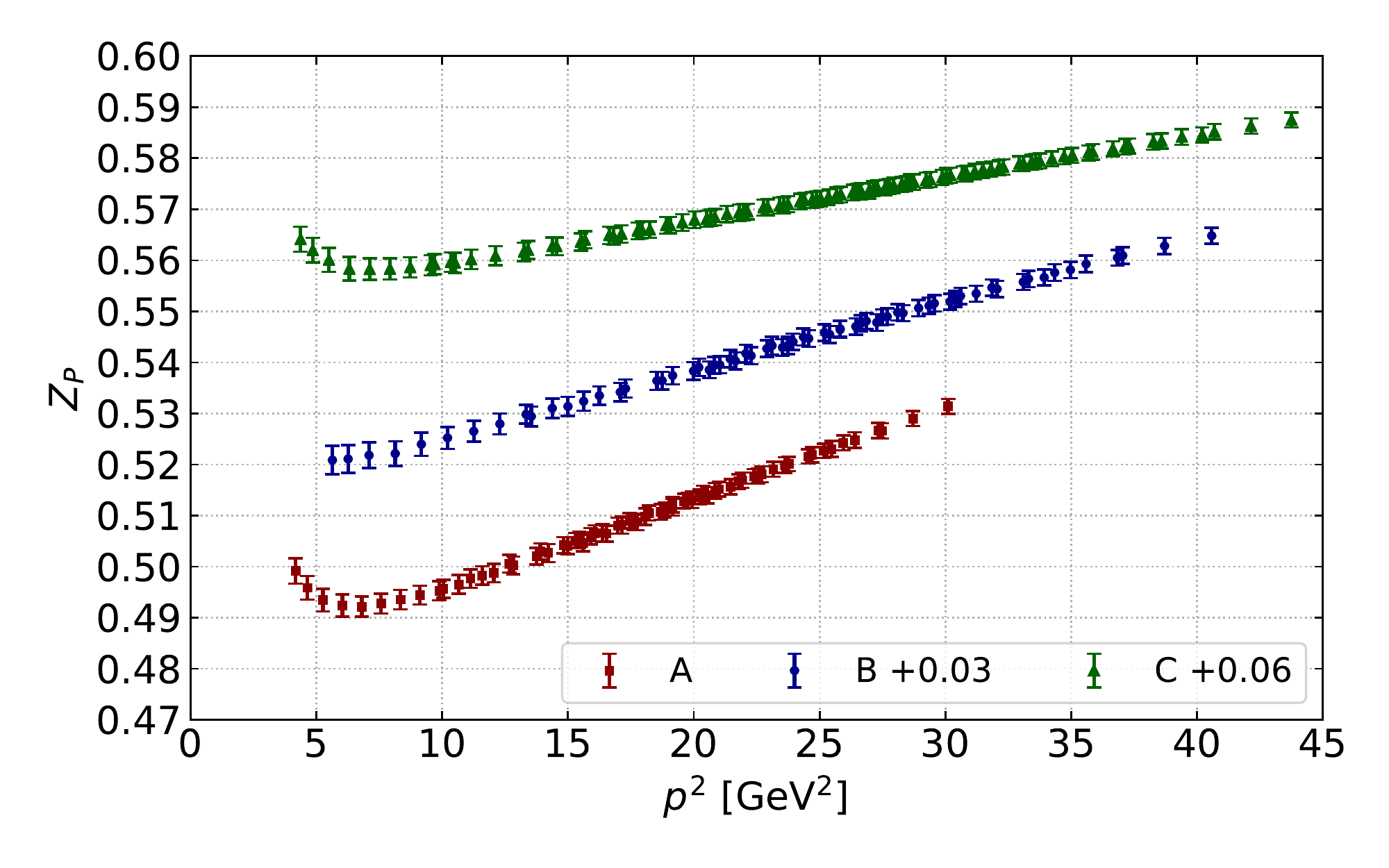}
\end{minipage}
\begin{minipage}{0.495\textwidth}
\includegraphics[width=1.\textwidth]{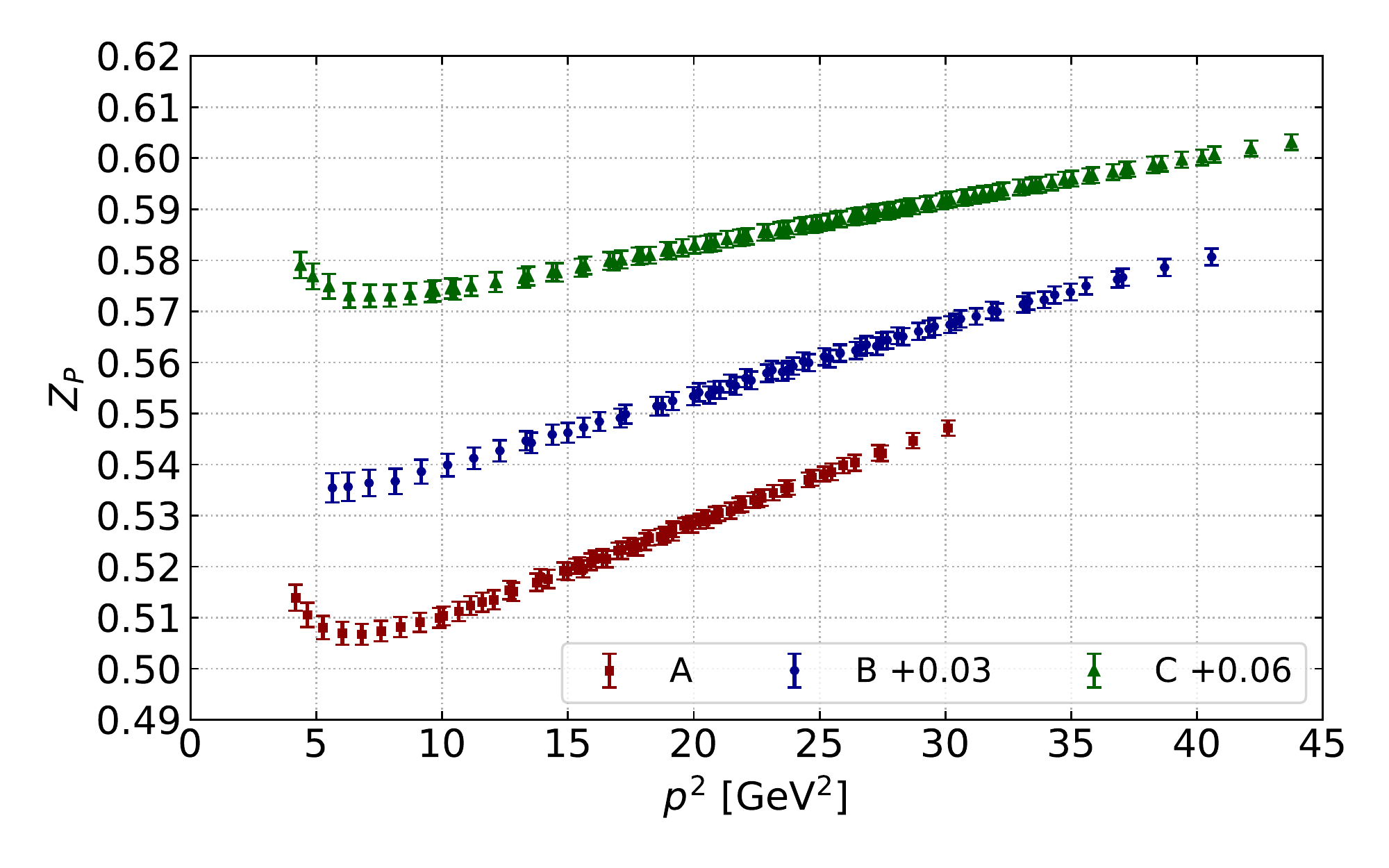}
\end{minipage}
\vspace{-0.5cm}
\caption{Dependence of $Z_P$ on the momentum $p^2$ for two values of the reference scale, $\mu_\mathrm{ref}^2=17$ GeV$^2$ (left panel) and $\mu_\mathrm{ref}^2=21$ GeV$^2$ (right panel). The vertical offsets shown in the inset are applied to the data of the corresponding ensembles. \label{fig:ZP_1721}}
\end{figure}

The improved $Z_P$ estimators, namely $Z_P^\mathrm{impr}$ obtained from Eq.~\eqref{eq:ZPimpr}, are evaluated at different values of  $p^2=\mu_0^2$ and extrapolated to the chiral limit, a step  which is discussed in detail in the following Section~\ref{sec:GBpole}.
Then, the chirally extrapolated lattice estimators of $Z_P$ are evolved to a common reference scale $p^2=\mu_\mathrm{ref}^2$ in the \RIp-MOM scheme using the anomalous dimension known up to N$^3$LO according to Ref.~\cite{Chetyrkin:1999pq} and adopting $\Lambda_{QCD}(N_f=4) = 294~(12)$ MeV from Ref.~\cite{Aoki:2019cca}.
The curves obtained at the three $\beta$ values are reported in Fig.~\ref{fig:ZP_1721} for two different values of the reference scale, namely $\mu_\mathrm{ref}^2=17$ GeV$^2$ and $\mu_\mathrm{ref}^2=21$ GeV$^2$. 

A dependence on $p^2$ of the $Z_P(\mu^2_\mathrm{ref})$ estimators in Fig.~\ref{fig:ZP_1721} is expected due to lattice artifacts, i.e. $\mathcal{O}(a^2p^2)$ terms, and possibly also to residual $\mathcal{O}(a^0)$ hadronic contaminations~\footnote{This point is discussed e.g. in the Appendix of Ref.~\cite{Martinelli:1994ty}.}, which, however, must vanish as $1/p^2$ at large $p^2$. The plots of $Z_P(\mu^2_\mathrm{ref})$ in Fig.~\ref{fig:ZP_1721} show a very good linearity in $p^2$  within the range $p^2\in[15,24]$ GeV$^2$, which is the one  relevant for our determination of $Z_P$ and hence of the renormalized quark masses. This fact indicates that lattice artifacts other than $a^2p^2$-terms and possible hadronic contamination effects are negligible within our small statistical errors. Such a property is explicitly checked at each $\beta$ value by performing a fit of $Z_P(\mu^2_\mathrm{ref}=17$ GeV$^2)$ (left panel of Fig.~\ref{fig:ZP_1721}) with the two Ans\"atze given by
\begin{equation}
\text{Ansatz 1:}\quad Z_P(\mu^2_\mathrm{ref}) = z_0 + z_1 p^2 + z_2 (p^2)^2~, \qquad
\text{Ansatz 2:}\quad Z_P(\mu^2_\mathrm{ref}) = z_0 + z_1 p^2 + \frac{z_{-1}}{p^2}~.
\label{eq:ZP_Ansatz12}
\end{equation}
The resulting best fit values for $z_i$ are given in Table~\ref{tab:ZPlinp2}. The coefficients $z_{2}$ and $z_{-1}$ are compatible with zero within statistical errors at all $\beta$ values, while the coefficients $z_1$ scale nicely with $a^2$ (see Tab.~\ref{tab:scale}). From this check we see that the systematic uncertainties on $Z_P$ are negligible within our small statistical errors. We will comment on the value of $z_0$ when we present our results in Sec.~\ref{sec:ZPresults}. 

\begin{table}[htb!]
\begin{tabular}{|c|c|c|c|c|c|c|}
\hline 
$\beta$ &  Ansatz & $z_{-1}$ & $z_0$ & $z_1$ & $z_2$ & $\chi^2$/d.o.f. \\
\hline
\multirow{2}{*}{1.726} & 1 &            & 0.4762(28) & 0.00190(23) & -0.00021(59) & 0.4  \\
 & 2  & -0.011(43) & 0.4782(52) & 0.00178(13) &              & 0.4  \\
\hline
\multirow{2}{*}{1.778} & 1  &            & 0.4828(39) & 0.00117(25) &  0.00053(56) & 0.3  \\
& 2  &  0.033(42) & 0.4772(40) & 0.00147(10) &              & 0.3  \\
\hline
\multirow{2}{*}{1.836} & 1  &            & 0.4888(29) & 0.00101(19) & -0.00032(43) & 0.03  \\
& 2 & -0.021(30) & 0.4922(34) & 0.00083(08) &              & 0.04  \\
\hline
\end{tabular}
\caption{Results of the two fits on $Z_P(\mu^2_\mathrm{ref}=17$ GeV$^2)$, according to the fit ansatz in Eq.~\eqref{eq:ZP_Ansatz12}.}
\label{tab:ZPlinp2}
\end{table}

Within the present study of renormalized quark masses, we follow two different methods for determining $Z_P$ in the \RIp~scheme and use data in two different $p^2$-ranges. The first method (M1) consists in fitting the $Z_P(\mu^2_\mathrm{ref})$ data linearly in $p^2$ in a given $p^2$-range  with the aim of removing ${\cal{O}}(a^2p^2)$ discretization effects, while in the second method (M2) the same data are fitted to a constant~\cite{Carrasco:2014cwa}. Method M2 is by construction much less sensitive than M1 to possible small residual hadronic contaminations but at the expense of leaving some  $\mathcal{O}(a^2)$ artifacts in the determination of $Z_P$. The ranges of $p^2$ used in the present analysis are $p^2\in[15,19]$ GeV$^2$ and $p^2\in[18,24]$ GeV$^2$ for determining $Z_P(17$ GeV$^2)$ and $Z_P(21$ GeV$^2)$, respectively. 

\begin{figure}[htb!]
\includegraphics[width=0.55\textwidth]{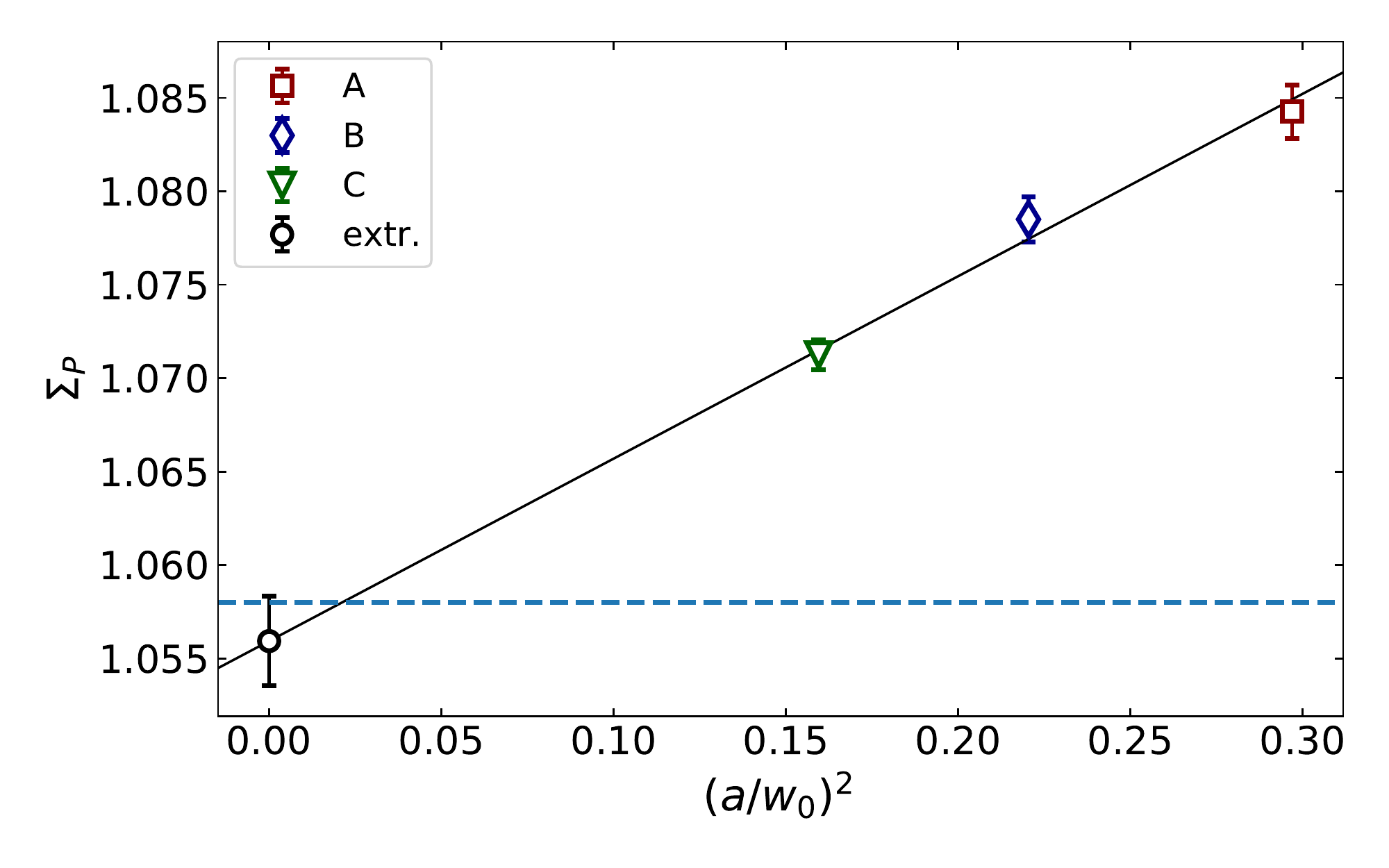}
\vspace{-0.5cm}
\caption{The scaling function $\Sigma_P(\mu_A^2,\mu_B^2)=Z_P(\mu_A^2)/Z_P(\mu_B^2)$ versus $(a/w_0)^2$ for the three $\beta$-values studied (red square for $\beta=1.726$, blue rhombus for $\beta=1.778$ and green triangle for $\beta=1.836$) as well as the continuum extrapolation (black circle). The dashed line shows the perturbative value, $\Sigma_P^\mathrm{pt}(\mu_A^2,\mu_B^2)=1.058$. \label{fig:ZP_stepwise}}
\end{figure}

As an additional check of our determination of $Z_P(\mu^2)$ in the \RIp-MOM scheme, we show in Fig.~\ref{fig:ZP_stepwise}  the results  for the non-perturbative step scaling function $\Sigma_P(\mu_A^2,\mu_B^2)=Z_P(\mu_A^2)/Z_P(\mu_B^2)$ versus $(a/w_0)^2$ for $(\mu_A^2,\mu_B^2)=(21.5,14.3)$ GeV$^2$. We see that the lattice QCD data exhibit small discretization errors and agree in the continuum limit with the perturbative counterpart $\Sigma_P^\mathrm{pt}(\mu_A^2,\mu_B^2)=1.058$, which is computed to N$^3$LO~\cite{Chetyrkin:1999pq}.

Moreover, as will be shown in Section~\ref{sec:mesons},  using the four $Z_P$ determinations corresponding to the methods M1 and M2 and at the two reference scales $\mu_\mathrm{ref}^2={17}$~GeV$^2$ and $\mu_{\rm ref}=21$~GeV$^2$ we obtain in the continuum limit consistent final results for the renormalized quark masses.

\subsection{Chiral extrapolation and Goldstone boson pole subtraction}
\label{sec:GBpole}

A crucial step in determining the renormalization constant $Z_P$ is the extrapolation of its lattice estimators to the chiral limit, where the mass-independent \RIp-MOM scheme is defined.

\subsubsection{Hadronic contaminations in the pseudoscalar vertex}

It is well-known that the pseudoscalar vertex $\mathcal{V}_P$ receives contributions at the non-perturbative level by hadronic contaminations whose leading term scales as ${\sim (p^2 m_\pi^2)^{-1}}$~\cite{Martinelli:1994ty}. Such Goldstone boson pole has to be identified and subtracted from the data.
In a unitary lattice setup for QCD with $N_f=4$ degenerate flavours of mass $m_q$, the lattice estimator of the vertex, $v_P(p^2, m_q)$, is expected to have the form
\begin{equation}
    \label{eq:nuP}
    v_P(p^2, m_q)\ =\ \mathcal{V}_P(p^2) + {h} \, \frac{\Lambda_\mathrm{QCD}^4}{p^2 m_\pi^2} + {h^\prime} \, \frac{\Lambda_\mathrm{QCD}^2}{p^2} + 
    {h^{\prime\prime}} \, m_q \frac{\Lambda_\mathrm{QCD}}{p^2}
    + \dots\,,
\end{equation}
where the dimensionless quantities $\mathcal{V}_P$ (our target vertex), ${h}$, $h^\prime$ and $h^{\prime\prime}$
depend in general on $a^2 p^2$, $a^2 \Lambda_\mathrm{QCD}^2$, $a^2 m_q \Lambda_\mathrm{QCD}$ and $a^2m_q^2$, while the ellipses stand for terms suppressed by higher powers of $1/p^2$ as $p^2\to\infty$. We note that terms linear in $m_q$ are either hadronic contaminations suppressed as $\sim 1/p^2$ at large $p^2$ or lattice artifacts of the form $\sim a^2m_q\Lambda_\mathrm{QCD}$, which are numerically tiny for the $am_q$ values of interest here.
Since close to the chiral limit $m_\pi^2\sim m_q$, an equivalent Ansatz for $v_P(p^2,m_q)$ can be written in the form~\footnote{Due to NLO terms in the chiral expansion of $m_\pi^2$, here $\kappa^\prime$ does not coincide with $h^\prime \Lambda_\mathrm{QCD}^2/p^2$.} \begin{equation}
    v_P(p^2,m_q)\ =\ \mathcal{V}_P(p^2) + \frac{1}{p^2}\left({\kappa}\, m_q^{-1} + \kappa' + \kappa'' \,m_q\right)+\, \dots\,,
    \label{eq:ansatz_unitary}
\end{equation}
where we separate the hadronic contaminations decreasing, for large $p^2$, like $1/p^2$ from the vertex of interest $\mathcal{V}_P(p^2)$.

\subsubsection{Choice of a partially quenched setup}
In a PQ setup, such as the one adopted in the present analysis (see Sec.~\ref{sec:ZP_analysis}), the lattice action is power counting renormalizable and the operator vertices evaluated at several values of valence ($\mu_\mathrm{val}$) and sea ($\mu_\mathrm{sea}$) quark masses  approach, as $(\mu_\mathrm{val},\mu_\mathrm{sea})\to(0,0)$, the corresponding unitary vertices from which the \RIp-MOM renormalization constants can be computed. As detailed in Sec.~\ref{sec:ZP_analysis} above, at all $\beta$ values we use nine  values of $\mu_\mathrm{val}$ for each of the four $\mu_\mathrm{sea}$ values. This allows us to have a good control on the mass dependence of the pseudoscalar vertex and to adopt, at fixed $\beta$, $p^2$ and $\mu_\mathrm{sea}$ values, the following fit Ansatz for the chiral fit in $\mu_\mathrm{val}$ 
\begin{equation}
    v_P(p^2,\mu_\mathrm{val},\mu_\mathrm{sea})\ =\ \mathcal{V}_P(p^2,\mu_\mathrm{sea}) + {H} \, \frac{\Lambda_\mathrm{QCD}^4}{p^2 [m_\pi^2]_\mathrm{val}} + {H^\prime} \, \frac{\Lambda_\mathrm{QCD}^2}{p^2} + 
    {H^{\prime\prime}} \, \mu_\mathrm{val} \frac{\Lambda_\mathrm{QCD}}{p^2}
    + \dots\,,
\end{equation}
where the quantities $\mathcal{V}_P(p^2,\mu_\mathrm{sea})$, ${H}$, $H^\prime$ and $H^{\prime\prime}$ depend, besides on $\mu_\mathrm{sea}$ (to a numerically negligible level, as we shall see below), also on $a^2 p^2$, $a^2 \Lambda_\mathrm{QCD}^2$, $a^2 m_q \Lambda_\mathrm{QCD}$ and $a^2m_q^2$, while the ellipses have the same meaning as in Eq.~\eqref{eq:nuP}. 
Noting that the hadronic contaminations in the three-point correlation function of quark, pseudoscalar bilinear and antiquark fields at fixed four-momenta (and in the derived quantity $v_P$) arise from the time orderings where the quark and antiquark fields are located at time distances both before or both after the pseudoscalar density~\cite{Martinelli:1994ty}, it follows that the Goldstone boson pole contamination is controlled by the mass $[m_\pi^2]_{\rm val}$ of the valence pion that appears as an intermediate state in the aforementioned time orderings.
Recalling also that, to leading order in PQ chiral perturbation theory, $[m_\pi^2]_\mathrm{val}\sim \mu_\mathrm{val}$~\cite{Sharpe:1997by}, we choose to use the equivalent Ansatz
\begin{eqnarray}
v_P(p^2,\mu_\mathrm{val},\mu_\mathrm{sea})\ =\ \mathcal{V}_P(p^2,\mu_\mathrm{sea}) + \frac{K}{p^2}\,\frac{1}{\mu_\mathrm{val}} + \frac{K^\prime}{p^2} + \frac{K^{\prime\prime}}{p^2}\, \mu_\mathrm{val} \, +\dots\,,
\label{eq:ansatz_PQ}
\end{eqnarray}
where again we separate the hadronic contaminations (suppressed like $1/p^2$ as $p^2\to\infty$) from the vertex of interest and the dimensionful coefficients $K$, $K^\prime$ and $K^{\prime\prime}$ in general depend on $\mu_\mathrm{sea}$ and may be affected by lattice artifacts. 
From the fit of $v_P$ in $\mu_\mathrm{val}$ at fixed $p^2$ and $\mu_\mathrm{sea}$, one can determine the coefficients $K$ and $K^{\prime\prime}$, but it is not possible to disentangle the vertex $\mathcal{V}_P(p^2,\mu_\mathrm{sea})$ from the hadronic contamination $K^\prime/p^2$. 
At this point one can safely take the limit $\mu_\mathrm{sea}\to 0$ and determine $\mathcal{V}_P(p^2,0)+K^\prime|_{\mu_\mathrm{sea}=0}/p^2$. 
After taking the full chiral limit, one can check that the residual hadronic contamination $K^\prime|_{\mu_\mathrm{sea}=0}/p^2$ 
is completely negligible in the range of $p^2$ used for the extraction of $Z_P$, as already detailed in Section~\ref{sec:ZP_analysis} (see the discussion around Tab.~\ref{tab:ZPlinp2}).

\subsubsection{Numerical data and intermediate analysis results}
\begin{figure}[htb!]
\centering
\includegraphics[width=.5\textwidth]{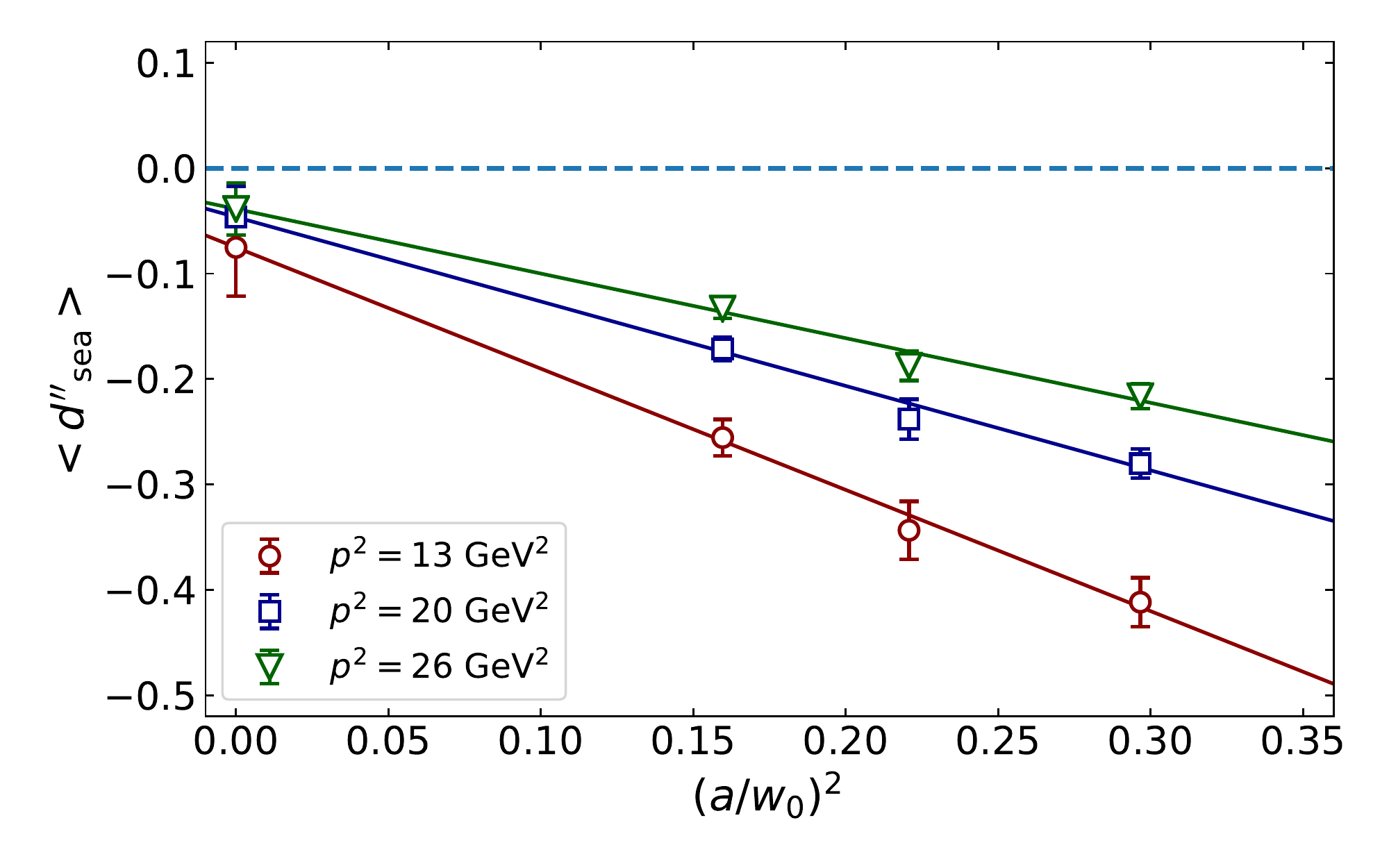}
\vspace{-0.5cm}
\caption{Scaling of the coefficient $d^{\prime\prime}\equiv K^{\prime\prime}/(w_0 p^2)$ averaged over $\mu_\mathrm{sea}$, $\langle d^{\prime\prime}\rangle_\mathrm{sea}$. The three curves correspond to different values of $p^2$, namely 13, 20 and 26 GeV$^2$. \label{fig:dPval}}
\end{figure}

Within the $p^2$ ranges used in the present analysis, our data for ${v}_P(p^2,\mu_\mathrm{val},\mu_\mathrm{sea})$ exhibit a tiny linear dependence on $\mu_\mathrm{val}$, which is compatible with $K^{\prime\prime}=\mathcal{O}(a^2)$ up to statistical errors. This is checked by fitting the data to the Ansatz of Eq.~\eqref{eq:ansatz_PQ} for each fixed $a\mu_\mathrm{sea}$ and $p^2$, which determines the quantities 
$[\mathcal{V}_P(p^2,\mu_\mathrm{sea}) +  {K^\prime}/{p^2}]$, $K/p^2$ and $K^{\prime\prime}/p^2$, and then studying the dimensionless ratio $d^{\prime\prime}\equiv K^{\prime\prime}/(w_0 p^2)$ as a function of $(a/w_0)^2$. At fixed $\beta$ value, we observe that $d^{\prime\prime}$ changes with the sea quark mass of a given gauge ensemble non monotonically in $\mu_\mathrm{sea}$ by the same amount as the statistical errors. Therefore, we average over the values of $d^{\prime\prime}$ at each $\mu_\mathrm{sea}$ for a fixed $\beta$ value,  obtaining the quantity $\langle d^{\prime\prime}\rangle_\mathrm{sea}$, which is shown in Fig.~\ref{fig:dPval} for three representative $p^2$ values, namely 13, 20 and 26~GeV$^2$. As one can see, the continuum limit of $d^{\prime\prime}$ is consistent with zero up to statistical errors and/or a small residual term, which even if present (given the numerical values of $a\mu_\mathrm{val}\lesssim0.02$) would alter $Z_P$ only to  ${\cal O}(10^{-4})$.
In view of the evidence that $K^{\prime\prime}$ is $\mathcal{O}(a^2)$ or numerically negligible in the $p^2$ range of interest here, we can perform the fit in $\mu_\mathrm{val}$ on the data for $v_P$ excluding the term linear in $\mu_\mathrm{val}$, namely we use
\begin{equation}
    v_P(p^2,\mu_\mathrm{val},\mu_\mathrm{sea})\ =\ \left[\mathcal{V}_P(p^2,\mu_\mathrm{sea}) +\frac{K'}{p^2}  \right] + \frac{K}{p^2}\,\frac{1}{\mu_\mathrm{val}}~.
    \label{eq:vP_valence_ansatz}
\end{equation}
This procedure has the advantage of yielding small statistical errors at the  price of including well-controlled $\mathcal{O}(a^2)$ artifacts in the numerical estimate of $[\mathcal{V}_P(p^2,\mu_\mathrm{sea}) +  {K^\prime}/{p^2}]$ and hence of $Z_P$. 
\begin{figure}[htb!]
\centering
\begin{minipage}{0.495\textwidth}
\includegraphics[width=1.\textwidth]{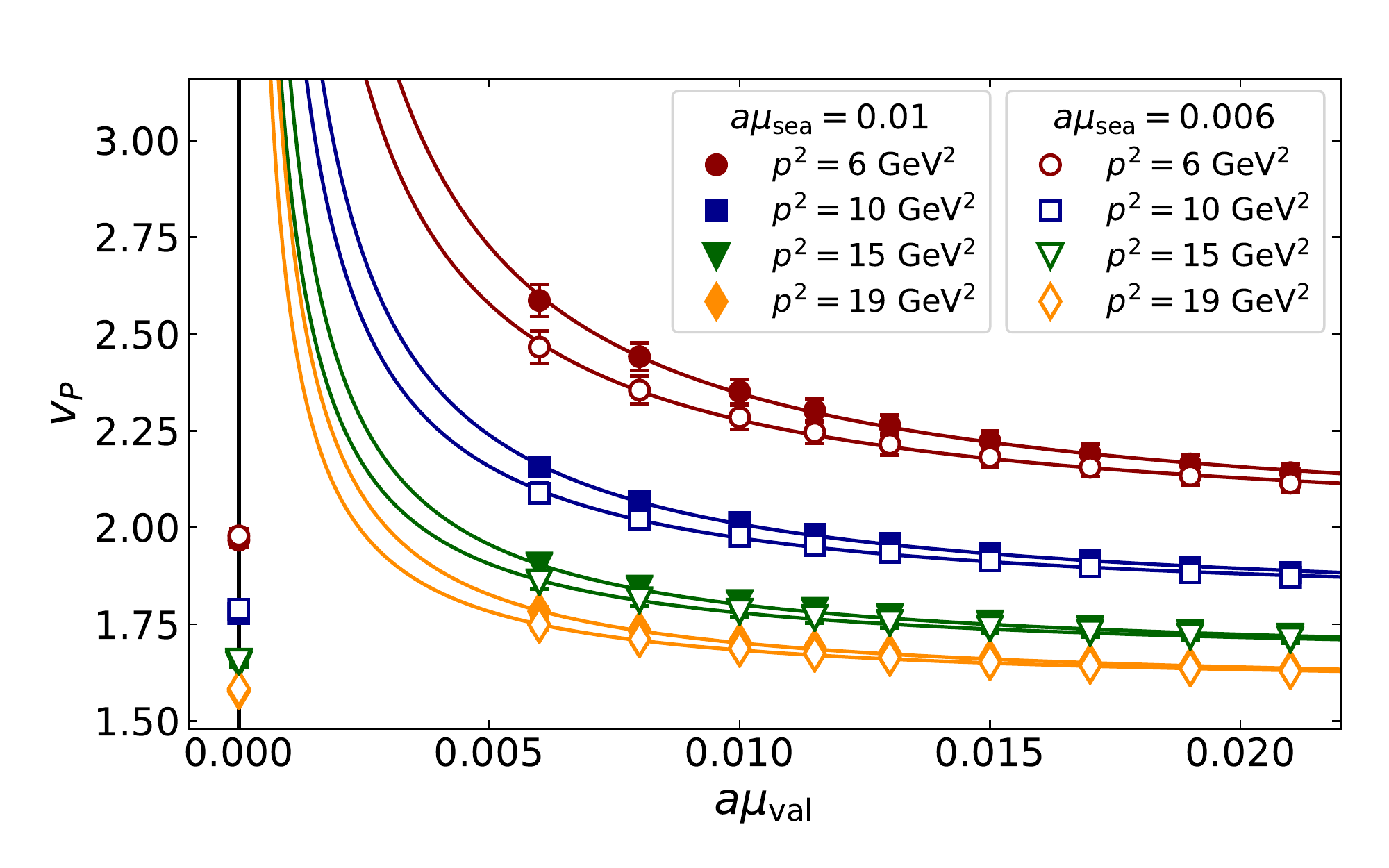}
\end{minipage}
\begin{minipage}{0.495\textwidth}
\includegraphics[width=1.\textwidth]{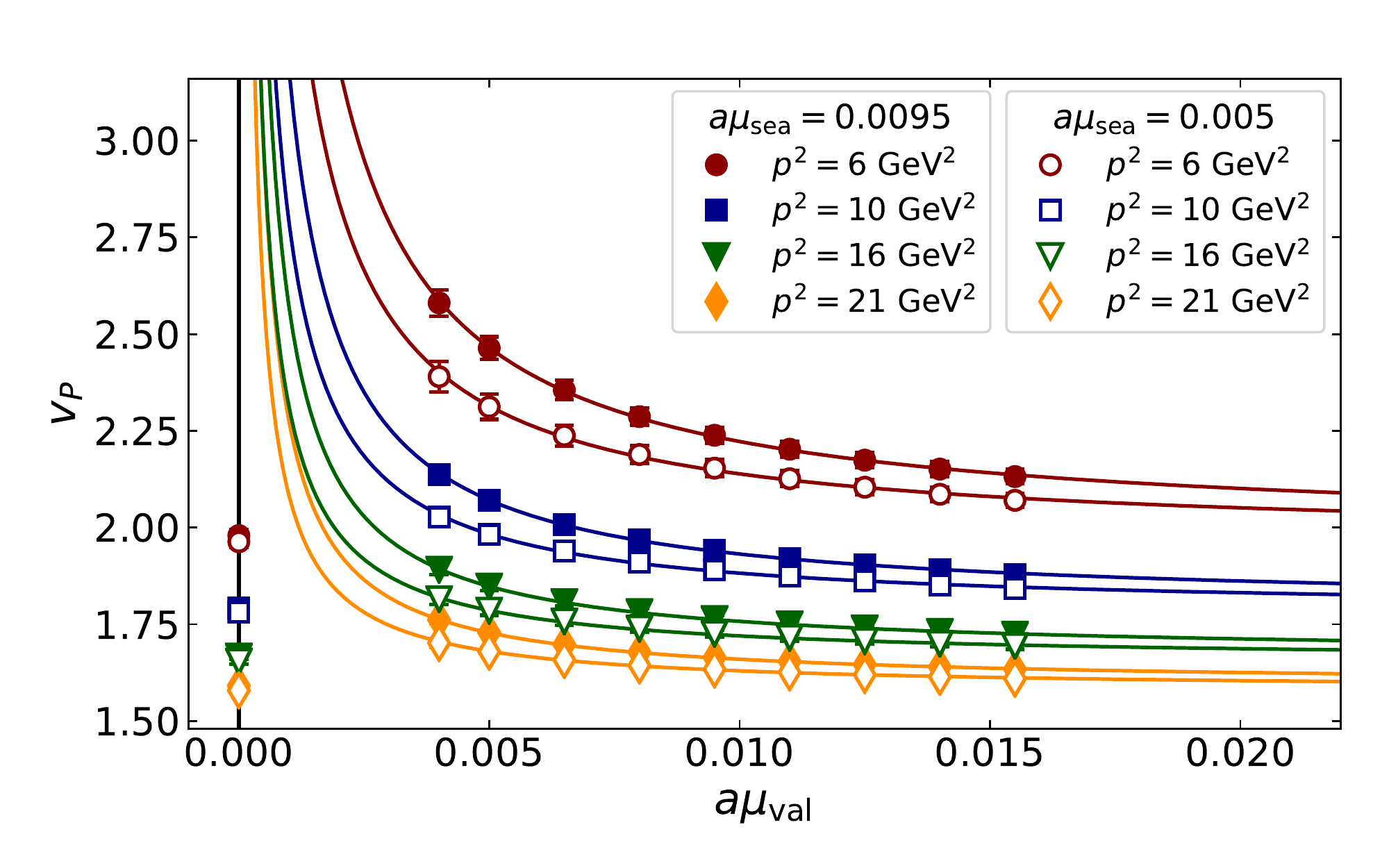}
\end{minipage}
\vspace{-0.5cm}
\caption{Fit of the pseudoscalar vertex lattice estimators $v_P(p^2,\mu_\mathrm{val},\mu_\mathrm{sea};a)$ according to the Ansatz of Eq.~\eqref{eq:vP_valence_ansatz} for $\beta=1.726$ (left panel) and $\beta=1.836$ (right panel). Different colours correspond to different values of $p^2$, while full and empty circles correspond to different values of the sea quark mass $a\mu_\mathrm{sea}$. The extrapolated values at $a\mu_\mathrm{val}=0$ correspond to the quantities $[\mathcal{V}_P(p^2,\mu_\mathrm{sea}) +  {K^\prime}/{p^2}]$ in Eq.~\eqref{eq:vP_valence_ansatz}. 
\label{fig:chir_extr_AC}}
\end{figure}

The results of the fit on $v_P$ for few  values of $p^2$ and the two extreme values of $\mu_\mathrm{sea}$ are shown in Fig.~\ref{fig:chir_extr_AC} for the cases  $\beta=1.726$ and $\beta=1.836$. Besides the very good quality of the fits we remark that the resulting estimates of $[\mathcal{V}_P(p^2,\mu_\mathrm{sea}) +  {K^\prime}/{p^2}]$ at $\mu_\mathrm{val}=0$ indeed shows a very tiny dependence on $\mu_\mathrm{sea}$ as mentioned above. Such a dependence on  $\mu_\mathrm{sea}$ turns out to be of the same size as the statistical errors (about $\sim 0.5\%$), non monotonic in $\mu_\mathrm{sea}$ at fixed $\beta$ and with different trends at different $\beta$'s.

\begin{figure}[htb!]
\centering
\includegraphics[width=0.5\textwidth]{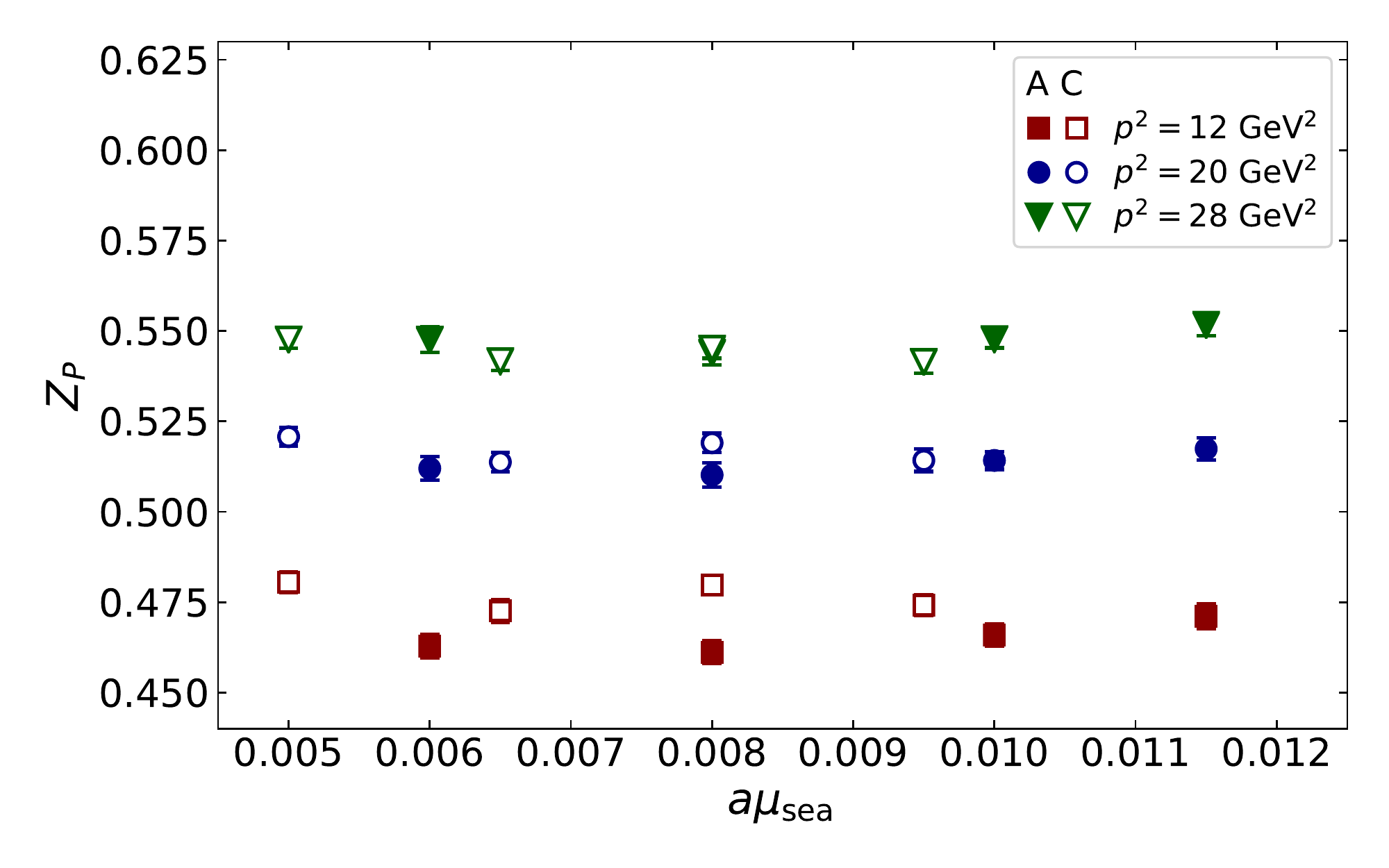}
\vspace{-0.5cm}
\caption{ 
Dependence of the renormalization constant $Z_P(\mu_\mathrm{sea},p^2)$ on the quark sea mass $\mu_\mathrm{sea}$ at $\beta=1.726$~(A) and $\beta=1.836$~(C) for different values of the momentum $p^2=(12,20,28)$~GeV$^2$. \label{fig:ZP_sea}}
\end{figure}

\noindent This feature is illustrated in Fig.~\ref{fig:ZP_sea}, where the resulting estimates of $Z_P(\mu_\mathrm{sea},p^2)$, obtained using the RI condition of Eq.~\eqref{eq:RIMOM_condition}, are shown at $\beta=1.726$ and $\beta=1.836$. Therefore, as the dependence on $\mu_\mathrm{sea}$ of $Z_P(\mu_\mathrm{sea},p^2)$ is not statistically significant, we  average them in order to estimate $Z_P(p^2)$ in the unitary chiral limit.

\subsection{Results for $Z_P$ in the \RIp-MOM and $\overline{\rm MS}$ schemes}
\label{sec:ZPresults}

\begin{table}[htb!]
\begin{minipage}{0.4\textwidth}
\begin{tabular}{|c|c|c|c|c|}
\hline
\multicolumn{5}{|c|}{\RIp-MOM}\\
\hline 
$\beta$ & M1a & M2a & M1b & M2b \\
\hline
1.726 & 0.4774(24) & 0.5079(24) & 0.4917(26) & 0.5301(24) \\
1.778 & 0.4812(32)  & 0.5042(26) & 0.4944(27) & 0.5255(23) \\
1.836 & 0.4899(26) & 0.5053(23) & 0.5046(27) & 0.5240(24)\\
\hline
\end{tabular}
\caption{Results for $Z_P$ in \RIp-MOM for each $\beta$ value 
for methods M1 (second and fourth columns) and M2 (third and fifth columns), computed  at  reference scales of $\mu_{\rm ref}=17$~GeV$^2$ (indicated by~``a'') and $\mu_{\rm ref}=21$~GeV$^2$ (indicated by~``b''). For cases M1a and M2a, results are extracted in the range $p^2\in (15,19)$~GeV$^2$, while for M1b and M2b, $p^2 \in (18,24)$~GeV$^2$.}
\label{tab:ZP_RI}
\end{minipage}\hfill
\begin{minipage}{0.55\textwidth}\vspace{-1.55cm}
\begin{tabular}{|c|c|c|c|c|}
\hline
\multicolumn{5}{|c|}{\RIp-MOM, \,\,\,\,$\mu_\mathrm{ref}^2$=19 GeV$^2$}\\
\hline 
$\beta$ & M1a & M2a & M1b & M2b \\
\hline
1.726 & 0.4849(24)(35) & 0.5159(24)(37) & 0.4851(26)(32)  & 0.5229(24)(34)  \\
1.778 & 0.4888(33)(35)  & 0.5121(26)(37) & 0.4877(27)(32)  & 0.5184(23)(34)  \\
1.836 & 0.4976(26)(36) & 0.5133(23)(37) & 0.4978(27)(33)  & 0.5169(24)(34) \\
\hline
\end{tabular}
\caption{Results for $Z_P$ in the \RIp-MOM scheme  but evolved to the common reference scale $\mu_\mathrm{ref}^2=19$ GeV$^2$. The notation is the same as that in Table~\ref{tab:ZP_RI} and we report separately the statistical error and the  systematic uncertainty related to the evolution factors.}
\label{tab:ZP_RI_19}
\end{minipage}
\end{table}

\begin{table}[htb!]
\begin{tabular}{|c|c|c|c|c|}
\hline
\multicolumn{5}{|c|}{$\overline{\rm MS}$, \,\,\,\,$\mu_\mathrm{ref}^2$=19 GeV$^2$}\\
\hline 
$\beta$ & M1a & M2a & M1b & M2b \\
\hline
1.726 & 0.569(3)(5) & 0.605(3)(5) & 0.569(3)(5)  & 0.614(3)(5)  \\
1.778 & 0.574(4)(5) & 0.601(3)(5) & 0.572(3)(5)  & 0.608(3)(5)  \\
1.836 & 0.584(3)(5) & 0.602(3)(5) & 0.584(3)(5)  & 0.607(3)(5) \\
\hline
\end{tabular}
\caption{Results for $Z_P$ at the scale $\mu_\mathrm{ref}^2=19$ GeV$^2$ converted from  \RIp-MOM (results of Table~\ref{tab:ZP_RI_19}) to the $\overline{\rm MS}$ scheme. 
}
\label{tab:ZP_MSb_19}
%
%
\end{table}

\begin{figure}[htb!]
\centering
\includegraphics[width=0.5\textwidth]{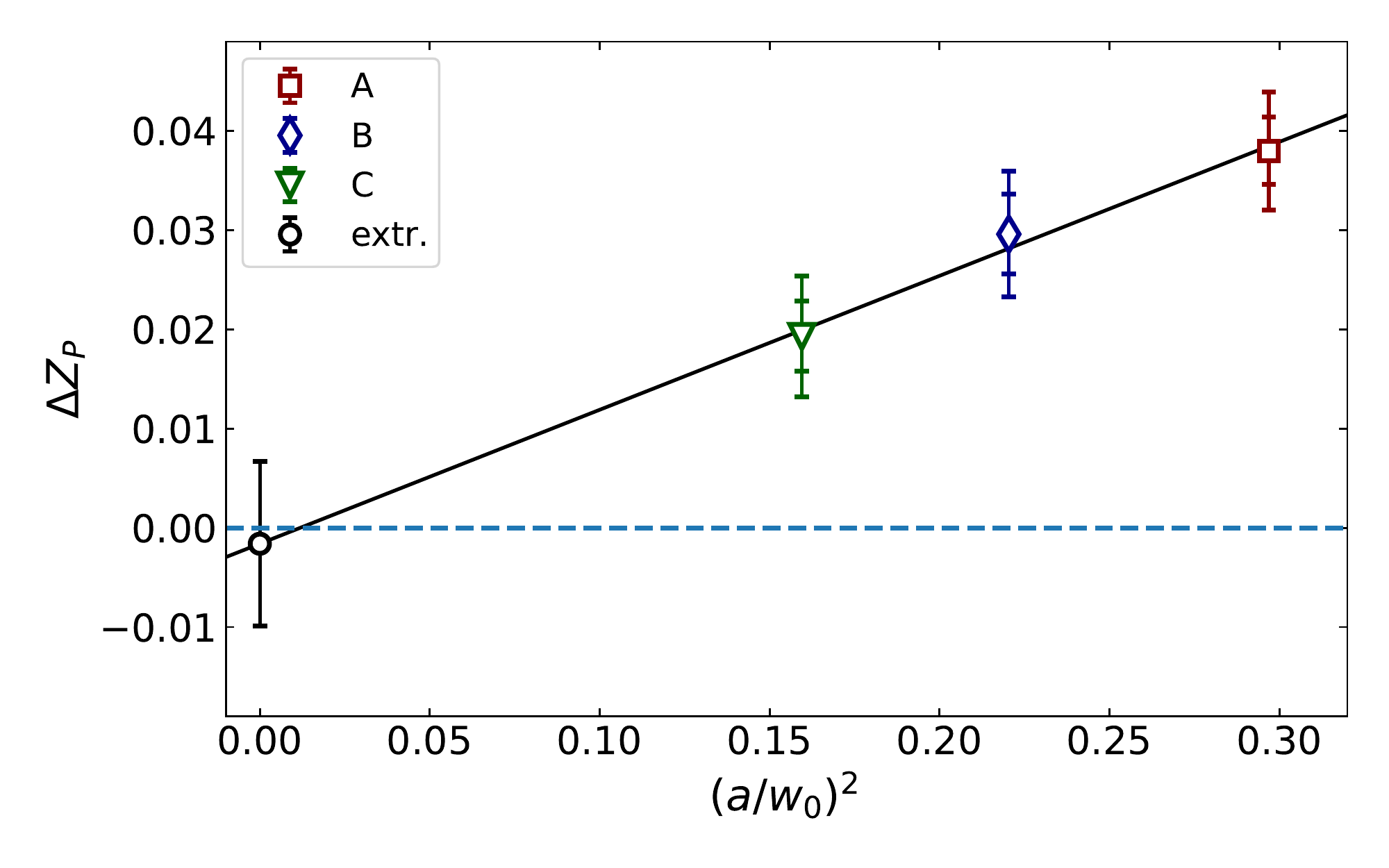}
\vspace{-0.5cm}
\caption{$\Delta Z_P = Z_P[\mathrm{M2b}]-Z_P[\mathrm{M1a}]$ versus $(a/w_0)^2$ and its (linear) continuum extrapolation. At each finite value of $a$ the smaller error bars correspond to the first (statistical) error in Tab.~\ref{tab:ZP_RI_19}, while the larger ones also include the second error in the same Table, which is due to the N$^3$LO evolution to 19 GeV$^2$ and is independent from the lattice spacing. Therefore the continuum limit value and its uncertainty are obtained by taking into account only the statistical errors at finite $a$.
\label{fig:ZP_a2_scaling}}
\end{figure}

In Table~\ref{tab:ZP_RI} we give the results of $Z_P$ determined in the \RIp-MOM scheme using methods M1 and M2 (see Sec.~\ref{sec:ZP_analysis}) for the two reference scales $\mu_\mathrm{ref}^2=17$ GeV$^2$ (labelled by ``a'') and $\mu_\mathrm{ref}^2=21$ GeV$^2$ (labelled by ``b''). We find that at each $\beta$ value, the parameter $z_0$ appearing in Eq.~(\ref{eq:ZP_Ansatz12}) is found to be compatible with the results for $Z_P(\mu^2_\mathrm{ref}=17$ GeV$^2)$ extracted using the method M1, as expected since this method corresponds to a linear fit Ansatz in $p^2$. The results are obtained by fitting the data in the momentum ranges $p^2\in (15,19)$~GeV$^2$ and $p^2 \in (18,24)$~GeV$^2$, respectively, for the reference scales $17$~GeV$^2$ and~$21$~GeV$^2$. 
The values of $Z_P$ given in Table~\ref{tab:ZP_RI} (in the \RIp-MOM scheme) are then evolved to the common reference scale  $\mu_\mathrm{ref}^2=19$~GeV$^2$ and are reported in Table~\ref{tab:ZP_RI_19}. 
The four determinations of $Z_P(\mu_\mathrm{ref}^2)$ (M1a, M2a, M1b and M2b) are equally good estimates of the renormalization constant that only differ by $\mathcal{O}(a^2)$ cut-off effects. This implies that using whichever of them leads to equivalent results for the renormalized quark masses and renormalized matrix elements of the pseudoscalar density in the continuum limit. As a check of the good accuracy to which this property is expected to be satisfied, we show in Fig.~\ref{fig:ZP_a2_scaling} the scaling behaviour of the difference $\Delta Z_P = Z_P[\mathrm{M2b}]-Z_P[\mathrm{M1a}]$ (in the \RIp-MOM scheme at $19$~GeV$^2$), for which all logarithmic divergences cancel and the continuum limit vanishes. A similar behaviour, but with smaller $\mathcal{O}(a^2)$ artifacts, is observed for all the analogous differences of the four $Z_P$ determinations in Tab.~\ref{tab:ZP_RI_19}.
Finally in Table~\ref{tab:ZP_MSb_19} we show the values of $Z_P(19~\mathrm{GeV}^2)$  converted to the $\overline{\rm MS}$ scheme.


Since quark masses are generally given in the $\overline{\rm MS}$ scheme at 2 or 3 GeV, we obtain the corresponding renormalization constants at these scales by using the following evolution factors
\begin{eqnarray}
Z_P^{\overline{\rm MS}}(4\, {\rm GeV}^2) &=& 0.83416(86) \, \, Z_P^{\overline{\rm MS}}(19\, {\rm GeV}^2)~,\\
Z_P^{\overline{\rm MS}}(9\, {\rm GeV}^2) &=& 0.92570(34) \, \, Z_P^{\overline{\rm MS}}(19\, {\rm GeV}^2)~, \\
Z_P^{\overline{\rm MS}}(16\, {\rm GeV}^2) &=& 0.98359(19)\  \, Z_P^{\overline{\rm MS}}(19\, {\rm GeV}^2)~. 
\end{eqnarray}
Our evolution function is accurate at N$^3$LO~\cite{Chetyrkin:1999pq}, i.e.~$\mathcal{O}(\alpha_\mathrm{s}^3)$, and therefore, we estimate the uncertainty due to higher orders as the last known term raised to the power $4/3$ (see the second error in the results of Table~\ref{tab:ZP_RI_19}). When computing conversion factors, which are ratios of evolution functions, we add in quadrature the error coming from the numerator and the denominator. We verified that this procedure provides a good estimate of the uncertainty due to higher orders when applied to the N$^2$LO conversion factors in order to estimate the (known) N$^3$LO results.

\section{Meson sector analysis}
\label{sec:mesons}
\input{Meson_analysis}

\section{Baryon sector analysis}
\label{sec:baryons}
\input{Baryon_analysis}

\section{Conclusions}
\label{sec:conclusions}
The focus of this work is the determination of the light, strange and charm quark masses. We perform an analysis of ten $N_f=2+1+1$ ensembles simulated at three lattice spacings smaller than 0.1~fm and pion masses in the range from about 350~MeV to 135~MeV. Having two ensembles simulated with the physical value of the pion mass at the two smallest lattice spacings enables us to extrapolate reliably to the physical and continuum limit. 

The extraction of the quark masses is done using observables from both the meson sector and the baryon sector. 
The iso-symmetric values of the pion, kaon and $D$-meson masses as well as of the pion decay constant are used for the determination of the lattice spacings and the quark masses in the meson analysis.
In the baryon sector, we use as inputs the nucleon and pion masses to obtain the lattice spacing and the average light-quark mass, while the masses of the $\Omega^{-}$ and $\Lambda_c$ baryons determine the strange and charm quark masses.

\begin{table}[htb!]
    \centering
    \begin{tabular}{l|c|c|c||c|c}
    \hline
         & $m_{ud}$ [MeV] & $m_{s}$ [MeV] & $m_{c}$ [MeV] & $m_{s}/m_{ud}$ & $m_{c}/m_{s}$ \\
    \hline
        Meson sector &$3.689(80)(66)$ &  $101.0(1.9)(1.4)$   & $1039(15)(8)$ & $27.30(24)(14)$  & $11.43(9)(10)$ \\
        Baryon sector & $3.608(58)(^{+32}_{-19})$ & $94.9(2.4)(^{+4.1}_{-1.0})$ & $1030(21)(^{+22}_{-5})$ & $26.30(61)(^{+1.17}_{-0.33})$ & $12.04(31)(^{+58}_{-15})$ \\
        \hline
        Average  & $3.636(66)(^{+60}_{-57})$ & $98.7(2.4)(^{+4.0}_{-3.2})$ & $1036(17)(^{+15}_{-8})$ &$27.17(32)(^{+56}_{-38})$ & $11.48(12)(^{+25}_{-19})$ \\\hline\hline
         FLAG 2019 & 3.410(43) &  93.44(68) & 988(7) & 27.23(10) & 11.82(16) \\
    \end{tabular}
    \caption{The renormalized quark masses determined in the meson sector (first row) and baryon sector (second row) in the $\overline{\rm MS}$ scheme. In the third row we give the average over the values obtained in the the meson and baryon sectors, while in the last row we give the latest FLAG averages~\cite{Aoki:2019cca} for $N_f = 2+1+1$. The light quark mass, $m_{ud}$ (second column), and the strange quark mass, $m_s$ (third column), are given at 2~GeV, while the charm quark mass, $m_c$ (fourth column), is given at 3~GeV. The second error of the quark masses includes a $0.5 \%$ uncertainty (added in quadrature) due to the uncertainty of the conversion of the RCs $Z_P$ to the $\overline{\rm MS}$ scheme. In the fifth and sixth columns we give the ratios $m_s/m_{ud}$ and $m_c/m_s$, respectively. In the meson sector the error on the ratio is determined in a jackknife analysis. In the baryon sector, since different ensembles are involved in the determination of the quark masses, the error on the ratio is propagated quadratically using the  errors on each of the quark masses.}
    \label{tab:recap_masses}
\end{table}
In Table~\ref{tab:recap_masses} we collect the values of the quark masses obtained in Sections~\ref{sec:mesons} and \ref{sec:baryons} for the light and strange quark masses in the $\overline{\rm MS}$ scheme at 2~GeV and for the charm quark mass at 3~GeV. Since the isospin and electromagnetic corrections to the nucleon mass are only known for the mass difference between the neutron and proton~\cite{Borsanyi:2014jba}, in our analysis we average over the mass of the proton and neutron. This defines a QCD prescription different from that used in the meson sector. Using the values of the lattice spacing extracted in the meson sector we obtain a nucleon mass in the continuum limit a few MeV smaller than the input value $m_{N, phys} = 0.9389$~GeV adopted in the baryon sector (see Section~\ref{sec:lattice spacing}). This results in less than a percent change in the values given in the Table~\ref{tab:recap_masses}, which is much smaller than our statistical errors. It is thus justifiable to average over the values obtained in the meson and baryon sectors to produce our final values. 

\begin{figure}[htb!]
    \centering
    \includegraphics[width=1.\textwidth]{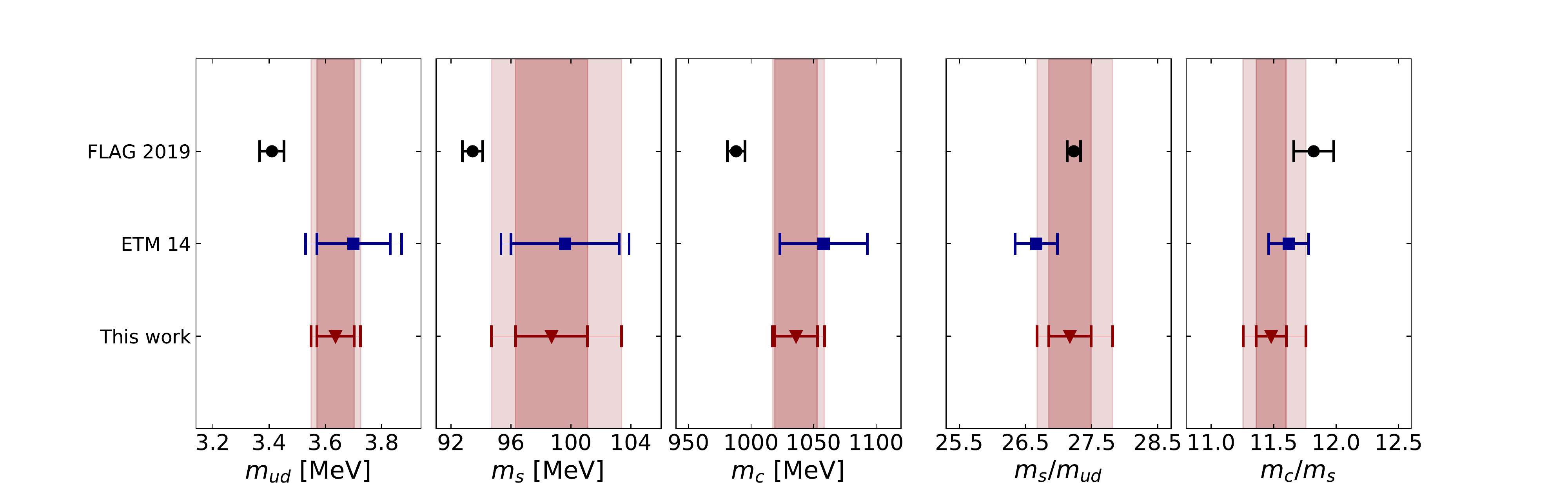}
    \vspace{-0.5cm}
    \caption{Comparison of the results average between the values determined in the meson and baryon sectors (red triangles) with the values obtained using twisted mass fermions in Ref.~\cite{Carrasco:2014cwa} (blue squares) and the $N_f=2+1+1$ averages given in the last FLAG report~\cite{Aoki:2019cca} (black circles). The shorter error bars take into account the statistical error only, while the larger represent the total error, obtained by summing in quadrature the statistical and the systematic errors.}
    \label{fig:final_res1}
\end{figure}

In order to perform the above average we adopt the weighted approach given in Eqs.~(\ref{eq:average}-\ref{eq:sigma_2}).
We assume the following weights
\begin{equation}
     w_M \propto 1 / \left( \sigma_M^{stat} \right)^2 ~ , \qquad w_B \propto 1 / \left( \sigma_B^{stat} \right)^2 ~ 
\end{equation}
for the quantities coming from the mesonic and the baryonic sectors, where $\sigma_{M(B)}^{stat}$ is the first error given in the corresponding rows of Table~\ref{tab:recap_masses}.
In this way we obtain
\begin{equation}
    \overline{x} \pm \sigma^{stat}  ~ (_{-\sigma^{syst, -}} ^{+\sigma^{syst, +}}  ) ~ , ~
\end{equation}
where
\begin{eqnarray}
    \overline{x} & = & w_M x_M + w_B x_B ~ , ~ \\[2mm] 
    (\sigma^{stat})^2  & = &  w_M (\sigma_M^{stat})^2 +  w_B (\sigma_B^{stat})^2 ~ , ~ \\[2mm]
    (\sigma^{syst, \pm})^2 & = & w_M \left[ (x_M -  \overline{x})^2 + (\sigma_M^{syst})^2 \right] 
     +  w_B \left[ (x_B -  \overline{x})^2 + (\sigma_B^{syst, \pm})^2 \right] ~ . ~
\end{eqnarray}
The results are given in the last row of Table~\ref{tab:recap_masses} and are compared in Fig.~\ref{fig:final_res1} with those of the ETM analysis of Ref.~\cite{Carrasco:2014cwa} and the ones entering the $N_f = 2+1+1$ averages in the latest FLAG report~\cite{Aoki:2019cca}.
The latter ones are based on the results of Refs.~\cite{Carrasco:2014cwa,Bazavov:2018omf} for the light-quark mass, Refs.~\cite{Carrasco:2014cwa,Bazavov:2018omf,Lytle:2018evc,Chakraborty:2014aca} for the strange mass, Refs.~\cite{Carrasco:2014cwa,Bazavov:2018omf,Lytle:2018evc,Chakraborty:2014aca,Alexandrou:2014sha} for the charm mass, Refs.~\cite{Carrasco:2014cwa,Bazavov:2014wgs,Bazavov:2017lyh} for the $m_s / m_{ud}$ ratio and Refs.~\cite{Carrasco:2014cwa,Chakraborty:2014aca,Bazavov:2018omf} for the $m_c / m_s$ ratio.

It can be seen that our results are larger by $\sim 2.5$ standard deviations in the case of $m_{ud}$ and by $\sim 2$ standard deviations in the case of $m_c$ with respect to the corresponding FLAG values.
Although for the strange quark mass our result coming from the meson sector is larger by $\sim 3$ standard deviations, our averaged result is consistent with the FLAG one within our final uncertainty.
A good agreement is observed for the mass ratios $m_s / m_{ud}$ and $m_c / m_s$.
We do not believe that these differences can be ascribed to possible uncontrolled effects on the mass renormalization constant $1 / Z_P$. Indeed, the detailed analysis carried out in this work concerning the pion pole subtraction and the residual hadronic contaminations in the RI-MOM determination of the renormalization constant $Z_P$ leaves little room for any significant leftover contribution from these terms. Our findings point to the fact  that hadronic contaminations are controlled at the level of few per mil. Therefore, we do not consider plausible that the observed tension with the FLAG values may be related to uncontrolled hadronic contaminations on the mass renormalization constant. In this respect, we are considering the possibility of repeating the determination of the quark masses using the same ETM gauge ensembles adopted in this work, but evaluating the mass renormalization in a different scheme, like RI-SMOM, while keeping the same level of control of the hadronic contaminations achieved in this work.

Our final results for the light, strange and charm quark masses as well for the mass ratios $m_s / m_{ud}$ and $m_c / m_s$ are consistent with our previous analysis of Ref.~\cite{Carrasco:2014cwa} (see also Fig.~\ref{fig:final_res1}), which was based on Wilson twisted-mass fermions far from the physical pion point.
The overall uncertainties for the light and charm quark masses are reduced by a factor of $\sim 1.7 - 2.0$, while in the case of the strange quark mass the uncertainty is almost unchanged, partly due to the difference between the mean values obtained in the meson and baryon sector, which is added to the systematic error. This is also reflected in the two ratios.
With respect to the quark mass analysis of Ref.~\cite{Carrasco:2014cwa} the main improvements are: ~ i) a better control of the chiral extrapolation thanks to gauge ensembles produced close to the physical pion point; ~ ii) a better control of hadronic contaminations in the calculations of the mass renormalization constant; ~ iii) the use of both mesonic and baryonic quantities, which requires simulations of different correlation functions.
For all the three masses the contribution from lattice systematics is important, in particular in the case of $m_s$.
Our plan is to add at least one further gauge ensemble at a fourth finer value of the lattice spacing at the physical point. This will allow a tightly controlled chiral and continuum extrapolations in both the meson and baryon sectors.

\begin{acknowledgements}
We would like to thank all members of ETMC for a very constructive and enjoyable collaboration. 

We acknowledge PRACE (Partnership for Advanced Computing in Europe) for awarding us access to the high-performance computing system Marconi and Marconi100 at CINECA (Consorzio Interuniversitario per il Calcolo Automatico dell'Italia Nord-orientale) under the grants Pra17-4394, Pra20-5171 and Pra22-5171, and CINECA for providing us CPU time under the specific initiative INFN-LQCD123. We also acknowledge PRACE for awarding us access to HAWK, hosted by HLRS, Germany, under the grant with Acid 33037.
The authors gratefully acknowledge the Gauss Centre for Supercomputing e.V.~(www.gauss-centre.eu) for funding the project pr74yo by providing computing time on the GCS Supercomputer SuperMUC at Leibniz Supercomputing Centre (www.lrz.de), the projects ECY00, HCH02 and HBN28 on the GCS supercomputers JUWELS and JUWELS Booster~\cite{JUWELS} at the J\"ulich Supercomputing Centre (JSC) and time granted by the John von Neumann Institute for Computing (NIC) on the supercomputers JURECA and JURECA Booster~\cite{Jureca}, also at JSC. Part of the results were created within the EA program of JUWELS Booster also with the help of the JUWELS Booster Project Team (JSC, Atos, ParTec, NVIDIA). We further acknowledge computing time granted on Piz Daint at Centro Svizzero di Calcolo Scientifico (CSCS) via the project with id s702.

Part of the statistics of the cA211.30.32 ensemble used in this work was generated on the Bonna cluster at the University of Bonn, access to which the authors gratefully acknowledge.

This work has been partially supported by the Horizon 2020 research and innovation program
of the European Commission under the Marie Sk\l{}odowska-Curie grant agreement No.~765048 (STIMULATE) as well as by the DFG as a project under the Sino-German CRC110.

R.F.~acknowledges the University of Rome Tor Vergata for the support granted to the project PLNUGAMMA.
F.S.~and S.S.~are supported by the Italian Ministry of Research (MIUR) under grant PRIN 20172LNEEZ.
F.S.~is supported by INFN under GRANT73/CALAT.
P.D.~and E.F.~acknowledge support form the European Unions Horizon 2020 research and innovation programme under the Marie Sk\l{}odowska-Curie grant agreement No.~813942 (EuroPLEx). P.D.~acknowledges support from INFN under the research project INFN-QCDLAT.
M.C.~acknowledges financial support by the U.S.~Department of Energy, Office of Nuclear Physics Early Career Award under Grant No.~DE-SC0020405. 
S.B.~and J.F.~are supported by the H2020 project PRACE 6-IP (grant agreement No.~82376) and the EuroCC project (grant agreement No.~951740). F.M.~and A.T.~are supported by the European Joint Doctorate program STIMULATE grant agreement No.~765048.
K.H.~and E.P.~are supported by the Cyprus Research and Innovation  Foundation under contract number POST-DOC/0718/0100.
F.P.~acknowledges support from project NextQCD, co-funded by the European Regional Development Fund and the Republic of Cyprus through the Research and Innovation Foundation (EXCELLENCE/0918/0129).
M.D.C.~is supported in part by UK STFC grant ST/P000630/1.

\end{acknowledgements}

\bibliography{refs}
 
\end{document}

%% file: Meson_analysis.tex
In this Section we describe the determination of the quark masses taking as input the iso-symmetric values of the pion, kaon and $D_{(s)}$-meson masses.
\subsection{Methodology}
For each ensemble, we compute the two point function
 \begin{equation}
     C(t) = \frac{1}{L^3} \sum\limits_{\vec{x}, \vec{z}} \left\langle 0 \right|J_{f f'} (\vec{x},t) J^\dag_{f f'} (\vec{z},0) \left| 0 \right\rangle ~,
    \label{eq:P5}
  \end{equation}
where $J_{f f'} (x) = \overline{q}_f(x) i\gamma_5  q_{f'}(x)$ is the meson interpolating field with $q_f$ being the valence quark field of flavor $f \in \{\ell,s,c\}$. By $\ell$ we denote the average up/down (light) quark.
The correlators for the pion and kaon are the same as those used in Ref.~\cite{Alexandrou:2021bfr}.
For all mesons the two valence quarks $q_f$ and $q_{f'}$ are always taken with opposite Wilson parameters, i.e.~$r_f = -r_{f'}=1$, as this choice is known to suppress O($a^2$) lattice artefacts~\cite{Frezzotti:2003ni,Frezzotti:2005gi}).
For the valence mass parameters, we evaluate correlators at $\mu_\ell$ values equal to its sea counterpart, as well as at three values of the quark mass parameter $\mu_s$ in the range of the strange quark masses and four values of the quark mass parameter $\mu_c$ in the range of the charm quark masses. The chosen values of valence quark masses are collected in Table~\ref{tab:valence_pion} and allow for a precise interpolation to the physical strange and charm  quark masses as determined by the kaon and $D$-meson masses in the isosymmetric QCD. The latter ones, following the FLAG report~\cite{Aoki:2016frl}, are given by
 \begin{eqnarray}
      \label{eq:MK_isoQCD}
      m_K^{\textrm{isoQCD}} & = & 494.2 ~ (3) ~ {\rm MeV} ~ , ~ \\[2mm]
      \label{eq:MD_isoQCD}
      m_D^{\textrm{isoQCD}} & = & 1867.0 ~ (4) ~ {\rm MeV} ~ , ~ \\[2mm]
      \label{eq:MDs_isoQCD}
      m_{D_s}^{\textrm{isoQCD}} & = &  1969.0 ~ (4) {\rm MeV} ~ .
\end{eqnarray}

From the correlator given in Eq.~(\ref{eq:P5}), the overlap $\mathcal{S} = | \langle PS | J_{f f'}  | 0 \rangle|^2$ can be extracted using an exponential fit at large time distances 
\begin{equation}
    C_{PS}(t)_{ ~ \overrightarrow{t  \gg a, ~ (T - t) \gg a} ~ } \frac{\mathcal{S}}{2m_{PS}} \left[ e^{ - m_{PS}  t}  + e^{ - m_{PS} (T - t)} \right] ~ , ~
    \label{eq:larget}
 \end{equation}
where $m_{PS}$ is the ground-state mass of a pseudoscalar (PS) meson made of the two valence quarks with flavor $f$ and $f'$. 
For maximally twisted quarks the value of the matrix element ${\mathcal{S}}$ determines the PS-meson decay constant with no need of any renormalization 
constant~\cite{Frezzotti:2000nk}, from the formula
 \begin{equation}
      af_{PS} = a (\mu_f+\mu_{f'}) \frac{\sqrt{a^4 \mathcal{S}}}{am_{PS} ~ \mbox{sinh}(am_{PS})} ~ .
    \label{eq:decaypi}
 \end{equation}
 The slight deviation from maximal twist of the ensemble cA211.12.48 is corrected according to Appendix~C of Ref.~\cite{Alexandrou:2021bfr}.

The global energy scale is set using the isosymmetric QCD inputs~(\ref{eq:pi_isoqcd}) and data at different lattice spacings are connected by exploiting the gradient-flow (GF) quantities $w_0$~\cite{Borsanyi:2012zs}, $\sqrt{t_0}$~\cite{Luscher:2010iy} and $t_0 / w_0$ measured in lattice units.
Their values have been already determined quite precisely in Ref.~\cite{Alexandrou:2021bfr}, namely\footnote{The result~(\ref{eq:w0_intro}) improves drastically our preliminary value of $w_0$ obtained in Ref.~\cite{ExtendedTwistedMass:2020tvp}. The improvement is mainly related to a better description of discretization effects on the decay constant $f_\pi$ (see below Eq.~(\ref{eq:cptfpiCh})) and to the replacement of $f_\pi$ with the quantity $(f_\pi m_\pi^4)^{1/5}$, which has been found to be less affected by lattice artifacts~\cite{Alexandrou:2021bfr}.}
\begin{eqnarray}
    \label{eq:w0_intro}
     w_0 & = & 0.17383~ (63) ~ {\rm fm} ~ , ~ \\[2mm]
    \label{eq:t0_intro}
     \sqrt{t_0} & = & 0.14436 ~ (61) ~ {\rm fm} ~ , ~ \\[2mm]
    \label{eq:t0w0_intro}
     t_0 / w_0 & = & 0.11969 ~ (62) ~ {\rm fm} ~ . ~
\end{eqnarray}
Nevertheless, in order to take properly into account all the correlations with the meson data the GF scales are determined again in the present analysis (see the next subsection), obtaining results well compatible with Eqs.~(\ref{eq:w0_intro}-\ref{eq:t0w0_intro}).

\subsection{Light quark mass}

The lattice QCD data on the pion mass and decay constant are computed in an unitary setup, i.e.~with $\mu_{sea}=\mu_{valence}=\mu_\ell$, the values used in this Section are reported in the Table~\ref{tab:valence_pion}.
\begin{table}[htb!]
    \centering
    \begin{tabular}{|cccccc|}
\hline
Ensemble & \(a\mu_\ell\) & \(a\mu_\ell\) & \(am_\pi\) &\(af_\pi\) & ~confs~ \tabularnewline
\hline
cA211.53.24 & 0.0053 & 0.0053 & 0.16626(51) &0.07106(36)& $~628$~\tabularnewline
cA211.40.24 & 0.0040 & 0.0040 & 0.14477(70) &0.06809(30)& $~662$~\tabularnewline
cA211.30.32 & 0.0030 & 0.0030 &0.12530(16) &0.06674(15)&$1237$~\tabularnewline
cA211.12.48 & 0.0012 & 0.0012 & 0.08022(18) &0.06133(33)&$~322$~\tabularnewline
\hline
cB211.25.32 & 0.0025 & 0.0025 & 0.10475(45) &0.05652(38)& $~400$~\tabularnewline
cB211.25.48 & 0.0025 & 0.0025 & 0.10465(14) &0.05726(12)& $~314$~\tabularnewline
cB211.14.64 & 0.0014 & 0.0014 & 0.07848(10) &0.05477(12) &$~437~$\tabularnewline
cB211.072.64 & 0.00072 & 0.00072 & 0.05659(8) &0.05267(14)&$~374$~\tabularnewline
\hline
cC211.20.48 & 0.0020 & 0.0020 & 0.08540(17) &0.04892(13)   &$~890$~\tabularnewline
cC211.06.80 & 0.0006 & 0.0006 & 0.04720(7) &0.04504(10) &$~401$~\tabularnewline
\hline
\end{tabular}
    \caption{Values of the bare valence quark mass parameters and the corresponding values of $m_\pi$ and $f_\pi$  for each of the ensembles used in the analysis in the pion sector.}
    \label{tab:valence_pion}
\end{table}

The lattice QCD data on the pion mass and decay constant are analyzed relying on SU(2) chiral perturbation theory (ChPT) using the formulae
\begin{align}
     \label{eq:cptmpi2Ch}
\hspace{-0.4cm}(m_\pi  w_0 )^2   =  2(B w_0 )(m_\ell w_0 )\left[ 1 + \xi_\ell \log \xi_\ell  + P_1 \xi_\ell  +  P_2  \,a^2/w_0^2 \right] K_{M^2}^{FSE} ~ ,  \\
    \label{eq:cptfpiCh}
\hspace{-0.4cm}    (f_\pi  w_0 )  =  (f w_0 )\left[ 1 - 2\xi_\ell \log \xi_\ell  + P_3 \xi_\ell  +  P_4 \,a^2/w_0^2 + a^2m_\ell P_5 \right] K_f^{FSE} ~ ,
\end{align} 
where the variable $\xi_\ell=2B m_\ell/(16 \pi^2 f^2)$ is related to the quark renormalized mass $m_\ell=\mu_\ell/Z_P$. The parameters $P_1$ and $P_3$ are related to the low-energy constants $\bar\ell_3$  and $\bar\ell_4$ by
\begin{gather}
 P_1=-\bar\ell_3-2\log{\left(m_\pi^{\textrm{isoQCD}}/(4\pi f)\right)}\,,\quad
  P_3=2\bar\ell_4+4\log{\left(m_\pi^{\textrm{isoQCD}}/(4\pi f)\right)}\,.
 \end{gather}
The quantities $K_{M^2}^{FSE}$ and $K_f^{FSE}$ represent the finite size effects (FSE) on the squared pion mass and the pion decay constant, respectively. 
In Ref.~\cite{Alexandrou:2021bfr} it was shown that SU(2) ChPT at NLO~\cite{GASSER1984142} adequately describes our lattice data, once discretization effects proportional both to $a^2$ and to $a^2 m_\ell$ are included in $f_\pi$ (see Eq.~(\ref{eq:cptfpiCh})), while in $m_\pi$ the leading lattice artefact is already directly proportional to $a^2m_\ell$ (see Eq.~(\ref{eq:cptmpi2Ch})).
The fit parameters are $B w_0$, $\bar\ell_3$, $P_2$, $f w_0$, $\bar\ell_4$, $P_4$ and $P_5$. We repeat the fit procedure adopting the values of the renormalization constant $Z_P$ determined using the methods M1a, M1b, M2a, M2b given in Table~\ref{tab:ZP_MSb_19}. 

To estimate possible systematics due to the scale setting and the chiral extrapolation, we repeat the analysis using:
\begin{itemize}
    \item  the  ratio $t_0/w_0$ to set the scale;
    \item  the GF scale  $\sqrt{t_0}$ to set the scale;
   \item only a combination of two lattice spacing\footnote{When considering the two finest lattice spacing  $\beta=1.778$ and $\beta=1.836$ we exclude the term $P_5$ from the fit of Eq.~($\ref{eq:cptfpiCh}$) because all the ensembles have $m_\pi<260$ MeV.}, namely
   \begin{itemize}
    \item $\beta=1.726$ and $\beta=1.778$,
    \item $\beta=1.726$ and $\beta=1.836$,
    \item $\beta=1.778$ and $\beta=1.836$;
   \end{itemize}    
    \item only ensembles with pion mass less than $190$~MeV and excluding the term $P_5$ in Eq.~(\ref{eq:cptfpiCh}) from the fit. 
\end{itemize}

\begin{figure}[htb!]
    \centering
    \begin{minipage}[c]{0.5\linewidth}
    \includegraphics[width=1.0\textwidth]{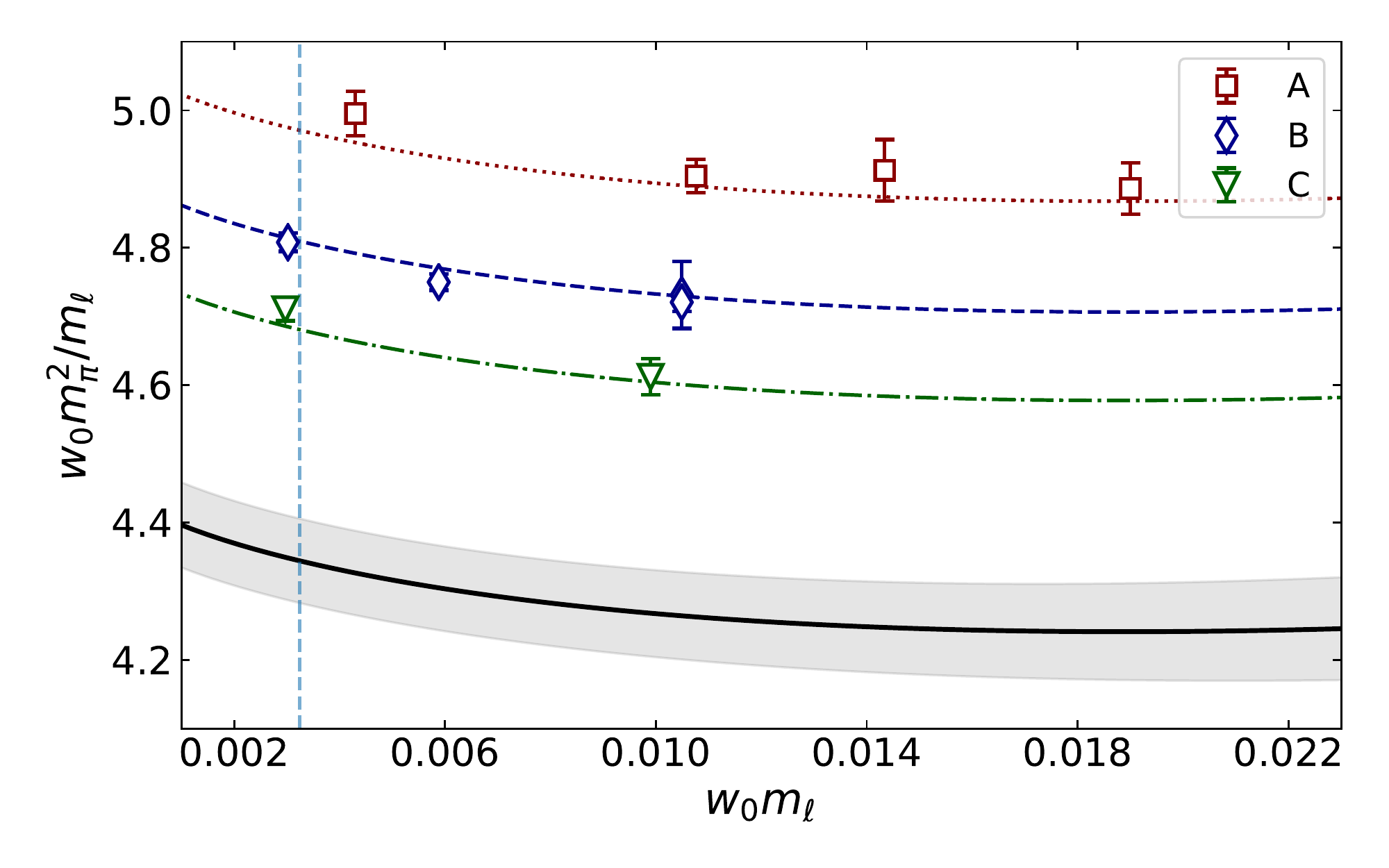}
    \end{minipage}%
    \begin{minipage}[c]{0.5\linewidth}
    \includegraphics[width=1.0\textwidth]{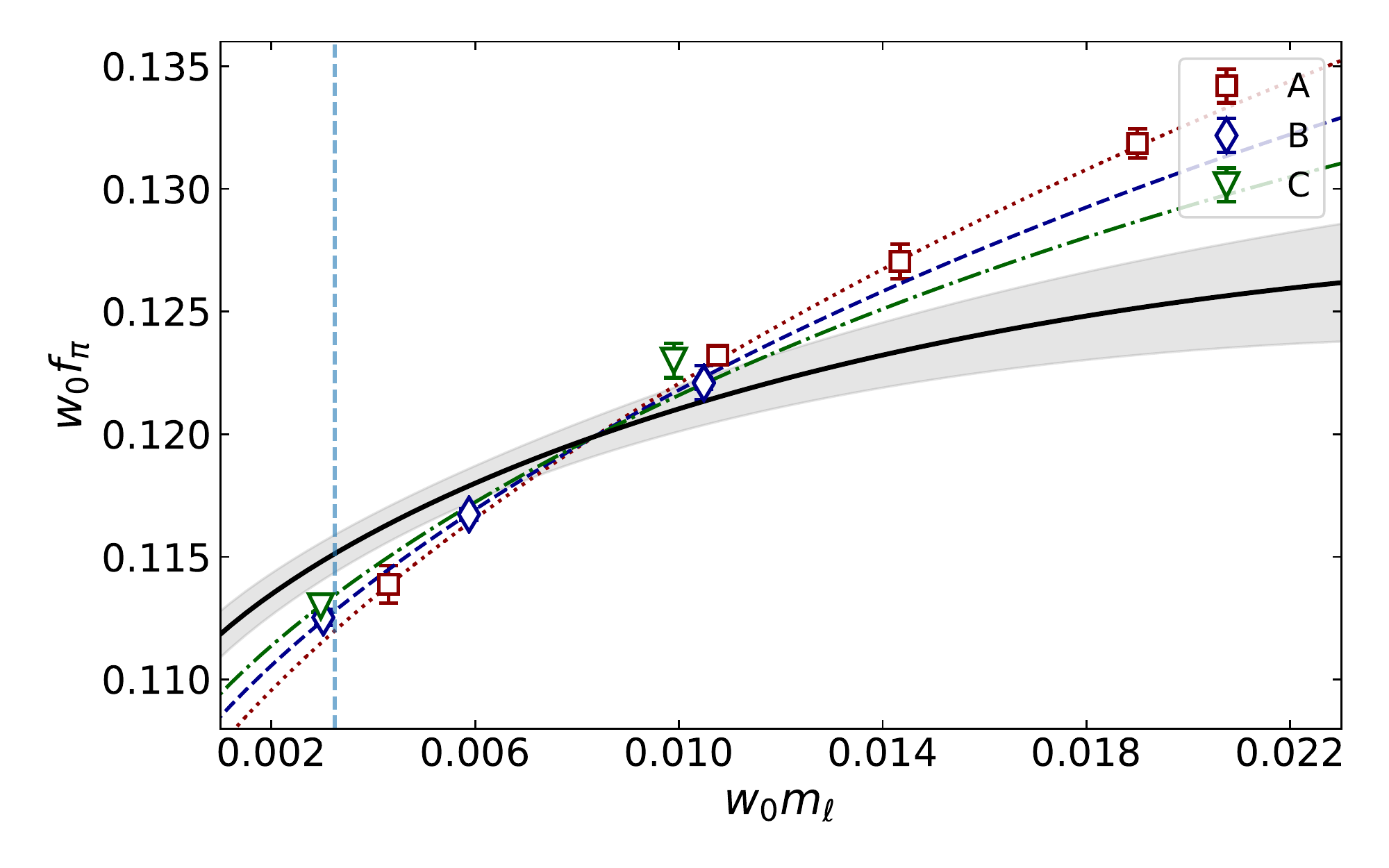}
    \end{minipage}
    \vspace{-0.5cm}
    \caption{Chiral and continuum extrapolation of $w_0 m_{\pi}^2/m_{\ell}$ (left) and $w_0 f_{\pi}$ (right) as function of $w_0\,m_\ell$ using Eqs.~(\ref{eq:cptmpi2Ch}) and (\ref{eq:cptfpiCh}) and $Z_P$ for the M2b method. Different colored bands correspond to different lattice spacings (red for the A ensembles, blue for the B and green for the C). The grey band is the extrapolation to the continuum limit. Note that for $w_0 f_{\pi}$ discretization effects proportional both to $a^2$ and to $a^2 m_\ell$ are visible (see text).} 
    \label{fig:Mpi_GL_NLO_am_w0_M2b}
\end{figure}

The results for the light quark mass $m_{ud}$ are reported in Table~\ref{tab:m_ud_pion}, where we also include the values of the leading low-energy constants $B$, $f$ and $\Sigma^{1/3}=(Bf^2/2)^{1/3}$ as well as the values of $\chi^2/{\rm d.o.f.}$ The chiral and continuum extrapolations are illustrated in Fig.~\ref{fig:Mpi_GL_NLO_am_w0_M2b}.

\begin{table}[htb!]
\centering
    \begin{tabular}{|c|ccccc|}
   \hline
\(Z_P\)  & \(m_{ud}\)[MeV] &
\(B\){[}MeV{]} & \(f\){[}MeV{]} & $\Sigma^{1/3}$ &\(\chi^2/{\rm d.o.f.}\)\tabularnewline
\hline
M1a & 3.677(65) & 2523(42) & 124.02(48) & 268.7(1.4) &
1.97\tabularnewline
M2a & 3.694(64) & 2512(40) & 124.04(50) & 268.3(1.4) &
1.56\tabularnewline
M1b & 3.677(66) & 2522(43) & 124.05(48) & 268.7(1.5) &
2.53\tabularnewline
M2b & 3.694(65) & 2512(41) & 124.02(51) & 268.3(1.4) &
1.15\tabularnewline
\hline
     &  \multicolumn{4}{c}{$t_0/w_0$}  &\\
\hline
M1a & 3.722(74) & 2493(47) & 124.40(49) & 268.2(1.6) &
2.91\tabularnewline
M2a & 3.766(73) & 2465(45) & 124.48(52) & 267.3(1.5) &
2.54\tabularnewline
M1b & 3.724(75) & 2492(48) & 124.42(49) & 268.2(1.6) &
3.72\tabularnewline
M2b & 3.771(74) & 2462(45) & 124.48(52) & 267.2(1.6) &
1.95\tabularnewline
\hline
 &  \multicolumn{4}{c}{$\sqrt{t_0}$}  &\\
\hline
M1a & 3.696(69) & 2510(45) & 124.19(48) & 268.5(1.5) &
2.48\tabularnewline
M2a & 3.726(68) & 2491(43) & 124.24(51) & 267.9(1.5) &
2.06\tabularnewline
M1b & 3.697(70) & 2509(45) & 124.22(48) & 268.5(1.5) &
3.18\tabularnewline
M2b & 3.729(69) & 2490(43) & 124.23(51) & 267.8(1.5) &
1.55\tabularnewline
\hline
 &\multicolumn{4}{c}{$w_0$,  $m_\pi<190$ MeV, $P_5=0$}  &\\
 \hline
M1a & 3.629(81) & 2572(55) & 122.71(45) & 268.5(2.3) &
4.72\tabularnewline
M2a & 3.656(81) & 2554(53) & 122.56(48) & 267.7(2.2) &
3.4\tabularnewline
M1b & 3.623(82) & 2577(56) & 122.72(45) & 268.7(2.3) &
6.32\tabularnewline
M2b & 3.663(82) & 2547(54) & 122.51(49) & 267.4(2.3) &
2.2\tabularnewline
\hline
 &\multicolumn{4}{c}{$w_0$,  $\beta=1.726$ and  $\beta=1.778$}  &\\
 \hline
M1a & 3.83(15) & 2420(93) & 124.54(83) & 265.8(3.2) &
0.0953\tabularnewline
M2a & 3.83(14) & 2425(84) & 124.57(88) & 266.0(2.7) &
0.0892\tabularnewline
M1b & 3.86(14) & 2402(86) & 124.54(84) & 265.1(2.8) &
0.0936\tabularnewline
M2b & 3.80(13) & 2446(81) & 124.57(88) & 266.7(2.6) &
0.0895\tabularnewline
\hline
 &\multicolumn{4}{c}{$w_0$,  $\beta=1.726$ and  $\beta=1.836$}  &\\
 \hline
M1a & 3.660(69) & 2539(44) & 122.30(49) & 266.8(1.5) &
0.104\tabularnewline
M2a & 3.682(67) & 2525(43) & 122.22(51) & 266.2(1.5) &
0.0861\tabularnewline
M1b & 3.657(70) & 2541(46) & 122.30(49) & 266.9(1.6) &
0.105\tabularnewline
M2b & 3.685(68) & 2523(43) & 122.21(52) & 266.1(1.5) &
0.0851\tabularnewline
\hline
 &\multicolumn{4}{c}{$w_0$,  $\beta=1.778$ and  $\beta=1.836$, $P_5=0$}  &\\
 \hline
M1a & 3.55(11) & 2614(77) & 122.64(16) & 269.9(2.8) &
0.356\tabularnewline
M2a & 3.587(97) & 2588(68) & 122.49(17) & 268.8(2.5) &
0.323\tabularnewline
M1b & 3.53(10) & 2631(73) & 122.66(16) & 270.5(2.6) &
0.362\tabularnewline
M2b & 3.612(98) & 2571(67) & 122.44(17) & 268.1(2.5) &
0.315\tabularnewline
\hline
    \end{tabular}
    \caption{The values of  the light quark mass, $m_{ud}$, $B$, $f$ and $\Sigma$ in  the $\overline{\textrm{MS}}$ scheme at  2~GeV obtained using the different determinations of $Z_P$, labeled M1a, M1b, M2a and M2b. Results using the GF scale $ w_0$ and all the ensembles of Table~\ref{tab:valence_pion}  are given in the top most panel, using
    $t_0/ w_0$ in the second panel, using $\sqrt{t_0}$ in the third panel, using $w_0$ and limiting $m_\pi<190$~MeV in the fourth panel, using $w_0$ and only the two coarser lattice spacings in the fifth panel,  using $w_0$ and only the coarser and finest lattice spacings in the sixth panel and using $w_0$ and only the two finest lattice spacings in the last panel.}
    \label{tab:m_ud_pion}
\end{table}

We need now to average the results coming from the different analyses collected in Table~\ref{tab:m_ud_pion}.
To this end we adopt a simple generalization of Eq.~(28) of Ref.~\cite{Carrasco:2014cwa}.
For a given observable $x$ we assume that its probability distribution $f(x)$ is given by
\begin{equation}
     f(x) = \sum_{i=1}^N w_i ~ f_i(x) ~ , ~
     \label{eq:distribution}
\end{equation}
where $w_i$ are weights to be specified and $f_i(x)$ are the probability distributions corresponding to the individual analyses (labelled with $i = 1, 2, ..., N$).
It is not necessary to specify the form of the individual distributions. It suffices to know that  $\overline{x}_i$ and $\sigma_i$ are the mean value and standard deviation of the distribution $f_i(x)$.

Thus, using Eq.~(\ref{eq:distribution}) we can represent the combination of the $N$ results of the various analyses in the form
\begin{eqnarray}
    \overline{x} \pm \sigma_{\textrm{stat}} \pm \sigma_{\textrm{syst}} ~, ~
\end{eqnarray}
where
\begin{eqnarray}
      \label{eq:average}
      \overline{x} & = & \sum_{i=1}^N w_i ~ \overline{x}_i ~, ~ \\[2mm]
      \label{eq:sigma_1}
       \sigma_{\textrm{stat}}^2 & = & \sum_{i=1}^N w_i ~ \sigma_i^2 ~ , ~ \\[2mm]
      \label{eq:sigma_2}
       \sigma_{\textrm{syst}}^2 & = &  \sum_{i=1}^N w_i ~ \left( \overline{x}_i  - \overline{x} \right)^2 ~ .      
\end{eqnarray}
Eq.~(\ref{eq:sigma_1}) represents the square of a ``statistical" error given by the weighted average of the individual variances, while Eq.~(\ref{eq:sigma_2}) corresponds to the square of a ``systematic" error related to the spread among the results of the different analyses. 
The total error $\sigma$ is given by the sum in quadrature of $\sigma_{\textrm{stat}}$ and $\sigma_{\textrm{syst}}$.

Given the limited number of data points, we refrain in using the values of $\chi^2$, shown in Table~\ref{tab:m_ud_pion}, as a quantitative estimate of the quality of the various fits. Instead, since the results of Table~\ref{tab:m_ud_pion} suggest the dominance of the ``statistical" uncertainties over the ``systematic" ones, a reasonable choice for the weights $w_i$ is $w_i \propto 1 / \sigma_i^2$, namely
\begin{eqnarray}
    w_i = \frac{1}{\sigma_i^2} \cdot \frac{1}{\sum_{j=1}^N 1 / \sigma_j^2} ~ . ~
\end{eqnarray}

Thus to obtain the value of $m_{ud}$ we combine the values of Table~\ref{tab:m_ud_pion} using Eq.~(\ref{eq:average}) for the mean and  Eqs.~(\ref{eq:sigma_1}-\ref{eq:sigma_2}) for the error excluding the analyses with $\chi^2/{\rm d.o.f.}>2.5$ (which leads to a total of 21 analyses). We get in this way
\begin{eqnarray}
  m_{ud} & = & 3.689 ( 80 )_{\textrm{stat}} ( 63 )_{\textrm{syst}} ~\mbox{MeV} = 3.689(102)~\mbox{MeV}\,, \\[2mm] 
  B & = & 2516 ( 51 )_{\textrm{stat}}  ( 42 )_{\textrm{syst}} ~\mbox{MeV} = 2516(67)~\mbox{MeV}\,, \\[2mm]
  f & = & 122.82 ( 32 )_{\textrm{stat}}  ( 65 )_{\textrm{syst}} ~\mbox{MeV} = 122.82(73)~\mbox{MeV}\,, \\[2mm]
  \Sigma^{1/3} & = & 267.6 ( 1.8 )_{\textrm{stat}} ( 1.1)_{\textrm{syst}} ~\mbox{MeV} = 267.6(2.1)~\mbox{MeV}\,.
 \end{eqnarray}

\subsection{Strange quark mass}
In this Section we present our determination of the strange quark mass $m_s$. For the valence mass parameters, we evaluate correlators for $\mu_\ell$ values equal to its sea counterpart, as well as 
at three values of the quark mass parameter $\mu_s$ in the  range of the strange quark masses shown in Table~\ref{tab:valence_kaon}.
\begin{table}[htb!]
    \centering
    \begin{tabular}{|cccccc|}
\hline
Ensemble & \(a\mu_\ell\) && \(a\mu_s\) && \(am_K\)  \tabularnewline
\hline
cA211.53.24 & 0.00530 && 0.0176 0.022 0.0264 && 0.24134(47) 0.26316(47) 0.28340(47)\\
cA211.40.24 & 0.00400 && 0.0176 0.022 0.0264 && 0.23419(51) 0.25650(52) 0.27709(52)\\
cA211.30.32 & 0.00300 && 0.0176 0.022 0.0264 && 0.22810(21) 0.25089(22) 0.27185(22)\\
cA211.12.48 & 0.00120 && 0.0176 0.022 0.0264 && 0.21789(26) 0.24153(29) 0.26319(34)\\
\hline
cB211.25.32 & 0.00250 && 0.0148 0.0185 0.0222 && 0.19212(45) 0.21143(46) 0.22920(47)\\
cB211.25.48 & 0.00250 && 0.0148 0.0185 0.0222 && 0.19141(19) 0.21067(20) 0.22838(23)\\
cB211.14.64 & 0.00140 && 0.0148 0.0185 0.0222 && 0.18484(13) 0.20460(15) 0.22268(17)\\
cB211.072.64 & 0.00072 && 0.0148 0.0185 0.0222 && 0.18038(14) 0.20044(16) 0.21875(18)\\
\hline
cC211.20.48 & 0.00200 && 0.0128 0.0161 0.0193 && 0.16179(16) 0.17878(17) 0.19390(18)\\
cC211.06.80 & 0.00060 && 0.0128 0.0161 0.0193 && 0.153321(99) 0.17098(11) 0.18656(13)\\
\hline
\end{tabular}
    \caption{Values of the bare valence quark mass parameters and the corresponding values of $am_K$   for each of the ensembles used in the analysis in the kaon. The number of configuration analysed for each ensemble is the reported in Table~\ref{tab:valence_pion}.}
    \label{tab:valence_kaon}
\end{table}

\begin{figure}[htb!]
     \centering
     \includegraphics[width=0.55\textwidth]{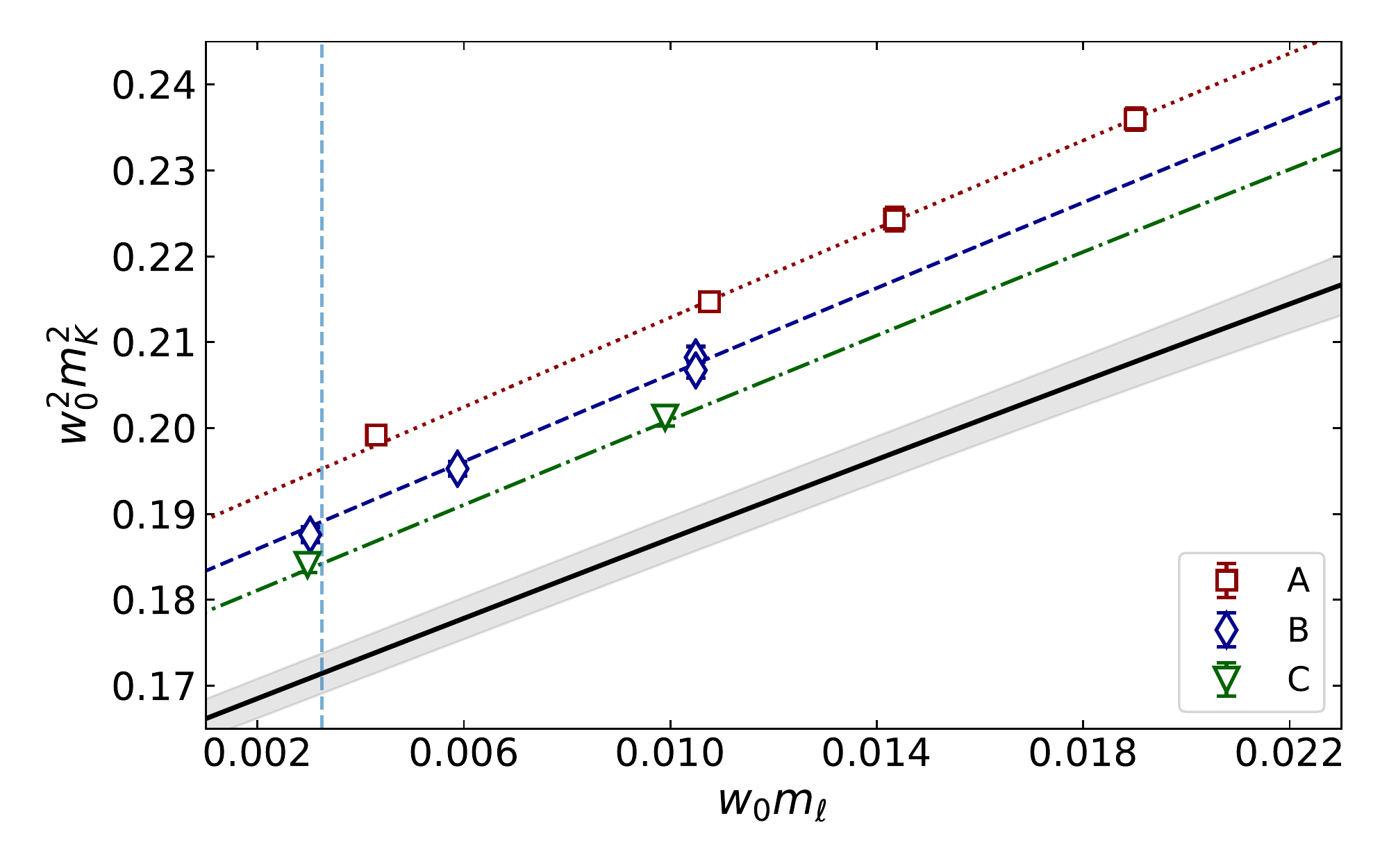}
     \vspace{-0.5cm}
     \caption{The red, blue and green solid lines show the resulting fits using Eq.~(\ref{eq:fit_MK})  for ensembles A, B and C respectively. The gray line shows the continuum extrapolation. We use  $(m_s w_0)_{ref}=0.080$ and the $Z_P$ computed with  method M2b.}
     \label{fig:MK}
 \end{figure}

For each ensemble we perform a linear interpolation of the kaon mass to three reference values of $(m_s w_0)_{ref}=0.064,0.080,0.095$ using the Ansatz
\begin{gather}
m_K^2 =a+b m_s w_0\, .
\label{eq:inter_MK_fk_mref}
\end{gather}
A similar interpolation is also performed  for the other GF scales $t_0/w_0$ and $\sqrt{t_0}$. Then for each value of $(m_s w_0)_{ref}$ we extrapolate to the continuum limit and to the isosymmentric QCD point using the value of $m_\ell = m_{ud}$ determined in the previous Section and our best fit to the data for $m_K$ according to the Ansatz
\begin{gather}
    (m_K w_0 )^2 =  P_0(m_\ell w_0+m_sw_0 )\left[ 1 +P_1 m_\ell w_0+P_2 m_\ell^2 w_0^2 +  P_3  \,a^2/w_0^2\right]\,.
    \label{eq:fit_MK}
\end{gather}
At  NLO order of SU(2) ChPT there are no finite volume effects on the kaon mass, and in Ref.~\cite{Alexandrou:2021bfr} it has been shown that the lattice QCD data on the kaon masses agree with this prediction.
The fit parameters in Eqs.~(\ref{eq:fit_MK}) are $P_0$, $P_1$, $P_2$, $P_3$, while the LO low-energy constants $f$ and $B$ are taken from our pion sector fit.
The quality of the resulting fit to Eq.~(\ref{eq:fit_MK}) is shown in Fig.~\ref{fig:MK} as an example for the specific determination of $Z_P$. Other determinations yield similar results.

 The last step of the analysis is an interpolation using Eq.~(\ref{eq:inter_MK_fk_mref}) to find the value of $m_s$ that reproduces $m_K^{\textrm{isoQCD}}=494.2(3)$ MeV given in Eq.~(\ref{eq:MK_isoQCD}).
 As in the case of the pion, to estimate the systematic errors related to the scale setting, in the chiral extrapolation and the continuum limit we repeat the analyses using two different GF scales,  excluding the ensembles with pion mass larger than $190$~MeV and the term proportional to $P_2$ in Eq.~(\ref{eq:fit_MK}), and with only pairs of values of the lattice spacing.  
 The results are shown in Table~\ref{tab:ms_kaon}.
  
\begin{table}[htb!]
     \centering
     \begin{tabular}{|c|ccccc|}
     \hline
 \(Z_P\)  && \(m_{s}\) [MeV] && ~\(m_s/m_{ud}\)~ & \(\chi^2/{\rm d.o.f.}\)\tabularnewline
\hline
M1a && 100.4(1.6) && 27.32(21) & 1\tabularnewline
M2a && 100.9(1.5) && 27.32(22) & 1\tabularnewline
M1b && 100.6(1.6) && 27.37(22) & 1.7\tabularnewline
M2b && 101.0(1.5) && 27.34(22) & 0.76\tabularnewline
\hline
     &  \multicolumn{5}{c|}{$t_0/w_0$}  \tabularnewline
\hline
M1a && 101.6(1.8) && 27.28(23) & 1.1\tabularnewline
M2a && 102.8(1.7) && 27.30(24) & 1.3\tabularnewline
M1b && 101.8(1.8) && 27.34(23) & 1.9\tabularnewline
M2b && 103.1(1.7) && 27.34(24) & 1\tabularnewline
\hline
 &  \multicolumn{5}{c|}{$\sqrt{t_0}$}  \tabularnewline
\hline
M1a && 101.0(1.7) && 27.31(22) & 1.1\tabularnewline
M2a && 101.8(1.6) && 27.32(23) & 1.2\tabularnewline
M1b && 101.2(1.7) && 27.37(22) & 1.8\tabularnewline
M2b && 102.0(1.6) && 27.35(23) & 0.9\tabularnewline
\hline
 &\multicolumn{5}{c|}{$w_0$,  $m_\pi<190$ MeV, $P_2=0$}  \tabularnewline
 \hline
M1a && 100.9(1.8) && 27.82(40) & 3.6\tabularnewline
M2a && 101.4(1.7) && 27.73(37) & 4\tabularnewline
M1b && 101.1(1.8) && 27.90(38) & 6.1\tabularnewline
M2b && 101.4(1.7) && 27.68(37) & 2.9\tabularnewline
\hline
 &  \multicolumn{5}{c|}{$w_0$, $\beta=1.726$ and $\beta=1.778$ only , $P_2=0$}  \tabularnewline
\hline
M1a && 103.4(3.7) && 26.96(29) & 0.34\tabularnewline
M2a && 103.1(3.2) && 26.94(31) & 0.46\tabularnewline
M1b && 104.1(3.4) && 26.95(29) & 0.42\tabularnewline
M2b && 102.3(3.0) && 26.94(31) & 0.52\tabularnewline
\hline
 &  \multicolumn{5}{c|}{$w_0$, $\beta=1.726$ and $\beta=1.836$ only , $P_2=0$}  \tabularnewline
\hline
M1a && 99.8(1.6) && 27.26(22) & 0.18\tabularnewline
M2a && 100.4(1.5) && 27.26(22) & 0.18\tabularnewline
M1b && 99.7(1.7) && 27.26(22) & 0.16\tabularnewline
M2b && 100.5(1.6) && 27.26(22) & 0.17\tabularnewline
\hline
 & \multicolumn{5}{|c|}{$w_0$, $\beta=1.778$ and $\beta=1.836$ only , $P_2=0$}  \tabularnewline
\hline
M1a && 97.7(2.8) && 27.54(29) & 0.21\tabularnewline
M2a && 98.8(2.4) && 27.55(30) & 0.32\tabularnewline
M1b && 97.1(2.6) && 27.53(29) & 0.28\tabularnewline
M2b && 99.5(2.5) && 27.55(30) & 0.37\tabularnewline
\hline
     \end{tabular}
      \caption{The values of  the strange quark mass, $m_{s}$, in the $\overline{\textrm{MS}}$ scheme at 2~GeV and the ratio $m_s/m_{ud}$ obtained using the different determinations of $Z_P$, labeled M1a, M1b, M2a and M2b. Results using the GF scale $ w_0$ and all the ensembles of Table~\ref{tab:valence_kaon} are given in the top most panel, using
    $t_0/ w_0$ in the second panel, using $\sqrt{t_0}$ in the third panel, using $w_0$ and limiting $m_\pi<190$~MeV in the fourth panel, using $w_0$ and only the two coarser lattice spacings in the fifth panel,  using $w_0$ and only the coarser and finest lattice spacings in the sixth panel and using $w_0$ and only the two finest lattice spacings in the last panel. To determine the ratio \(m_s/m_{ud}\) we use the values of $m_{ud}$ from Table~\ref{tab:m_ud_pion}. } 
     \label{tab:ms_kaon}
 \end{table}
  
 We use the same procedure as for $m_{ud}$ to obtain the mean value, statistical and systematic errors for $m_s$ using Eqs.~(\ref{eq:average}-\ref{eq:sigma_2}) excluding the analyses with $\chi^2/{\rm d.o.f.}> 2.5$. We find in the $\overline{\textrm{MS}}$ scheme at  2~GeV
  
\begin{eqnarray}
  m_{s} & = & 101.0 ( 1.9 )_{\textrm{stat}} ( 1.3 )_{\textrm{syst}} ~\mbox{MeV} = 101.0(2.3)~\mbox{MeV}\,, \\[2mm]
  \frac{m_{s}}{m_{ud}} & = & 27.30 ( 24 )_{\textrm{stat}} ( 14 )_{\textrm{syst}} = 27.30(28) \, .
 \end{eqnarray}

\subsection{Charm quark mass}

In this Section we present our determination of the mass of the charm quark obtained by analyzing  $D$- and $D_s$-meson masses, following a strategy similar to the one presented for the determination of $m_s$.
For the valence mass parameters, we evaluate correlators for $\mu_\ell$ values equal to its sea counterpart, as well as at four values of the quark mass parameter $\mu_c$ in the  range of the charm mass. In the case of the $D_s$ meson we also use three values of the quark mass parameter $\mu_s$ equal to the values used in the kaon analysis (see Table~\ref{tab:valence_kaon}).  The values for the $D$-meson masses are given in Table~\ref{tab:valence_D}, while the ones for the $D_s$-meson in Table~\ref{tab:valence_Ds}.
The $D$- and $D_s$-meson correlators are computed using both smeared and local interpolating fields. Using the four combinations of smeared-smeared, smeared-local and local-local correlators we construct a $2\times2$ matrix and  perform a GEVP analysis~\cite{Blossier:2009kd} to extract the mass of the $D$- and $D_s$-mesons~\footnote{As discussed in Ref.~\cite{Blossier:2009kd}, the mass of the ground state is estimated through an average over the values in the plateau region using the principal correlator corresponding to the smallest eigenvalue $\lambda_0(t,t_0)$ obtained from the GEVP for a suitable choice of the reference time $t_0$. In other words,  we fit 
$ \lambda_0(t,t_0)=C \left(e^{ - m_{D,D_s}  (t-t_0)}  + e^{ - m_{D,D_s}  ((T - t)-t_0)}\right)$.}. 
We employ Jacobi smearing for the quark fields~\cite{Allton:1993wc}, combined with APE smearing of the gauge links~\cite{FALCIONI1985624} used in the Jacobi smearing function. 
The values of $m_D$ and $m_{D_s}$ used in this analysis are reported in Table~\ref{tab:valence_D} and Table~\ref{tab:valence_Ds}.

\begin{table}[htb!]
    \centering
    \begin{tabular}{|cccccc|}
\hline
Ensemble & \(a\mu_\ell\) && \(a\mu_c\) && \(am_D\)  \tabularnewline
\hline
cA211.53.24 & 0.00530 && 0.2077 0.2336 0.2596 0.2856 && 0.7694(12) 0.8207(12) 0.8703(13) 0.9179(14)\\
cA211.40.24 & 0.00400 && 0.2077 0.2336 0.2596 0.2856 && 0.7676(13) 0.8190(15) 0.8686(16) 0.9164(18)\\
cA211.30.32 & 0.00300 && 0.2077 0.2336 0.2596 0.2856 && 0.76284(67) 0.81428(73) 0.86387(80) 0.91167(86)\\
cA211.12.48 & 0.00120 && 0.2077 0.2336 0.2596 0.2856 && 0.7567(17) 0.8078(20) 0.8570(23) 0.9044(26)\\
\hline
cB211.25.48 & 0.00250 && 0.1745 0.1962 0.2181 0.2399 && 0.6477(11) 0.6919(12) 0.7349(12) 0.7762(13)\\
cB211.14.64 & 0.00140 && 0.1745 0.1962 0.2181 0.2399 && 0.64373(70) 0.68816(79) 0.73133(87) 0.77284(96)\\
cB211.072.64 & 0.00072 && 0.1745 0.1962 0.2181 0.2399 && 0.6415(13) 0.6860(15) 0.7292(16) 0.7709(18)\\
\hline
cC211.20.48 & 0.00200 && 0.1526 0.1716 0.1907 0.2098 && 0.55934(60) 0.59821(64) 0.63589(68) 0.67237(73)\\
cC211.06.80 & 0.00060 && 0.1526 0.1716 0.1907 0.2098 && 0.5533(11) 0.5921(12) 0.6297(13) 0.6660(15)\\
\hline
\end{tabular}
    \caption{Values of the bare valence quark mass parameters and the corresponding values of $am_D$  from the GEVP analysis~\cite{Blossier:2009kd}  for each of the ensembles used in the analysis in the $D$ meson. The number of configuration analysed for each ensemble is the reported in Table~\ref{tab:valence_pion}.}
    \label{tab:valence_D}
\end{table}

\begin{table}[htb!]
    \centering
    \begin{tabular}{|cccccc|}
\hline
Ensemble & \(a\mu_\ell\) && \(a\mu_c\) && \(am_{D_s}\)  \tabularnewline
\hline
cA211.53.24 & 0.0176 && 0.2077 0.2336 0.2596 0.2856 && 0.79724(78) 0.84790(81) 0.89693(85) 0.94414(88)\\
              & 0.0220 && 0.2077 0.2336 0.2596 0.2856 && 0.80696(72) 0.85738(75) 0.90619(78) 0.95323(81)\\
              & 0.0264 && 0.2077 0.2336 0.2596 0.2856 && 0.81656(67) 0.86673(69) 0.91534(72) 0.96220(75)\\
cA211.40.24 & 0.0176 && 0.2077 0.2336 0.2596 0.2856 && 0.79838(58) 0.84918(63) 0.89825(69) 0.94561(75)\\
              & 0.0220 && 0.2077 0.2336 0.2596 0.2856 && 0.80789(50) 0.85844(54) 0.90729(59) 0.95447(64)\\
              & 0.0264 && 0.2077 0.2336 0.2596 0.2856 && 0.81733(44) 0.86763(48) 0.91626(52) 0.96326(56)\\
cA211.30.32 & 0.0176 && 0.2077 0.2336 0.2596 0.2856 && 0.79610(28) 0.84675(30) 0.89567(32) 0.94289(34)\\
              & 0.0220 && 0.2077 0.2336 0.2596 0.2856 && 0.80573(26) 0.85613(27) 0.90484(28) 0.95189(30)\\
              & 0.0264 && 0.2077 0.2336 0.2596 0.2856 && 0.81527(24) 0.86544(25) 0.91395(26) 0.96082(27)\\
\hline
cA211.12.48 & 0.0176 && 0.2077 0.2336 0.2596 0.2856 && 0.79416(39) 0.84469(42) 0.89347(45) 0.94055(48)\\
              & 0.0220 && 0.2077 0.2336 0.2596 0.2856 && 0.80385(35) 0.85417(37) 0.90278(39) 0.94973(42)\\
              & 0.0264 && 0.2077 0.2336 0.2596 0.2856 && 0.81345(32) 0.86355(33) 0.91199(35) 0.95878(37)\\
cB211.25.48 & 0.0148 && 0.1745 0.1962 0.2181 0.2399 && 0.67488(25) 0.71849(26) 0.76093(27) 0.80181(29)\\
              & 0.0185 && 0.1745 0.1962 0.2181 0.2399 && 0.68310(20) 0.72652(21) 0.76881(22) 0.80956(23)\\
              & 0.0222 && 0.1745 0.1962 0.2181 0.2399 && 0.69127(17) 0.73451(17) 0.77664(18) 0.81726(19)\\
\hline
cB211.14.64 & 0.0148 && 0.1745 0.1962 0.2181 0.2399 && 0.67415(20) 0.71771(21) 0.76010(22) 0.80094(24)\\
              & 0.0185 && 0.1745 0.1962 0.2181 0.2399 && 0.68243(18) 0.72579(19) 0.76803(20) 0.80873(21)\\
              & 0.0222 && 0.1745 0.1962 0.2181 0.2399 && 0.69064(16) 0.73381(17) 0.77589(18) 0.81646(18)\\
cB211.072.64 & 0.0148 && 0.1745 0.1962 0.2181 0.2399 && 0.67351(22) 0.71707(24) 0.75948(27) 0.80035(31)\\
              & 0.0185 && 0.1745 0.1962 0.2181 0.2399 && 0.68188(19) 0.72526(21) 0.76752(23) 0.80826(26)\\
              & 0.0222 && 0.1745 0.1962 0.2181 0.2399 && 0.69016(17) 0.73336(18) 0.77547(21) 0.81608(23)\\
\hline
cC211.20.48 & 0.0128 && 0.1526 0.1716 0.1907 0.2098 && 0.58322(25) 0.62162(26) 0.65893(27) 0.69511(28)\\
              & 0.0161 && 0.1526 0.1716 0.1907 0.2098 && 0.59050(22) 0.62874(23) 0.66592(24) 0.70199(25)\\
              & 0.0193 && 0.1526 0.1716 0.1907 0.2098 && 0.59752(20) 0.63561(21) 0.67266(21) 0.70863(22)\\
cC211.06.80 & 0.0128 && 0.1526 0.1716 0.1907 0.2098 && 0.58181(21) 0.62019(23) 0.65748(25) 0.69365(28)\\
              & 0.0161 && 0.1526 0.1716 0.1907 0.2098 && 0.58919(19) 0.62741(20) 0.66457(22) 0.70063(24)\\
              & 0.0193 && 0.1526 0.1716 0.1907 0.2098 && 0.59629(17) 0.63436(18) 0.67139(20) 0.70734(21)\\
\hline
\end{tabular}
    \caption{Values of the bare valence quark mass parameters and the corresponding values of $am_{D_s}$ from the GEVP analysis~\cite{Blossier:2009kd}    for each of the ensembles used in the analysis in the $D_s$ meson. The strange quark masses $m_s$ are the same used in the kaon sector Table~\ref{tab:valence_kaon}. The number of configuration analysed for each ensemble is the reported in Table~\ref{tab:valence_pion}.}
    \label{tab:valence_Ds}
\end{table}
Analogously to the case of the analysis for the strange quark mass determination, we interpolate the $D$ and $D_s$ masses to three reference values given by 
$(m_cw_0)_{ref}=0.94, 1.04,  1.08$ using  the Ansatz
\begin{gather}
m_{D_{s}} =a+b m_c w_0\, . 
\label{eq:inter_Md_fD_mref}
\end{gather}

For the $D_s$ meson we also perform an interpolation to the the mass $m_s$ given in Table~\ref{tab:ms_kaon} . 
At each of the reference charm quark masses, we extrapolate to the continuum and to the isospin-symmetric QCD (isoQCD) light quark mass $m_\ell = m_{ud}$ using the following polynomials in $m_\ell$
\begin{gather}
    m_D=P_{0}+P_{1} m_\ell w_0+P_{2} a^2/w_0^2 \, , \label{eq:m_D_fit}\\
    m_{D_s}=P_{0}^s+P_{1}^s m_\ell w_0+P_{2}^s a^2/w_0^2 \, , \label{eq:m_Ds_fit}
\end{gather}
where $P_j$ and $P_{j}^s,\, j=0,1,2 $ are fit parameters.
For each reference mass $(m_c w_0)_{ref}$ we compute the masses $m_D$ and $m_{D_s}$ in the continuum limit at the isoQCD value of $m_\ell = m_{ud}$ given in Table~\ref{tab:m_ud_pion}.
We then perform an interpolation in $m_c$ with the Ansatz given in Eq.~(\ref{eq:inter_Md_fD_mref}) to compute the value of $m_c$ that reproduces the isoQCD masses of the $D$ of $D_s$ mesons, given in Eqs.~(\ref{eq:MD_isoQCD}) and (\ref{eq:MDs_isoQCD}). We note that the analysis is done separately using either the 
 $D$  or the $D_s$ meson.  
 
\begin{figure}[htb!]
    \centering
    \includegraphics[width=0.5\textwidth]{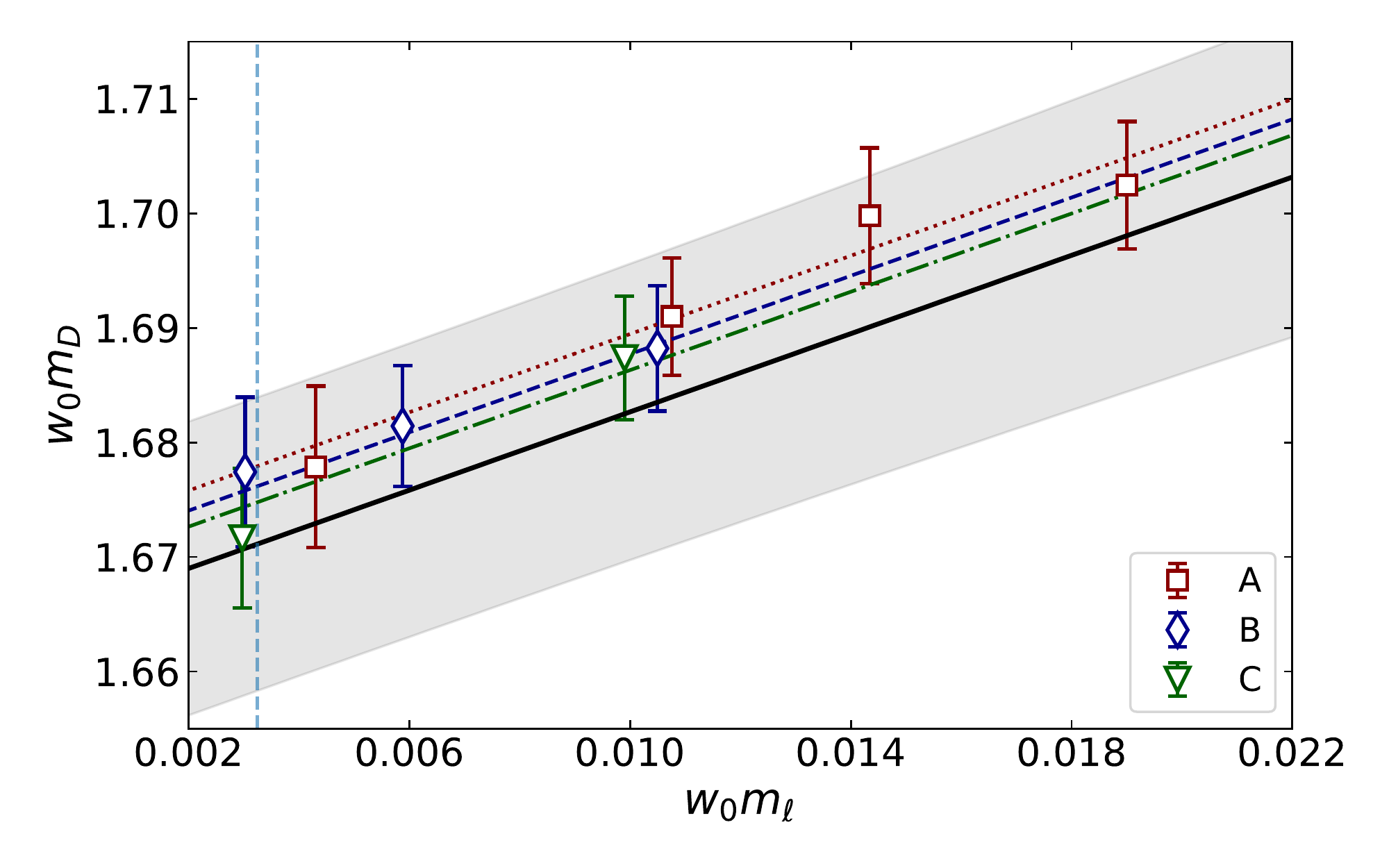}%
    \includegraphics[width=0.5\textwidth]{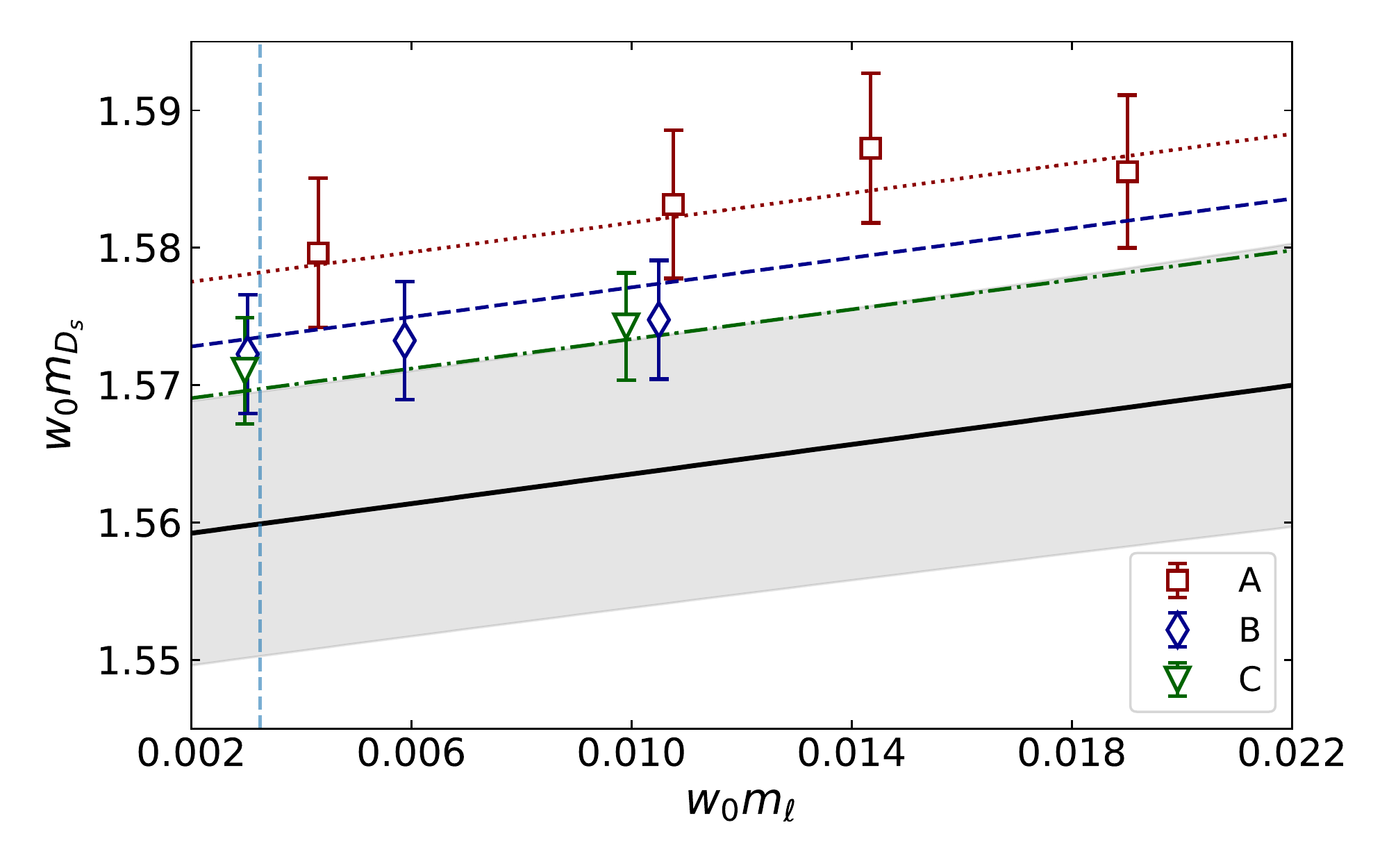}
    \vspace{-0.5cm}
    \caption{Results obtained by fitting to Eq.~(\ref{eq:m_D_fit})  for the $D$ meson (left panel)  and  Eq.~(\ref{eq:m_Ds_fit}) for the $D_s$ meson (right panel)  using  the value of $Z_P$ extracted from the M2b method and using $w_0$ to set the scale. The notation is the same as that of Fig.~\ref{fig:MK}. }
    \label{fig:D_Ds}
\end{figure}

The resulting fits to Eqs.~(\ref{eq:m_D_fit}) and (\ref{eq:m_Ds_fit}) for the $D$ and $D_s$ mesons are shown in Fig.~\ref{fig:D_Ds} for the case where $Z_P$ is determined from the M2b method. 

The values in physical units that we obtain for the  $m_c$ mass are shown in Table~\ref{tab:mc_D_Ds}.

\begin{table}[htb!]
     \centering
     \begin{tabular}{|c|ccc|ccc|}
     \hline
     \multicolumn{1}{|c|}{}  &  \multicolumn{3}{c|}{$D$} & \multicolumn{3}{c|}{$D_s$} \\
     \hline
          \(Z_P\) &\(m_{c}\) [MeV]  & \(m_c/m_{s}\) & $\chi^2/{\rm d.o.f.}$ & \(m_{c}\) [MeV]  
  & \(m_c/m_{s}\) & $\chi^2/{\rm d.o.f.}$ \tabularnewline
 \hline
 M1a & 1041(14) & 11.504(76) & 0.25 & 1039(13) & 11.476(70) &0.62\tabularnewline
M2a & 1041(13) & 11.443(76) & 0.16 & 1038(12) & 11.414(69) &0.45\tabularnewline
M1b & 1043(14) & 11.496(77) & 0.45 & 1040(13) & 11.463(71) &1.2\tabularnewline
M2b & 1039(13) & 11.418(78) & 0.14 & 1037(12) & 11.397(71) &0.21\tabularnewline
  \hline
     &  \multicolumn{6}{c|}{$t_0/w_0$}  \tabularnewline
\hline
M1a & 1042(15) & 11.385(84) & 0.12 & 1038(13) & 11.342(79) &0.36\tabularnewline
M2a & 1045(14) & 11.283(85) & 0.15 & 1041(12) & 11.235(78) &0.25\tabularnewline
M1b & 1043(15) & 11.375(86) & 0.2 & 1039(14) & 11.322(81) &0.76\tabularnewline
M2b & 1045(14) & 11.248(86) & 0.25 & 1041(13) & 11.206(80) &0.1\tabularnewline
  \hline
     &  \multicolumn{6}{c|}{$\sqrt{t_0}$}  \tabularnewline
\hline
M1a & 1041(14) & 11.448(79) & 0.16 & 1038(13) & 11.412(74) &0.48\tabularnewline
M2a & 1043(13) & 11.368(80) & 0.13 & 1039(12) & 11.329(73) &0.35\tabularnewline
M1b & 1043(14) & 11.439(81) & 0.29 & 1039(14) & 11.396(75) &0.96\tabularnewline
M2b & 1042(13) & 11.338(81) & 0.18 & 1039(12) & 11.306(75) &0.15\tabularnewline
  \hline
     &  \multicolumn{6}{c|}{$w_0$,  $m_\pi<190$ MeV, $P_2=0$}  \tabularnewline
\hline
M1a & 1042(16) & 11.46(14) & 0.099 & 1037(13) & 11.396(96) &1.3\tabularnewline
M2a & 1041(15) & 11.39(14) & 0.028 & 1035(12) & 11.333(97) &0.8\tabularnewline
M1b & 1043(16) & 11.45(14) & 0.38 & 1037(14) & 11.380(96) &2.6\tabularnewline
M2b & 1039(15) & 11.37(14) & 0.25 & 1034(12) & 11.316(98) &0.26\tabularnewline
\hline
&  \multicolumn{6}{c|}{$w_0$, $\beta=1.726$ and $\beta=1.778$ only , $P_2=0$}  \tabularnewline
\hline
M1a & 1057(31) & 11.34(16) & 0.12 & 1062(29) & 11.41(15) &0.043\tabularnewline
M2a & 1046(26) & 11.26(16) & 0.13 & 1052(24) & 11.32(15) &0.045\tabularnewline
M1b & 1063(28) & 11.34(16) & 0.11 & 1069(26) & 11.40(15) &0.04\tabularnewline
M2b & 1037(24) & 11.25(16) & 0.13 & 1042(22) & 11.31(15) &0.046\tabularnewline
\hline
 &  \multicolumn{6}{c|}{$w_0$, $\beta=1.726$ and $\beta=1.836$ only , $P_2=0$}  \tabularnewline
\hline
M1a & 1038(14) & 11.545(76) & 0.23 & 1036(13) & 11.522(66) &0.097\tabularnewline
M2a & 1039(13) & 11.487(77) & 0.24 & 1036(12) & 11.456(67) &0.096\tabularnewline
M1b & 1037(14) & 11.546(76) & 0.21 & 1035(14) & 11.523(66) &0.086\tabularnewline
M2b & 1039(13) & 11.474(77) & 0.23 & 1036(12) & 11.441(68) &0.095\tabularnewline
\hline
 &  \multicolumn{6}{c|}{$w_0$, $\beta=1.778$ and $\beta=1.836$ only , $P_2=0$}  \tabularnewline
\hline
M1a & 1029(28) & 11.68(12) & 0.082 & 1019(25) & 11.576(87) &0.0033\tabularnewline
M2a & 1037(24) & 11.65(12) & 0.12 & 1027(22) & 11.534(88)  &0.0045\tabularnewline
M1b & 1022(26) & 11.68(12) & 0.097 & 1014(23) & 11.585(86) &0.0045\tabularnewline
M2b & 1044(24) & 11.64(12) & 0.13 & 1033(21) & 11.520(89)  &0.005\tabularnewline
\hline
     \end{tabular}
     \caption{Values of the charm quark mass  $m_{c}$ in the $\overline{\textrm{MS}}$ scheme at  3~GeV and the ratio 
     \(m_c/m_{s}\) from the analysis of the $D$  and $D_s$ meson,
    for different determinations of $Z_P$ and GF-scales.  Results using the GF scale $ w_0$ and all the ensembles of Table~\ref{tab:valence_D} are given in the top most panel, using
    $t_0/ w_0$ in the second panel, using $\sqrt{t_0}$ in the third panel, using $w_0$ and limiting $m_\pi<190$~MeV in the fourth panel, using $w_0$ and only the two coarser lattice spacings in the fifth panel,  using $w_0$ and only the coarser and finest lattice spacings in the sixth panel and using $w_0$ and only the two finest lattice spacings in the last panel. To determine the ratio \(m_c/m_{s}\) we use the values of $m_{s}$ from Table~\ref{tab:ms_kaon}. 
    }
     \label{tab:mc_D_Ds}
 \end{table}

We combine all the values given in Table~\ref{tab:mc_D_Ds} excluding the analyses with $\chi^2/{\rm d.o.f.}>2.5$ as in the case of the pion and kaon with Eqs.~(\ref{eq:average}-\ref{eq:sigma_2}).
 We find 
\begin{eqnarray}
  m_{c} & = & 1039 ( 15 )_{\textrm{stat}} ( 6 )_{\textrm{syst}} ~ \mbox{MeV} = 1039(16) ~ \mbox{MeV}\,, \\[2mm]
  \frac{m_{c}}{m_{s}} & = & 11.43 ( 9 )_{\textrm{stat}} ( 10 )_{\textrm{syst}} = 11.43(13) \,, 
\end{eqnarray}
where the charm quark mass is given in the $\overline{\textrm{MS}}$ at 3~GeV.

%% file: Baryon_analysis.tex
In the baryon sector, we use the nucleon and pion masses to set the scale and determine the light quark mass. We use the $\Omega^-\,(sss)$ and the  $\Lambda_c\,(udc)$ masses to determine, respectively, the strange and charm quark masses. The range of validity of ChPT in the baryon sector is more limited as compared to that in the pion sector and thus we restrict ourselves to using pion masses up to 260~MeV.   

\subsection{Methodology}

In order to compute the baryon masses we construct the following two-point correlation functions at zero momentum, defined as
\begin{equation}
C_B^{\pm}(t) =  \sum_{\vec{x}} \left\langle0\right| \frac{1}{4}\,{\rm Tr}\left[ (1\pm\gamma_0)  J_{B(qq^\prime q{\prime\prime})}(\vec{x},t) \bar{J}_{B(qq^\prime q^{\prime\prime})}(\vec{0},0) \right] \left|0\right\rangle ,
\label{eq:corrfunc_baryons}    
\end{equation}
where $J_{B(qq^\prime q^{\prime\prime})}$ is the interpolating operator for the baryon $B(q,\,q^\prime,q^{\prime\prime})$ with $q,\,q^\prime$ and $q^{\prime\prime} \in \{l,s,c\}$. In this work, we increase statistics by considering both $\frac{1}{2}(1\pm\gamma_0)$ projectors.  For the interpolating fields of the nucleon, the $\Omega$ and the $\Lambda_c$ we take, respectively
\begin{eqnarray}
J_N&=&\epsilon^{abc}\left(u^T_a C\gamma_5 d_b\right)u_c, \nonumber \\ J_\Omega&=&\epsilon^{abc}\left(s_aC\gamma_\mu s_b\right)s_c, \\
J_{\Lambda_c}&=&\frac{1}{\sqrt{6}} \epsilon^{abc}\left[2(u^{T}_a C\gamma_5 d_b)c_c + (u^{T}_a C\gamma_5 c_a)d_c-(d^{T}_a C\gamma_5 c_b)u_c\right], \nonumber
\end{eqnarray}
where latin indices refer to colour, $\epsilon^{abc}$ is the antisymmetric tensor and $C$ is the charge conjugation matrix. 

In order to suppress contributions from excited states  we apply Gaussian smearing to each quark field $q(\vec{x},t)$. The smeared quark field is given by $q_{\rm smear}(\vec{x},t) =\sum_{\vec{y}} F(\vec{x},\vec{y};U(t))q(\vec{y},t)$, where $F$ is the gauge invariant smearing function
\begin{equation}
F (\vec{x}, \vec{y}; U (t)) = (1 + \alpha H )^n (\vec{x}, \vec{y}; U (t)), 
\label{smear}
\end{equation}
constructed from the hopping matrix understood as a matrix in coordinate, color and spin space,
\begin{equation}
H(\vec{x},\vec{y}, U(t))=\sum_{i=1}^3 \left(U_i(\vec{x}, t)\delta_{\vec{x},\vec{y}-a\hat{i}} + U^\dagger(\vec{x}-a\hat{i},t)\delta_{\vec{x},\vec{y}+a\hat{i}}\right)
\end{equation}
In addition, we apply APE smearing to the spatial links that enter the hopping matrix $H$. Different Gaussian smearing is applied to the light and strange quarks. The parameters of the Gaussian and APE smearing for each ensemble for the light and strange quarks are given in Table~\ref{tab:smearing}.
The charm quark interpolating fields are not smeared.  

\begin{table}[htb!]
    \centering
\begin{tabular}{|l|cc|cc|}
	\multicolumn{1}{c}{}                     & \multicolumn{2}{c}{Light}  & \multicolumn{2}{c}{Strange}   \\
	\hline
	Ensemble     & $n$ & $\alpha$ & $n$ & $\alpha$ \\ \hline
	cA211.30.32  &                 40          &     1.0      & 25 & 4.0       \\
	cA211.12.48  &               50         &    1.0        & 25 & 4.0       \\ \hline
	cB211.25.32  &              40          &    1.0       & 25 & 4.0        \\
	cB211.25.48  &               40         &    1.0        & 25 & 4.0       \\
	cB211.14.64  &                 70        &    1.0       & 25 & 4.0        \\
	cB211.072.64 &                 125      &      0.4   & 25 & 4.0          \\ \hline
	cC211.20.48  &                   40       &     1.0    & 25 & 1.0          \\
	cC211.06.80  &                   140       &     1.0    & 25 & 1.0          \\ \hline
\end{tabular}
    \caption{Parameters $n$ and $\alpha$ entering  the  Gaussian smearing in Eq.~(\ref{smear}) per ensemble for the light and strange quark interpolating fields. The parameters for the APE smearing are kept the same  for all ensembles. They are $n_{\rm APE}=50$ and $\alpha_{\rm APE}=0.5$. }
    \label{tab:smearing}
\end{table}

Two-point correlators for the $\Omega$ and $\Lambda_c$ are computed for each ensemble at three different values of the valence strange and charm quark masses $a\mu_s$ and $a\mu_c$. For each value, an analysis of the the two-point correlator is carried out  in order to determine the masses $m_{\Omega}$ and $m_{\Lambda_c}$ as a function of $\mu_s$ and $\mu_c$, respectively. 
The effective mass
\begin{equation}
    \label{eq:baryons_effmass}
    am_{B}^{\rm eff} = \log\left(\frac{C_B(t)}{C_B(t+a)}\right)\,,
\end{equation}
can be written using the spectral decomposition of the two-point correlators as 
\begin{equation}
    \label{eq:baryons_effmass_exstates}
    am_{B}^{\rm eff} \approx am_{B} + \log\left(\frac{1+\sum_{j=1}^{K} c_j e^{-\Delta_j t}}{1+\sum_{j=1}^{K} c_j e^{-\Delta_j(t+a)}}\right)\,, 
\end{equation}
where $\Delta_j$ is the mass difference of the $j$-th excited state with respect to the ground state mass $m_{B}$. We consider one-, two- and three-state fits by taking $K=0,1,2$ in Eq.~\eqref{eq:baryons_effmass_exstates}. This allows us to check the consistency in our determination of the ground state mass $m_B$.
Since statistical errors are larger for baryons as compared to those of mesons and grow rapidly with increasing time separation $t$, it is important to identify the ground state for as small time separation as possible, so that we can be confident that excited are sufficiently suppressed. Our procedure for identifying $m_B$ is as follows:
\begin{enumerate}
\item We keep the upper time used in the fit constant. The upper time is chosen so that statistical errors are reliably evaluated. 
\item We fit the effective mass keeping two excited states, i.e., we take $K=2$  in Eq.~\eqref{eq:baryons_effmass_exstates} and vary the lower time used in the fit $t_{\rm low}^{3st}/a$ from one to three. We choose the parameters of the fit that has  the smallest $t_{\rm low}^{3st} $ for which $\chi^2/{\rm d.o.f.}\lesssim 1$. This determines  $m_B^{3st}$.
\item 
 We then fit the effective mass including one excited state, i.e., we set $K=1$ in Eq.~\eqref{eq:baryons_effmass_exstates} and vary  $t_{\rm low}^{2st}$  for $t_{\rm low}^{2st}>t_{\rm low}^{3st}$ until the extracted mass $m_B^{2st}$ satisfies the criterion  $|m_B^{2st}-m_B^{3st}|<\delta m_B^{2st} $ where $\delta m_B^{2st}$ is the statistical error on $m_B^{2st}$ i.e. the difference in the central values of the baryon masses extracted using one and two excited states  are compatible within the statistical error of $m_B^{2st}$. 
    \item Having determined $m_B^{2st}$ we make a single state fit, i.e., we set $K=0$ in Eq.~\eqref{eq:baryons_effmass_exstates} and vary the lower value of $t$. We choose $t_{\rm low}^{1st}>t_{\rm low}^{2st}$ and take the smallest value that satisfies $|m_B^{1st}-m_B^{2st}|<\delta m_B^{1st} $, where $\delta m_B^{1st}$ is the statistical error on $m_B^{1st}$, provided $\chi^2/{\rm d.o.f.}\sim 1$. We used $m_B^{1st}$ as our final value for $m_B$.
\end{enumerate} 
\begin{figure}[htb!]
    \centering
    \includegraphics[width=0.8\linewidth]{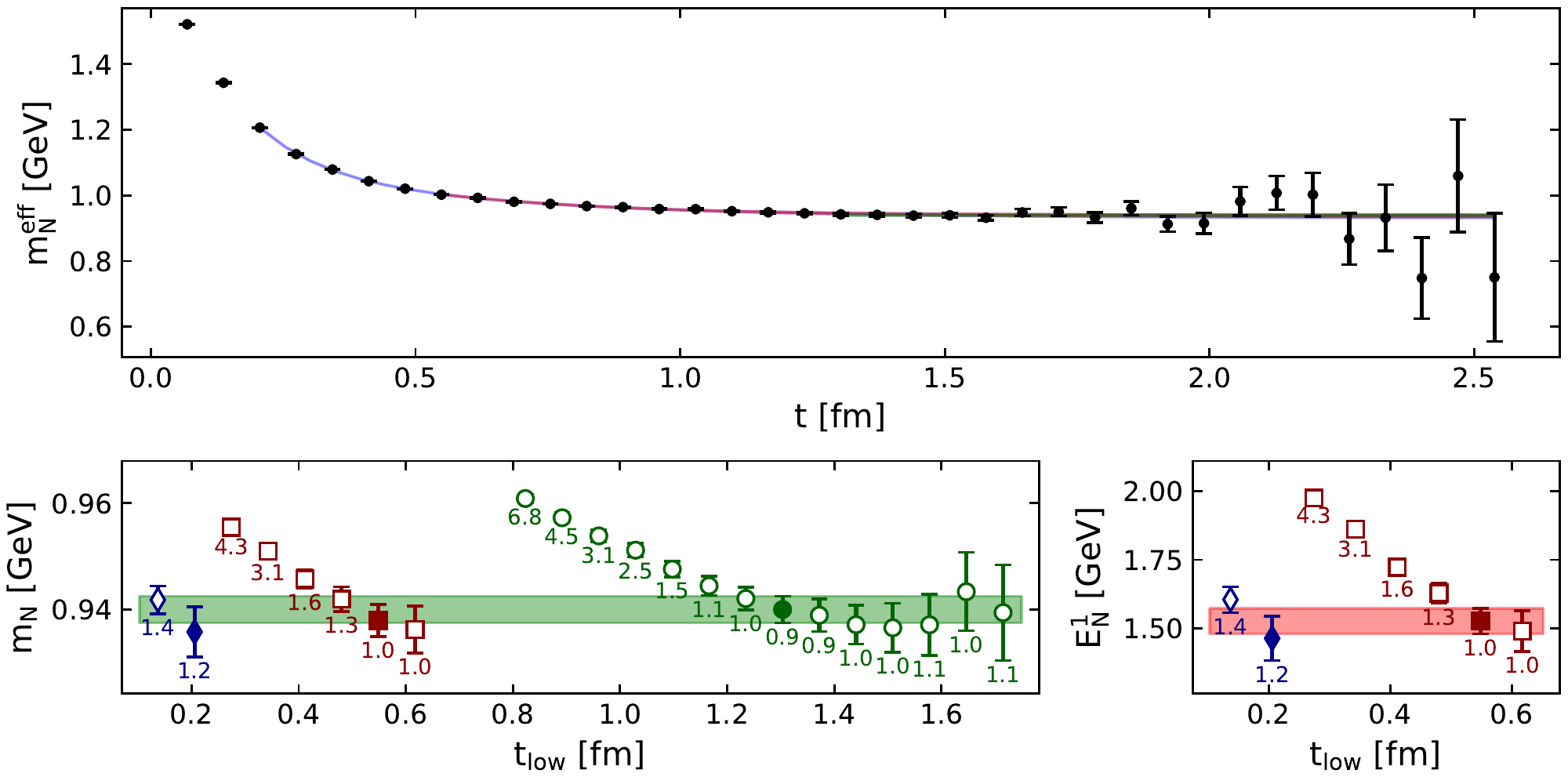}
    \vspace{-0.5cm}
    \caption{Upper panel: We show the nucleon effective mass $m_N^{\rm eff}(t)$ as a function of $t$ for the cC211.06.80 ensemble. Lower left panel: We show the convergence of the extracted value of $m_N$  as a function of the lowest time $t_{\rm low}$ used in the fit when we include one-state in the fit (green open circles), when we include two-states (open red squares) and when we include three-states (open blue rhombus). Lower right panel: The same as the lower left    panel but for the values extracted for the mass of the first excited state. The filled symbols and green and red bands show the values we pick for $m_N$ and for the mass of the excited state $m_{\rm Roper}$, respectively. For each point we give the $\chi^2/{\rm d.o.f.}$ of the fit. }
    \label{fig:nucl_fit}
\end{figure}
 We illustrate our analysis for the extraction of the masses by giving representative examples for the nucleon, $\Omega^-$ and $\Lambda_c$. In all cases we use correlation functions with smeared sources and for the ensembles listed in Table~\ref{tab:nucl_fit}. In Fig.~\ref{fig:nucl_fit} we show an example of the results obtained using the nucleon correlators for the  cC211.06.80 ensemble and in Table~\ref{tab:nucl_fit} we give the number of configurations and source positions used, the fit ranges for the one-, two- and three-state fit, as well as the extracted nucleon mass and the $\chi^2/{\rm d.o.f.}$. 
As can be seen, the mass of the first excited state converges to a value compatible with the mass of the Roper for the physical point ensembles. The nucleon-pion state, although it  has lower energy, it is volume suppressed. In our two-state fits to extract the energy of the first excited state we find that the coefficient of the second exponential compared to that of the ground state is of order 1. This is to be contrasted with the chiral perturbation theory analysis of  Ref.~\cite{Bar:2015zwa} which predicts a few percent for two-particle states. This indicates that the contribution of two-particles is suppressed.

\begin{table}[htb!]
    \centering
\begin{tabular}{|l|c|ccc|ccc|ccc|ccc|}
	\multicolumn{5}{c}{}                     & \multicolumn{3}{c}{one-state fit}  & \multicolumn{3}{c}{two-state fit}  & \multicolumn{3}{c}{three-state fit}       \\ \hline
	Ensemble  & $am_\pi$  & $n_{\rm conf}$ & $n_\text{srcs}$ & $t_{\rm max}/a$ & $t_{\rm low}/a$ & $\bar{\chi}^2$ & $am_N$ & $t_{\rm low}/a$ & $\bar{\chi}^2$ & $am_N$ & $t_{\rm low}/a$ & $\bar{\chi}^2$ & $am_N$ \\ \hline
 cA211.30.32 &  0.12525(13) & 287 & 121 & 25 & 12 & 0.9 & 0.5075(18)  & 4 & 0.8 & 0.5077(16)  & 2 & 0.8 & 0.5068(22) \\
 cA211.12.48 &  0.080281(75) & 325 & 160 & 27 & 15 & 1.1 & 0.4566(22)  & 7 & 1.4 & 0.4561(33)  & 2 & 1.4 & 0.4589(23) \\\hline
 cB211.25.32 &   0.10521(21) & 395 & 121 & 27 & 16 & 0.5 & 0.4325(38)  & 7 & 0.5 & 0.4288(39)  & 2 & 0.8 & 0.4301(47) \\
 cB211.25.48 &  0.104408(59) & 281 & 128 & 33 & 17 & 0.7 & 0.4307(18)  & 5 & 1.5 & 0.43097(94) & 2 & 1.7 & 0.4305(11) \\
 cB211.14.64  &   0.078429(38) & 194 & 128 & 33 & 18 & 2.4 & 0.4015(28)  & 7 & 1.7 & 0.3977(33)  & 2 & 1.8 & 0.3990(42) \\
 cB211.072.64 &  0.056578(20) & 751 & 264 & 34 & 18 & 1.4 & 0.3822(17)  & 7 & 1.1 & 0.3813(15)  & 2 & 1.3 & 0.3823(16) \\\hline
 cC211.20.48 &   0.086098(86) & 205 & 121 & 33 & 17 & 1.2 & 0.3664(17)  & 8 & 1.4 & 0.3669(20)  & 2 & 1.2 & 0.3658(18) \\
 cC211.06.80  &  0.047248(19) & 401 & 650 & 38 & 19 & 0.9 & 0.32679(88) & 8 & 1.0   & 0.3261(11)  & 3 & 1.2 & 0.3253(16) \\\hline
\end{tabular}
    \caption{We give for each ensemble the resulting values of $a m_N$, when using one-state (fourth main column), two-state (fifth main column) and three-state (sixth main column) fits,   $\bar{\chi}^2 \equiv\chi^2/{\rm d.o.f.}$ is the reduced $\chi^2$, $n_{\rm conf}$ is the number of configurations analyzed, $n_{\rm srcs}$ is the number of two-point functions generated per configuration at different source positions and $[t_{\rm low}, t_{\rm max}]$ is the fitting range. We also show the values for the pion mass computed on the same statistics, $am_\pi$, noticing that they are compatible with those given in Table~\ref{tab:params}. }
    \label{tab:nucl_fit}
\end{table}

\begin{figure}[htb!]
    \centering
    \includegraphics[width=1.0\textwidth]{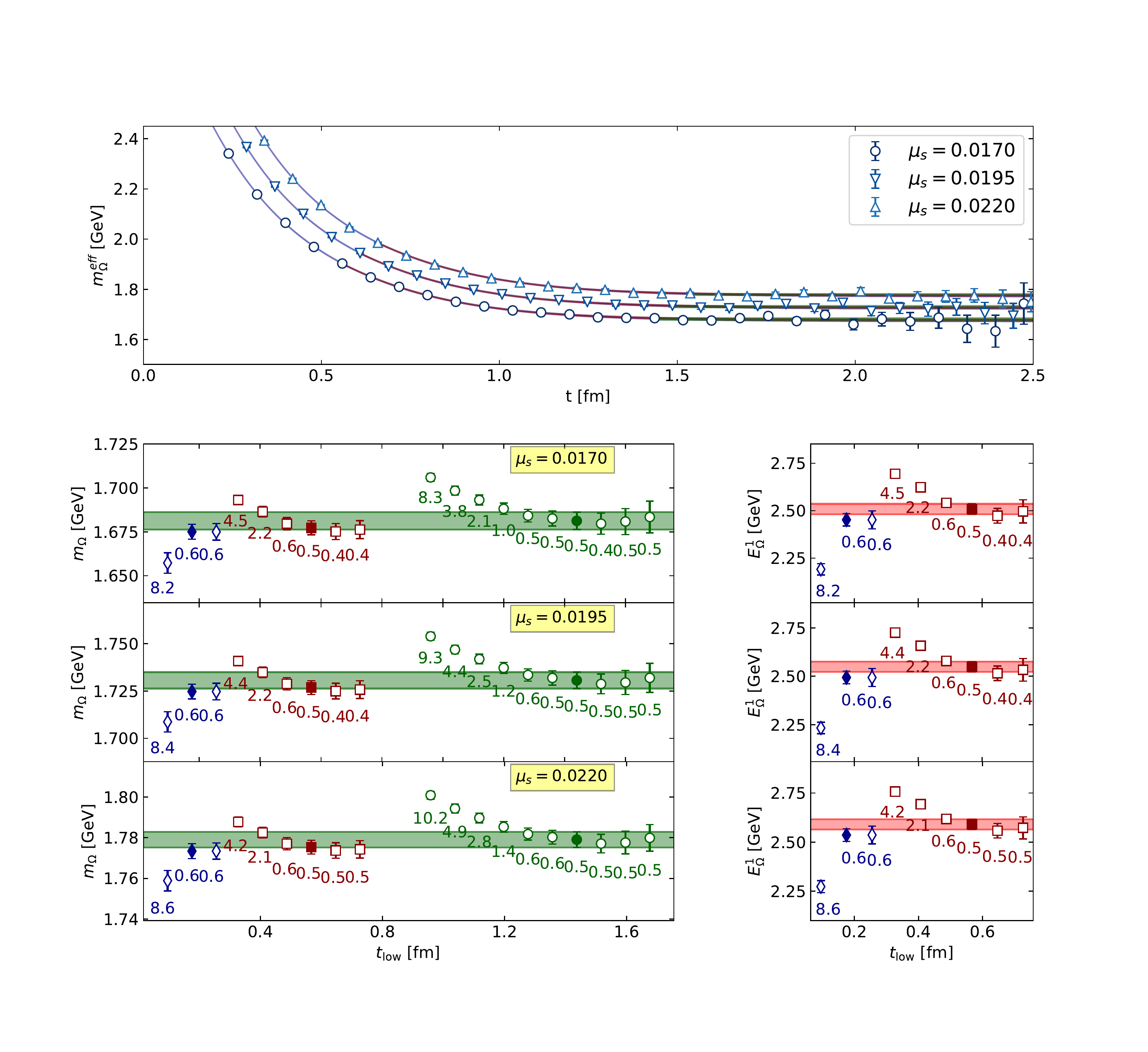}
    \vspace{-0.5cm}
    \caption{The same as in Fig.~\ref{fig:nucl_fit} but for the case of $\Omega^-$ using the cB211.072.64 ensemble for the three values of $\mu_s$ given the in the figure legend.}
    \label{fig:Extract_mass_example}
\end{figure}

We analyze in a similar way the effective mass defined by the $\Omega$ correlator given in Eq.~\eqref{eq:corrfunc_baryons}.  
In Fig.~\ref{fig:Extract_mass_example} we show an example of the effective mass $m^{\rm eff}_{\Omega}$  for the cB211.072.64 ensemble at $\mu_s=0.017$, $0.0195$ and $0.022$. As can be seen, we obtain accurate results that allow us to perform a fit including up to the second excited state. We fix the maximum time for these fits to be $t_{\rm max}/a=34$. The convergence of the effective mass for $\Omega^-$ as we vary $t_{\rm low}$ is demonstrated when using one-, two- and three-state fits. In a similar manner, the convergence of the first excited energy $E_\Omega^1$ is demonstrated by varying $t_{\rm low}$. We employ the criterion described above to choose the value of $m_\Omega$ from the one-state fit at each $\mu_s$. We note that for all the three values of $\mu_s$ we find the same $t_{\rm low}$. The masses extracted are given in Table~\ref{tab:table_baryon_mus}, where we also quote the reduced-$\chi^2$, $\bar{\chi}^2 \equiv \chi^2/{\rm d.o.f.}$, of the various fits.

\begin{table}[htb!]
    \centering
    \input{tab_mus_masses}
    \caption{We give the mass of  $\Omega^-$ in lattice units using one- (fourth main column that includes $t_{\rm low}/a$ and  the reduced $\bar{\chi}^2$), two- (fifth main column) and three- (sixth main column) state fits to the effective mass. In the second main column we give the number of configurations $n_{\rm conf}$, the number of two-point function per configuration $n_{\rm srcs}$ and $t_{\rm max}/a$. In the last main column we give the fit parameters  $A_\Omega$ in lattice units and $B_\Omega$ defined in  Eq.~\eqref{eq:baryonmass_vs_mus} using the $\Omega^-$ mass from the one state fit. }
    \label{tab:table_baryon_mus}
\end{table}
\begin{figure}[htb!]
    \centering
    \includegraphics[width=0.9\textwidth]{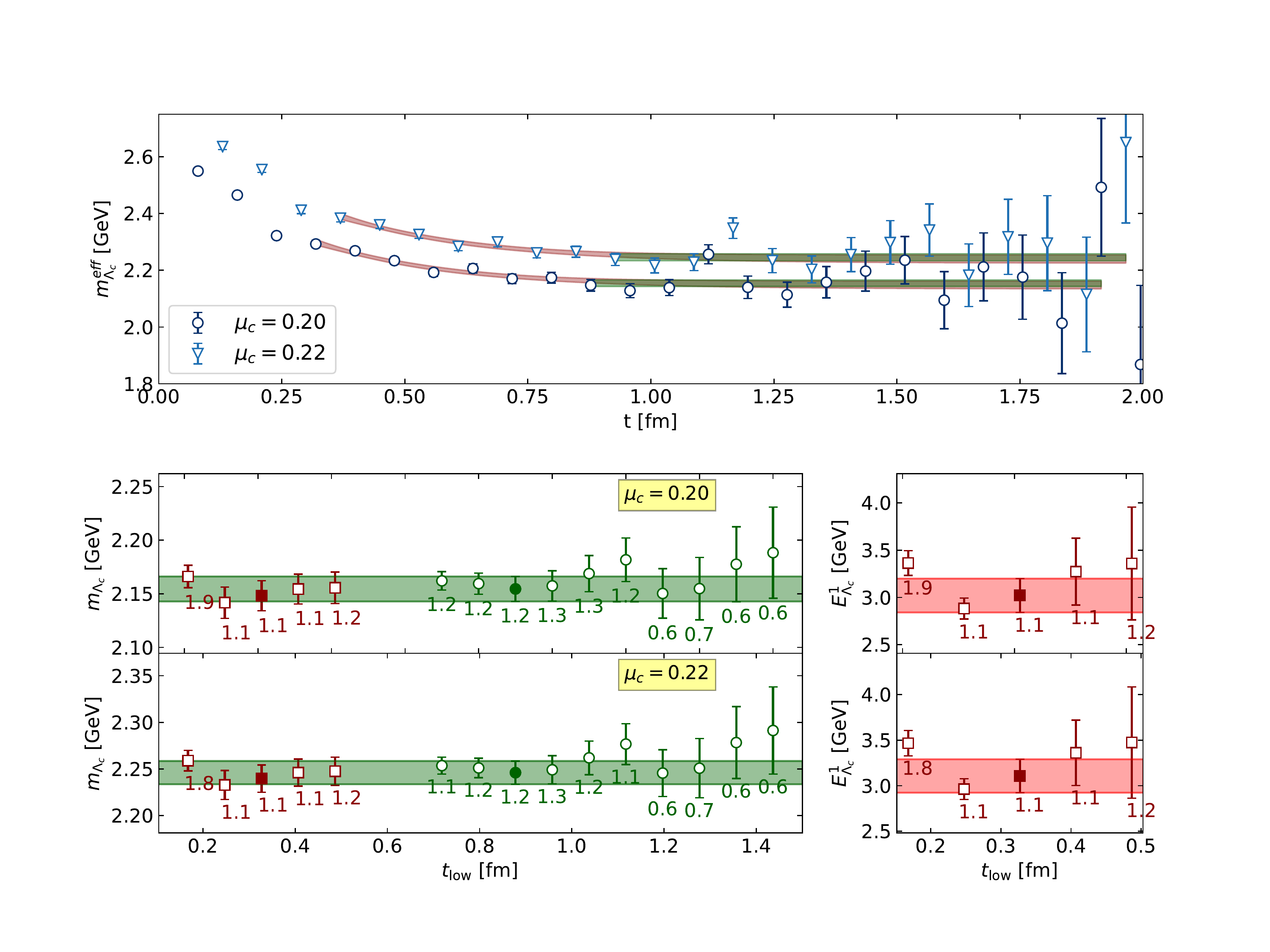}
    \vspace{-0.5cm}
    \caption{The same as in Fig.~\ref{fig:nucl_fit} but for the case of $\Lambda_c$ using the cB211.072.64 ensemble for the two values of $\mu_c$ given in the figure legend.}
    \label{fig:extr_mLambdac}
\end{figure}
\begin{table}[htb!]
    \centering
    \input{tab_muc_masses}
    \caption{The values of  $m_{\Lambda_c}(\mu_c)$ mass, statistics used and fit parameters defined in Eq.~\eqref{eq:baryonmass_vs_mus} using a similar notation as that in Table~\ref{tab:table_baryon_mus}.}
    \label{tab:table_baryon_muc}
\end{table}
The analysis of the two-point correlator for the $\Lambda_c$ proceeds in an analogous manner. We illustrate the results for the cB211.072.64 ensemble in Fig.~\ref{fig:extr_mLambdac} for two different values of the charm mass parameter $\mu_c$. From the study of the $\Omega^-$ mass we find that there is strong correlation among the data for the three values of $\mu_s$ as demonstrated in Fig.~\ref{fig:extr_mOmega} and, thus, for $\Lambda_c$ we opt to use two different values of $\mu_c$ in the interpolation. Since $\Lambda_c$ is heavier and decays faster, a three-state fit is not possible and we limit ourselves to comparing one- and two-state fits. The masses extracted, the statistics used and the value of $\bar{\chi}^2$ are given in Table~\ref{tab:table_baryon_muc}

In what follows we will use the values of $m_N$, $m_\Omega$ and $m_{\Lambda_c}$ extracted from the one-state fit given in Tables~\ref{tab:nucl_fit}, \ref{tab:table_baryon_mus} and \ref{tab:table_baryon_muc}, respectively, to determine the light, strange and charm quark masses. In order to estimate the systematic error due to the fit range we will also use the values for the masses extracted from  the one-state fit at $t_{\rm low}/a+1$.

\subsection{Light quark mass}
\begin{figure}[htb!]
    \centering
    \includegraphics[width=0.45\linewidth]{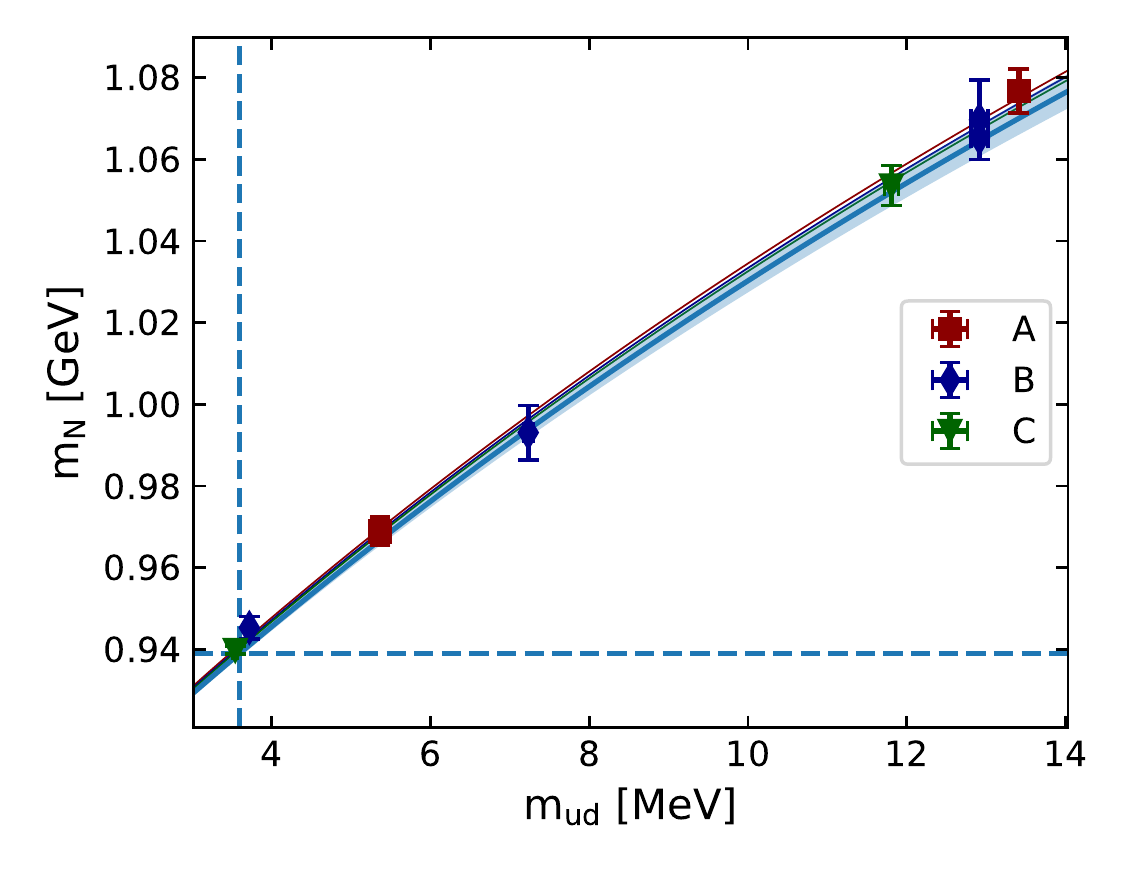}
    \includegraphics[width=0.45\linewidth]{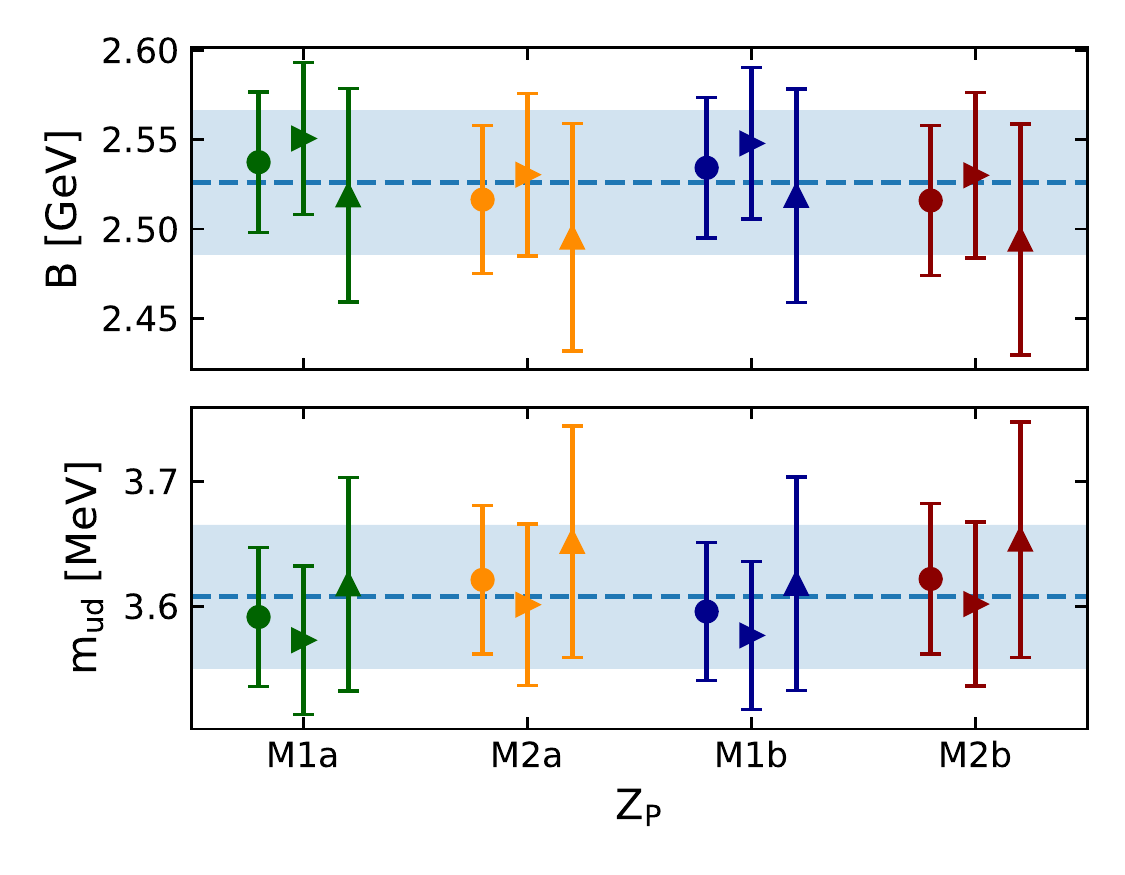}
    \vspace{-0.5cm}
    \caption{Left panel: We show the nucleon mass $m_N$ for the A- (red), B- (blue) and C- (green) ensembles. The blue band shows the continuum extrapolation according to Eq.~(\ref{eq:pion_mass_nucl}). Right panel: The values of the parameter $B$ (top) and light quark mass (bottom) for different determinations of $Z_P$. Circles show the results using the mass of the nucleon from one-state fit in Table~\ref{tab:nucl_fit}. Right-pointing triangles show results when $t_{\rm low}$ is increased by one unit, namely $t_{\rm low}/a+1$. Up-pointing triangles show  results when the chiral extrapolation is done using ensembles with pion mass lower than 190 MeV. The dashed blue line is our final value obtained by the results listed in  Table~\ref{tab:mud_nucl}.}
    \label{fig:mud_nucl}
\end{figure}

We use the ChPT expression of Eq.~\eqref{eq:nucl_fit} to extrapolate to the physical point. To one-loop order in ChPT (up to which the nucleon mass is expanded in Eq.~\eqref{eq:nucl_fit}) we can substitute the pion mass by $m^2_\pi = 2 B m_{ud}(1+c_2 a^2)$ to obtain the expansion
\begin{equation}
    m_N(m_{ud}) = m^0_{N} - 4 c_1 \left(2 B m_{ud}(1+c_2 a^2)\right) - \frac{3g^2_A}{16\pi f_\pi^2}\left(2 B m_{ud}(1+c_2 a^2)\right)^{3/2}
\label{eq:pion_mass_nucl}
\end{equation}
consistent to the order we are working and including $\mathcal{O}(a^2)$ effects both in the pion expansion in Eq.~\eqref{eq:pion_mass_nucl} with the coefficient $c_2$ and in the nucleon expansion in Eq.~\eqref{eq:nucl_fit}. We thus have two fit parameters, $B$ and $c_2$, while the lattice spacings, $m_N^0$ and $c_1$ are determined from Eqs.~\eqref{eq:nucl_fit}-\eqref{eq:c1}. The fit procedure is performed for the four values of the renormalization constant from Table~\ref{tab:ZP_MSb_19}, checking for consistency and estimating systematic effects in the determination of $Z_P$.
Since the values we obtain using the different methods of extracting $Z_P$ are in very good agreement we average over them. The statistical error in the lattice spacing is taken into account in the jackknife analysis. The fit results are reported in Table~\ref{tab:mud_nucl} and depicted in Fig.~\ref{fig:mud_nucl}. The final value of the light quark mass using the nucleon and pion mass, given in the $\overline{\rm MS}$ scheme at 2~GeV, is
\begin{equation}
 m_{ud}=3.608(58)(^{+32}_{-19})\text{~MeV}\,,
 \label{eq:value_mud_baryons}
\end{equation}
obtained by averaging the values in Table~\ref{tab:mud_nucl}. The systematic error is computed as in Eq.~(\ref{eq:sigma_2}) but in the sum we only take into account the mean values not included in the computation of the average. Namely, the  systematic error reflects the choice of the fitting range estimated by  increasing $t_{\rm low}$ by one unit and the sensitivity due to the chiral extrapolation estimated by using ensembles with pion mass smaller than 190~MeV. We will follow this procedure also for the computation of the systematic errors also for the strange and charm quark masses.
\begin{table}[htb!]
    \centering

\begin{tabular}{|c|ccc|c|}
\hline
    $Z_P$ & $\bar{\chi}^2$ &  $B$ [GeV] & $c_2$ [GeV${}^2$] &  $m_{ud}$ [MeV]  \\
\hline
 M1a & 0.7 & 2.537(39) & 0.155(98) & 3.591(56) \\
 M2a & 0.7 & 2.516(41) & 0.49(11)  & 3.621(60) \\
 M1b & 0.8 & 2.534(39) & 0.154(98) & 3.596(56) \\
 M2b & 0.7 & 2.516(42) & 0.57(11)  & 3.622(60) \\
\hline
\end{tabular}

    \caption{Fit results for the extraction of the light quark mass with the nucleon mass using the four different estimation of the renormalization constants. The fit parameters are $B$ and $c_2$. The light quark mass $m_{ud}$ is obtained as $m_\pi^{isoQCD}/(2B)$. The values are given in the $\overline{\rm MS}$ scheme at 2~GeV. }
    \label{tab:mud_nucl}
\end{table}

\subsection{Strange and charm quark masses}\label{sec:baryons_method_musc}
We determine the strange and charm quark masses using the experimental value of the $\Omega\,(sss)$ and $\Lambda_c\,(udc)$ masses and the lattice spacings determined from the nucleon mass. Namely, we use $m_{\Omega}^{(phys.)}=1672.5(3)$ and $m_{\Lambda_c}^{(phys)}=2286.5(1)$ from the PDG~\cite{10.1093/ptep/ptaa104}.  We use the renormalization constants $Z_P$ given in Table~\ref{tab:ZP_MSb_19}.

\begin{figure}[htb!]
    \centering
    \includegraphics[width=0.5\textwidth]{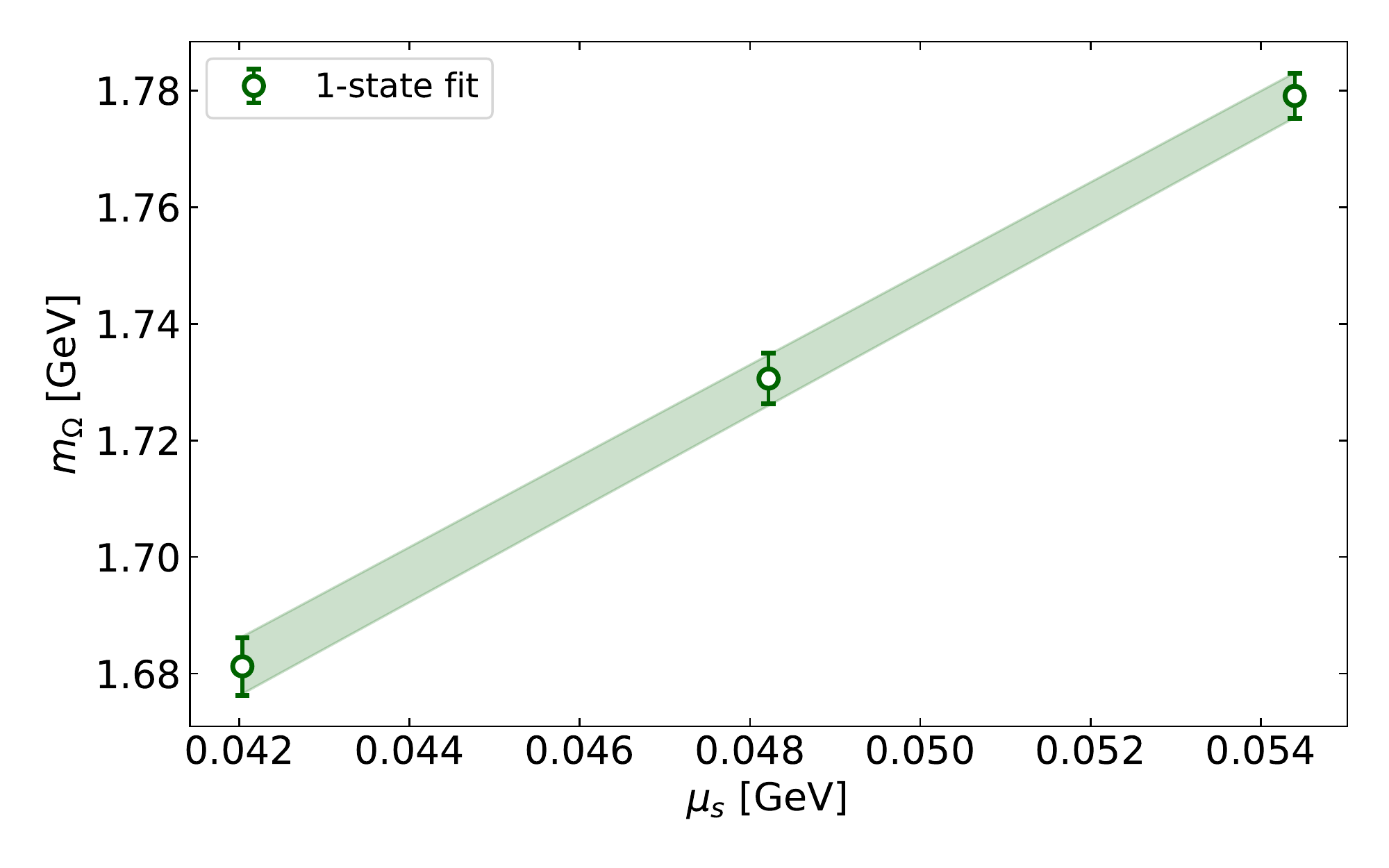}
    \vspace{-0.5cm}
    \caption{Dependence of $m_{\Omega}$ on the $\mu_s$ bare quark mass for the cB211.072.64 ensemble. Data from the 1-state are reported, together with the linear extrapolation.}
    \label{fig:extr_mOmega}
\end{figure}

We parametrize the $\Omega^-$ and $\Lambda_c$ mass dependence on the strange and charm quark mass by expanding around $\tilde{m}_s$ and $\tilde{m}_c$, that we chose to be in the same ballpark of the physical quark masses. In particular we use $\tilde{m}_s=95$ MeV and $\tilde{m}_c=1.2$ GeV, and we interpolate around these reference points using  
\begin{align}
    \label{eq:baryonmass_vs_mus}
    m_{\Omega} = A_{\Omega} +  B_{\Omega}\, (m_s-\tilde{m}_s),\\
    m_{\Lambda_c} = A_{\Lambda_c} +  B_{\Lambda_c}\, (m_c- \tilde{m}_c) \,.
    \label{eq:baryonmass_vs_muc}
\end{align}
We first discuss our procedure for the determination of the strange quark mass from the $\Omega$ mass. We then apply the same procedure for determining $m_c$ using the $\Lambda_c$ mass.

\subsubsection{Strange quark mass}

The knowledge of the $\Omega^-$ mass at three values of the valence strange quark mass parameter $\mu_s$ allows us to determine the $\Omega$ mass as a function of $\mu_s$ using the linear Ansatz of Eq.~(\ref{eq:baryonmass_vs_mus}). We show a representative example of the resulting fit in  Fig.~\ref{fig:extr_mOmega} for the ensemble cB211.072.64. The same analysis is carried out for all the ensembles listed in Table~\ref{tab:table_baryon_mus}, where we give the values of  $A_{\Omega}$ and $B_{\Omega}$, defined in Eq.~\eqref{eq:baryonmass_vs_mus}.

We employ two methods to determine $m_s$: In method I  we  perform a chiral and continuum extrapolation of the $A_\Omega$ and $B_\Omega$ parameters separately. Namely, we expand to leading order in ChPT and include  ${\cal O}(a^2)$ cut-off effects as follows 
\begin{align}
    \label{eq:expansion_m0}
    A_{\Omega} (a,m_{\pi}^2)   = c_1 + c_2 m_{\pi}^2 + c_3 a^2 \,,\\
    B_{\Omega} (a,m_{\pi}^2) = c^\prime_1 + c^\prime_2 m_{\pi}^2 +c^\prime_3a^2 \,,
    \label{eq:expansion_Zb}
\end{align}
and we limit ourselves to ensembles with $m_{\pi}<260 {\rm MeV}$ so that these leading order expression are reliable. 

\begin{figure}[htb!]
    \begin{minipage}[c]{0.5\linewidth}
       \centering
        \includegraphics[width=1.\textwidth]{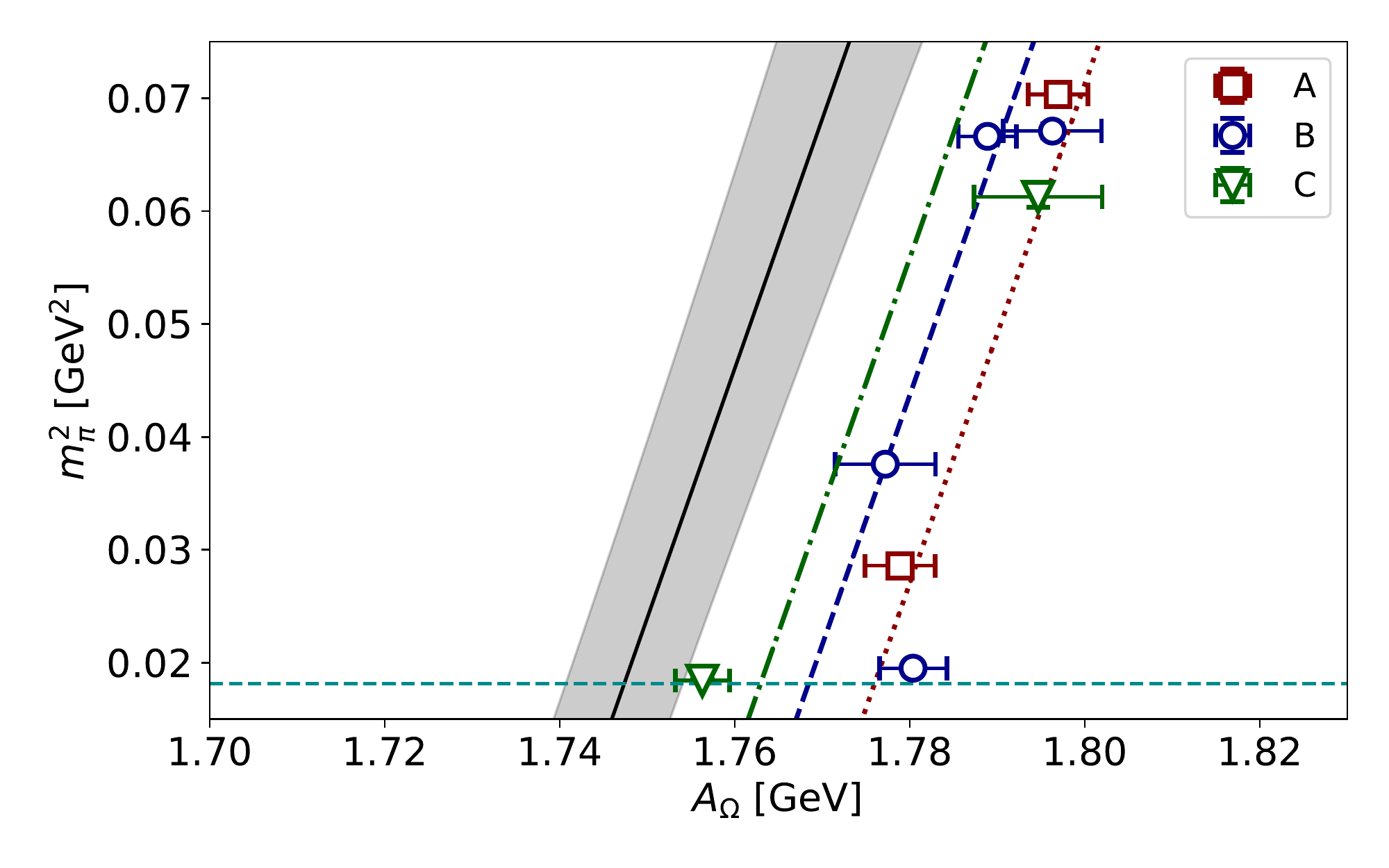}
    \end{minipage}%
    \begin{minipage}[c]{0.5\linewidth}
       \centering
        \includegraphics[width=1.\textwidth]{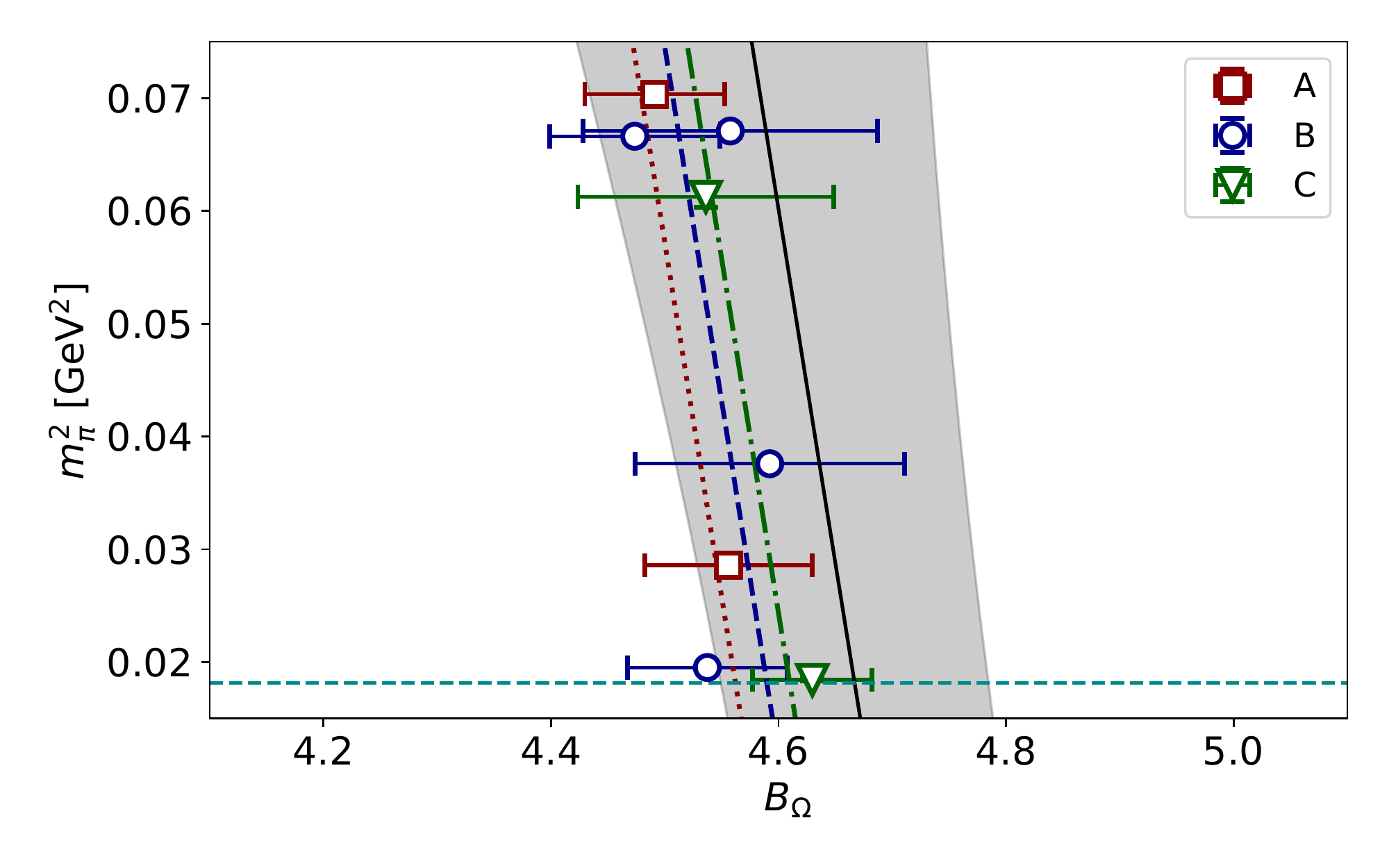}
    \end{minipage}
    
    \caption{Continuum and chiral extrapolation for the $\Omega^-$ in order to determine the coefficients of Eq.~\eqref{eq:expansion_m0} and Eq.~\eqref{eq:expansion_Zb}. We show the pion mass squared as a function of $A_\Omega$ (left) and $B_\Omega$ (right) for the A- (red), B- (blue) and C- (green) ensembles. The grey bands show the continuum extrapolation.}
    \label{fig:baryon_mus_method1}
\end{figure}

In Fig.~\ref{fig:baryon_mus_method1} we illustrate the chiral and continuum extrapolation for the parameters $A_{\Omega}$ and $B_{\Omega}$ using the value of $Z_P$ from method M1a (see Table~\ref{tab:ZP_MSb_19}). We note that the values of $A_{\Omega}$ and $B_{\Omega}$ using the cB211.025.32 and cB211.025.48 ensembles are  compatible, demonstrating that finite size effects are small.
Using the values of the parameters $A_\Omega$ and $B_\Omega$  at the physical pion mass and continuum limit  we can extract the strange quark mass  in the continuum limit and at the physical pion mass from

\begin{equation}
    m_s = \tilde{m}_s + \frac{m_{\Omega}^{(phys.)} - A_{\Omega}(0,m_{\pi}^{(phys.)})}{ B_{\Omega}(0,m_{\pi}^{(phys.)})} \,.
    \label{eq:qmass_method1}
\end{equation}

\begin{figure}[htb!]
    \centering
    \includegraphics[width=0.5\textwidth]{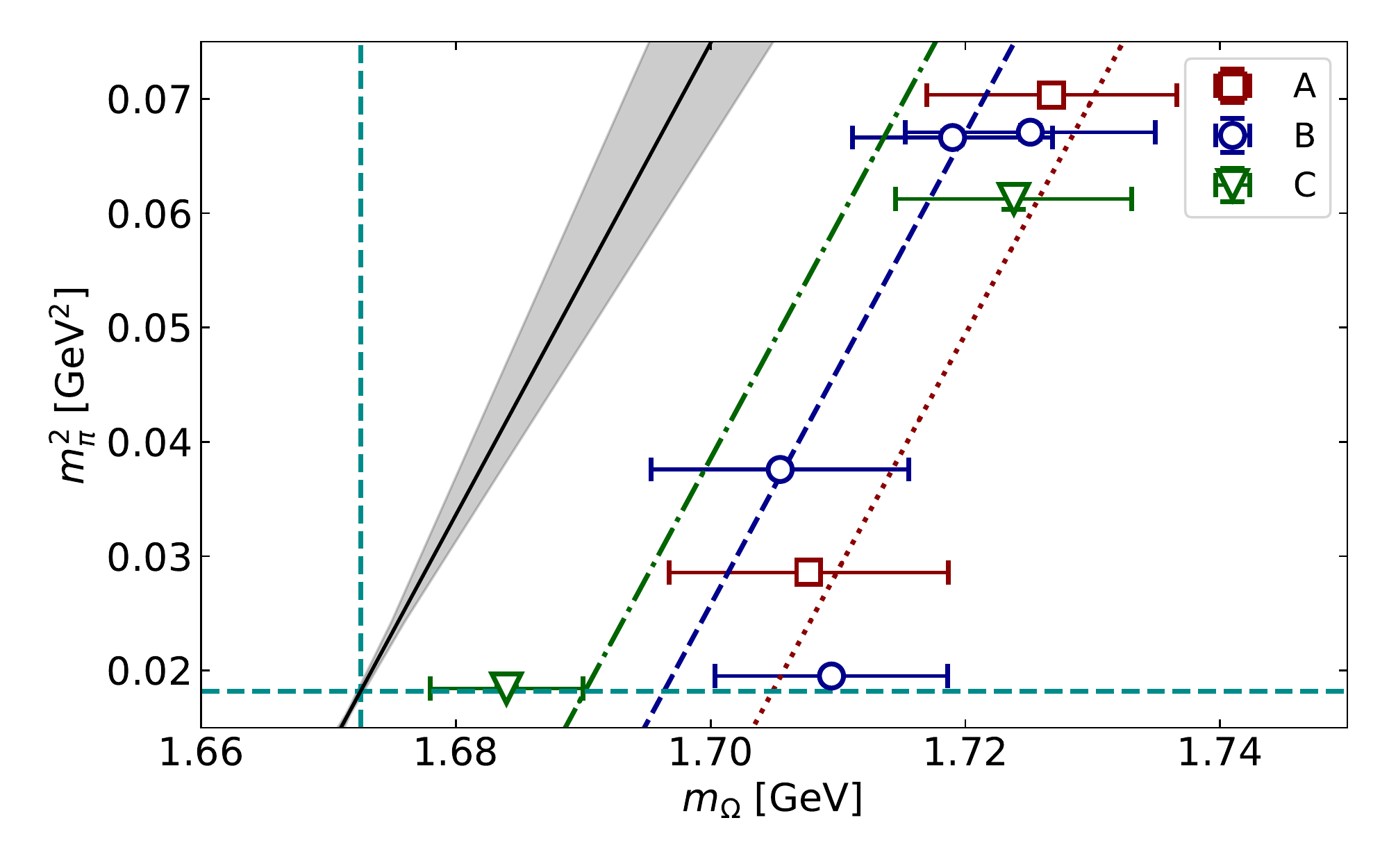}
    \vspace{-0.5cm}
    \caption{We show the mass of the $\Omega^-$, $m_{\Omega}$ at different pion mass squared for  $m_s=94.6(20)$~MeV, set by reproducing the physical mass of the $\Omega$ at the continuum limit as described in   method II. The dotted lines show the chiral extrapolation for the A- (red), B- (blue) and C- (green) ensembles. The solid black line shows the  continuum extrapolation using Eq.~\eqref{eq:qmass_method2} with the associated error (grey band). The horizontal and vertical dashed light blue lines represent, respectively, the physical pion and $\Omega$ masses.}
    \label{fig:baryon_mus_method2}
\end{figure}
In method II we adopt an iterative strategy: Namely, we start by fixing a value of the renormalized strange quark mass $m_s$ in physical units for all the ensembles. We use Eq.~\eqref{eq:baryonmass_vs_mus} to interpolate to the given $m_s$. We then extrapolate to the continuum limit and physical point using the ChPT result 
\begin{equation}
m_{\Omega} = m_{\Omega}^{(0)} - 4 c_{\Omega}^{(1)} m_{\pi}^2 + d_{\Omega}^{(2)} a^2\,.
\label{eq:qmass_method2}
\end{equation}
We iterate this procedure changing the value of $m_s$ until the resulting value of $m_\Omega$ given in Eq.~(\ref{eq:qmass_method2}) at the physical point and continuum limit matches the physical value $m_{\Omega}^{(phys.)}$. 
In Fig.~\ref{fig:baryon_mus_method2} we illustrate the analysis.


\begin{table}[htb!]
    \centering
    \input{tab_ms_method1}
    \caption{Results using method I and different values of of $Z_P$ as denoted in the first column. The second, third, fourth and fifth columns give the reduced ${\chi}^2$ of the fit to Eq.~\eqref{eq:expansion_m0} and the values of the fit parameters $c_1$, $c_2$ and $c_3$, respectively that determine $A_\Omega$ in the continuum and chiral limit.  The sixth, seventh, eight, and ninth columns give the corresponding  values for $B_\Omega$ of Eq.~\eqref{eq:expansion_Zb}. In the last column we give the extracted values for $m_s$ in $\overline{\rm MS}$ scheme at 2~GeV.}
    \label{tab:Baryon_ms_res_method1}
\end{table}

\begin{table}[htb!]
    \centering
    \input{tab_ms_method2}
    \caption{Results using method II and different values of of $Z_P$ as denoted in the first column. In the second column we give the reduced $\chi^2$ of the fit an in columns three, four, and five the fit parameters of Eq.~(\ref{eq:qmass_method2}). In the last column we give the extracted values for $m_s$ in $\overline{\rm MS}$ scheme at 2~GeV.}
    \label{tab:Baryon_ms_res_method2}
\end{table}

\begin{figure}[htb!]
    \centering
    \includegraphics[width=0.75\textwidth]{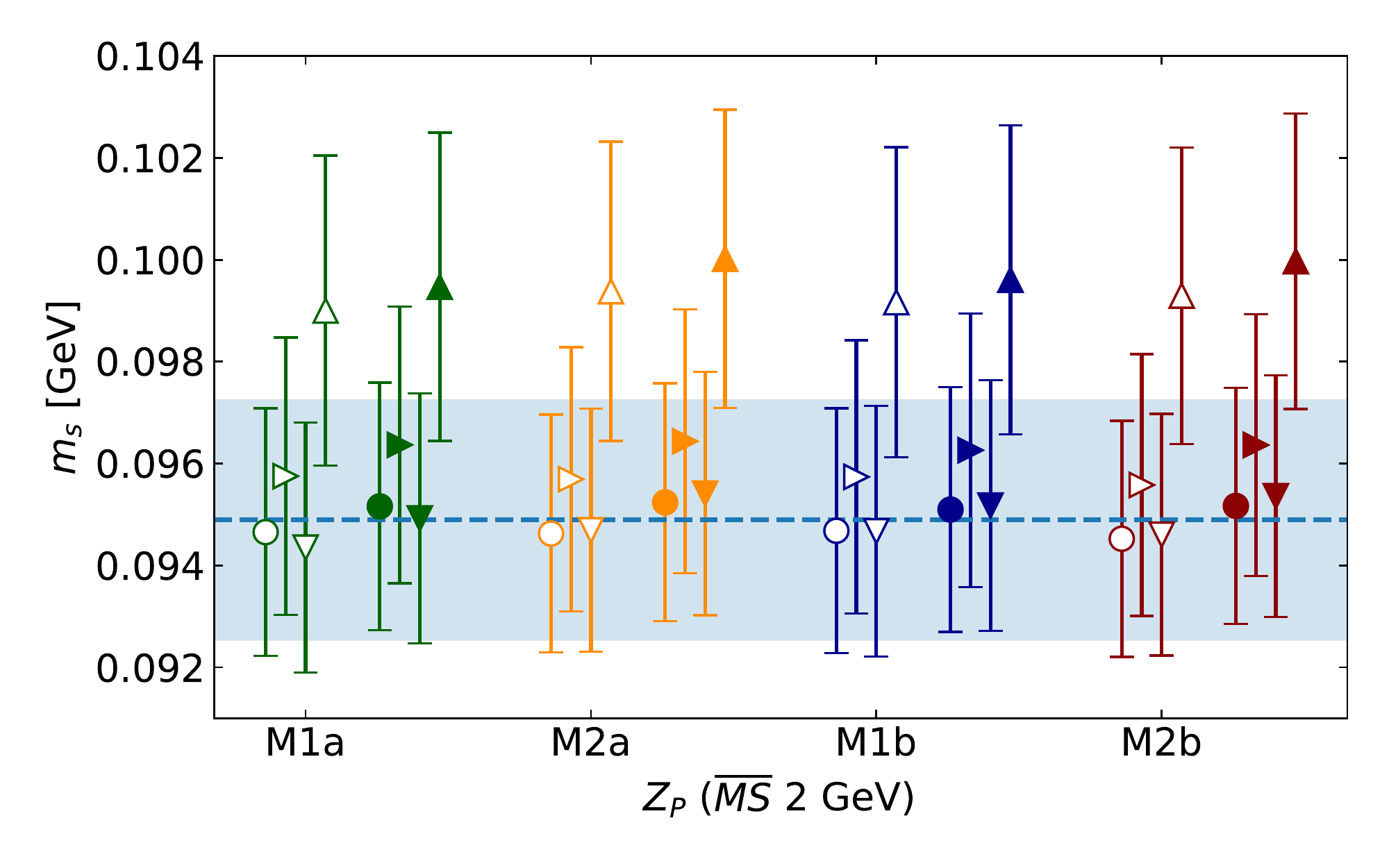}
    \vspace{-0.5cm}
    \caption{The renormalized strange quark mass for different values of $Z_P$ in the $\overline{\rm MS}$ at 2~GeV.  Symbols in green are obtained using method M1a, in orange method M2a, blue method M1b and red method M2b. For each value of $Z_P$ we give with open symbols the determination using method I to extract $m_s$ and with filled symbols using method II. 
    Circles show the results using the mass of the $\Omega$ from the  one-state fit given in    Table~\ref{tab:table_baryon_mus}. Triangles quantify systematic errors due to the selection of $t_{\rm low}$ (right pointing triangles when $t_{\rm low}/a+1$),  errors on the lattice spacing (down pointing triangles increasing by a standards deviation the lattice spacings set by the nucleon mass) and errors due to chiral extrapolation ( up pointing triangles obtained using only ensembles with $m_\pi<190$~MeV). The dashed line is the average over the values from method I and II and at different $Z_P$ given in Table~\ref{tab:Baryon_ms_res_method1} and Table~\ref{tab:Baryon_ms_res_method2}. } 
    \label{fig:baryon_mus_summary}
\end{figure}

The results for the renormalized strange quark mass in the $\overline{\rm MS}$ scheme at 2~GeV are provided in Table~\ref{tab:Baryon_ms_res_method1} using the values of $Z_P$ given in Table~\ref{tab:ZP_MSb_19}. We compare the different values by plotting them in Fig.~\ref{fig:baryon_mus_summary}. As can be seen, despite the different values of $Z_P$ at finite lattice spacing, in the continuum limit we obtain very good agreement among different estimates of $m_s$. A similar agreement is also obtained between methods I and II discussed in this Section for the determination of $m_s$. Since the error on the lattice spacing cannot be taken into account in a jackknife analysis because we used different statistics, we estimate the change in the value of $m_s$ by varying the lattice spacing by a standard deviation. As can be seen in Fig.~\ref{fig:baryon_mus_summary} this gives a very small change compared to the statistical error. By increasing $t_{\rm low}$ of the one-state fit by one lattice unit gives an estimate of the systematic error in the extraction of the mass of the $\Omega$. The change in the value of $m_s$ is well within the statistical error, as can be seen in Fig.~\ref{fig:baryon_mus_summary}. We, thus, average over all the values obtained using the four different determination of $Z_P$ and analysis methods I and II (values are given in Table~\ref{tab:Baryon_ms_res_method1} and Table~\ref{tab:Baryon_ms_res_method2}). The systematic error is computed  according to Eq.~(\ref{eq:sigma_2}) but excluding the values over which we average. As for the case of $\mu_{ud}$, the  systematic error reflects systematics due to the choice of the fitting range by letting $t_{\rm low}/a \to t_{\rm low}/a+1$ and systematics  due to the chiral extrapolation by using ensembles with $m_\pi<190$~MeV. Since the error in the lattice spacing cannot be included in the jackknife analysis due to using different statistics for the $\Omega$, we include an additional term in Eq.~(\ref{eq:sigma_2}) computed as the difference in the mean when we change the lattice spacing within its error. Using as input the $\Omega^-$ mass we obtain for $m_s$ in the $\overline{\rm MS}$ scheme at 2~GeV and the ratio $m_s/m_{ud}$ the values
\begin{equation}
    m_s = 94.9(2.4)(^{+4.1}_{-1.0})~\text{MeV}\,, \hspace*{0.3cm} 
m_s/m_{ud} = 26.30(61)(^{+1.17}_{-33})\,,
    \label{eq:ms_avgZPs}
\end{equation}
where for the ratio we use $m_{ud}$ from Eq.~\eqref{eq:value_mud_baryons}. The error on the ratio is computed by combining in quadrature the errors on $m_{ud}$ and $m_s$.

\subsubsection{Charm quark mass}

\begin{table}[htb!]
    \centering
    \input{tab_mc_method1}
    \caption{The same as in Table~\ref{tab:Baryon_ms_res_method1} but for the $A_{\Lambda_c}$ and $B_{\Lambda_c}$ parameters. In the last two columns, the values for $m_c$ in $\overline{\rm MS}$ scheme at 3 GeV are reported. }
    \label{tab:Baryon_mc_res_method1}
\end{table}

\begin{table}[htb!]
    \centering
    \input{tab_mc_method2}
    \caption{The same as in Table~\ref{tab:Baryon_ms_res_method2} but for the ${m_{\Lambda_c}}$ extrapolation, according to Eq.~\eqref{eq:qmass_charm_method2}. In the last column, we give the values for $m_c$ in the $\overline{\rm MS}$ scheme at 3 GeV.}
    \label{tab:Baryon_mc_res_method2}
\end{table}
\begin{figure}[htb!]
    \begin{minipage}[c]{0.5\linewidth}
       \centering
        \includegraphics[width=1.\textwidth]{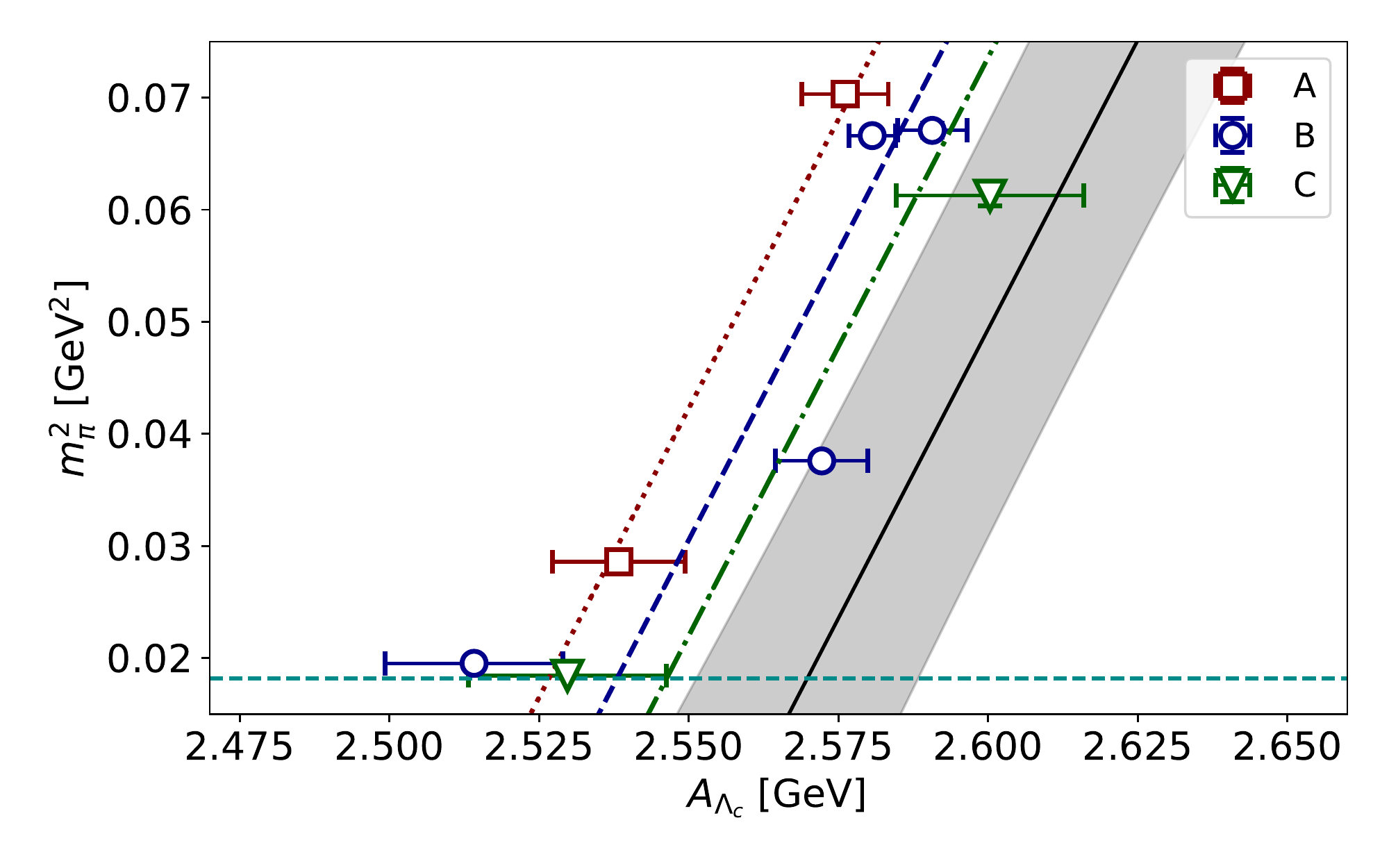}
    \end{minipage}%
    \begin{minipage}[c]{0.5\linewidth}
       \centering
        \includegraphics[width=1.\textwidth]{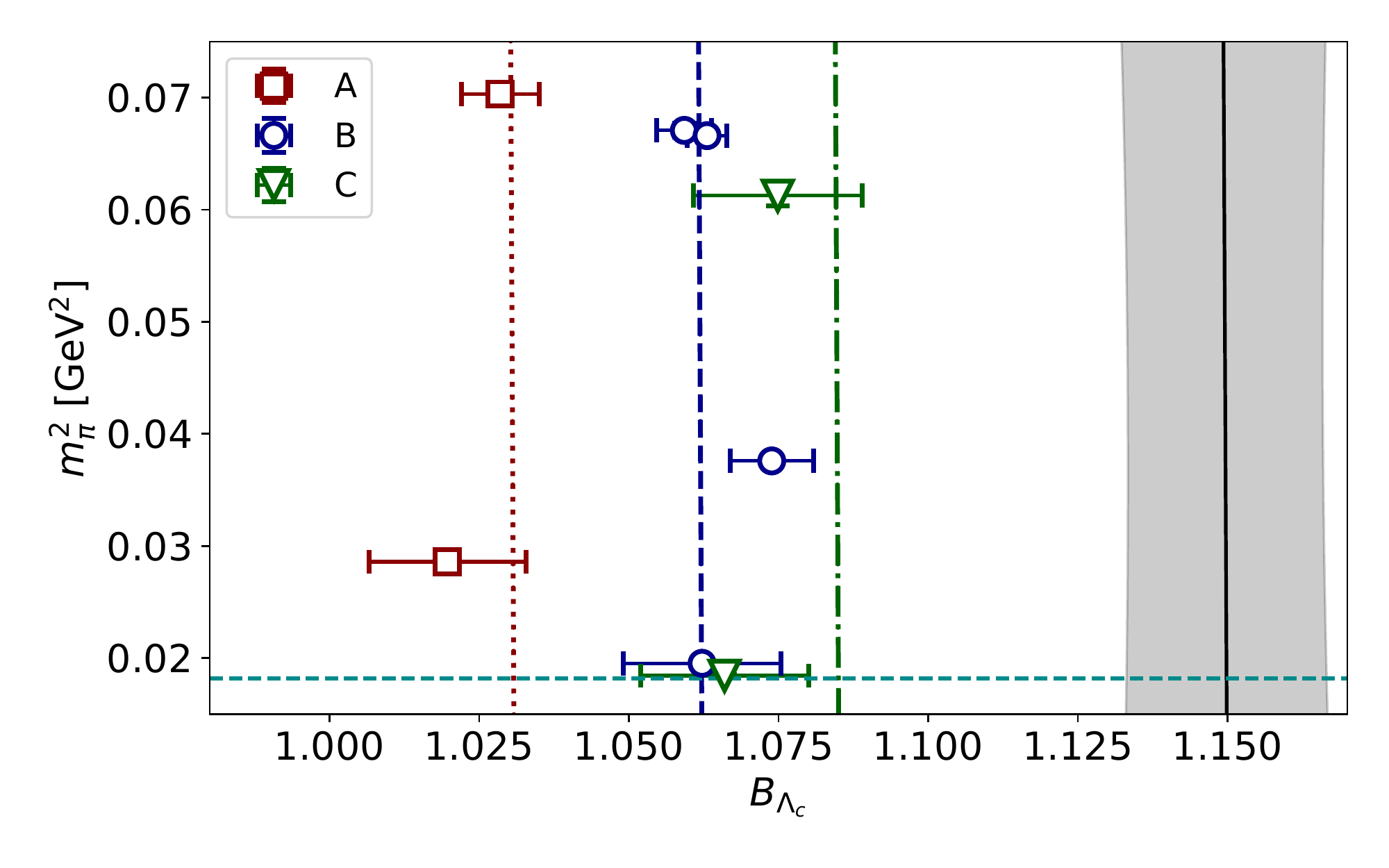}
    \end{minipage}%
    \vspace{-0.5cm}
    \caption{ The same as for Fig.~\ref{fig:baryon_mus_method1} but for the case of $\Lambda_c$.}
    \label{fig:baryon_muc_method1}
\end{figure}
\begin{figure}[htb!]
    \centering
    \includegraphics[width=0.5\textwidth]{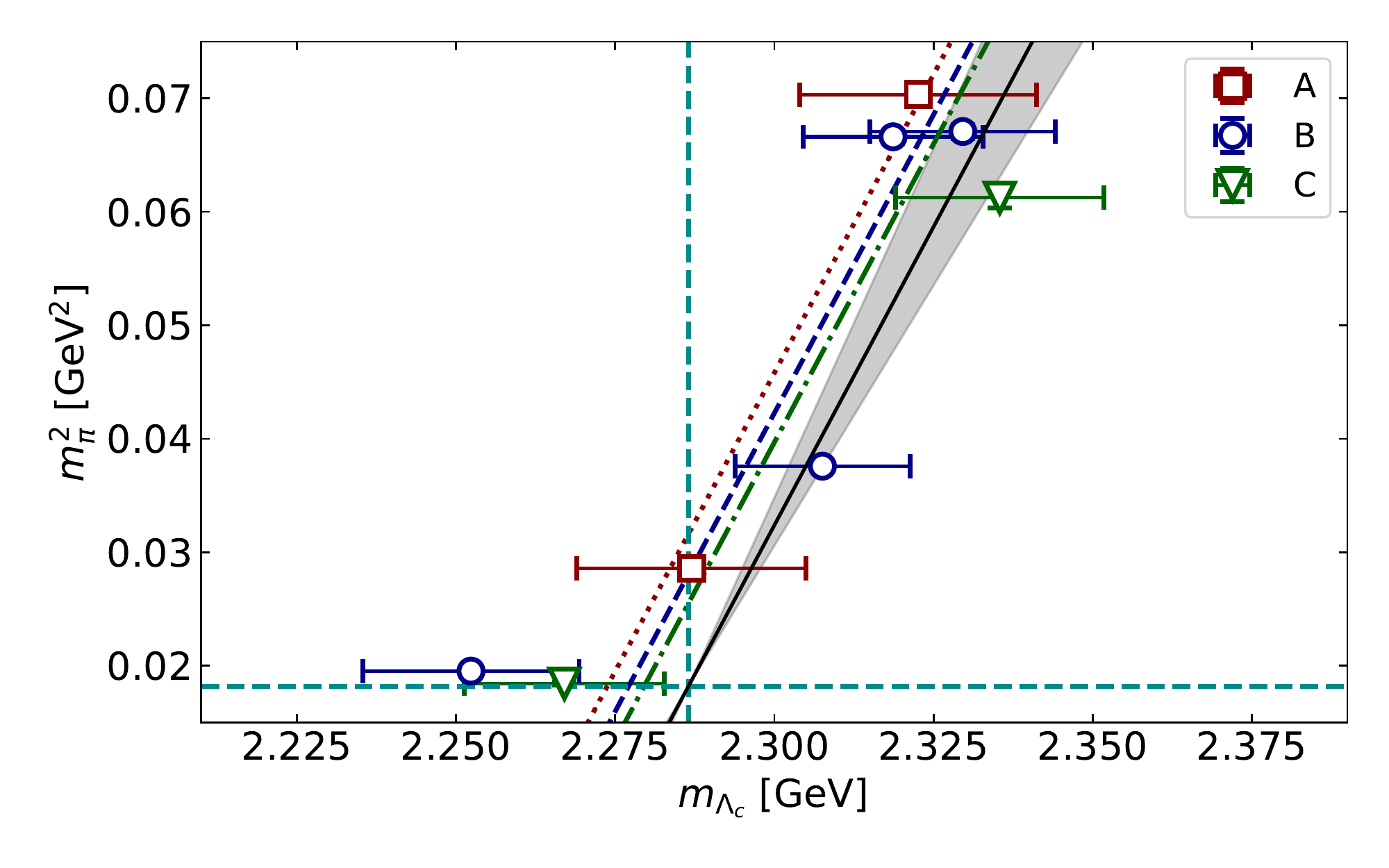}
    \vspace{-0.5cm}
    \caption{ The same as for Fig.~\ref{fig:baryon_mus_method2} but for the case of $\Lambda_c$.}
    \label{fig:baryon_muc_method2}
\end{figure}
\begin{figure}[htb!]
    \centering
    \includegraphics[width=0.75\textwidth]{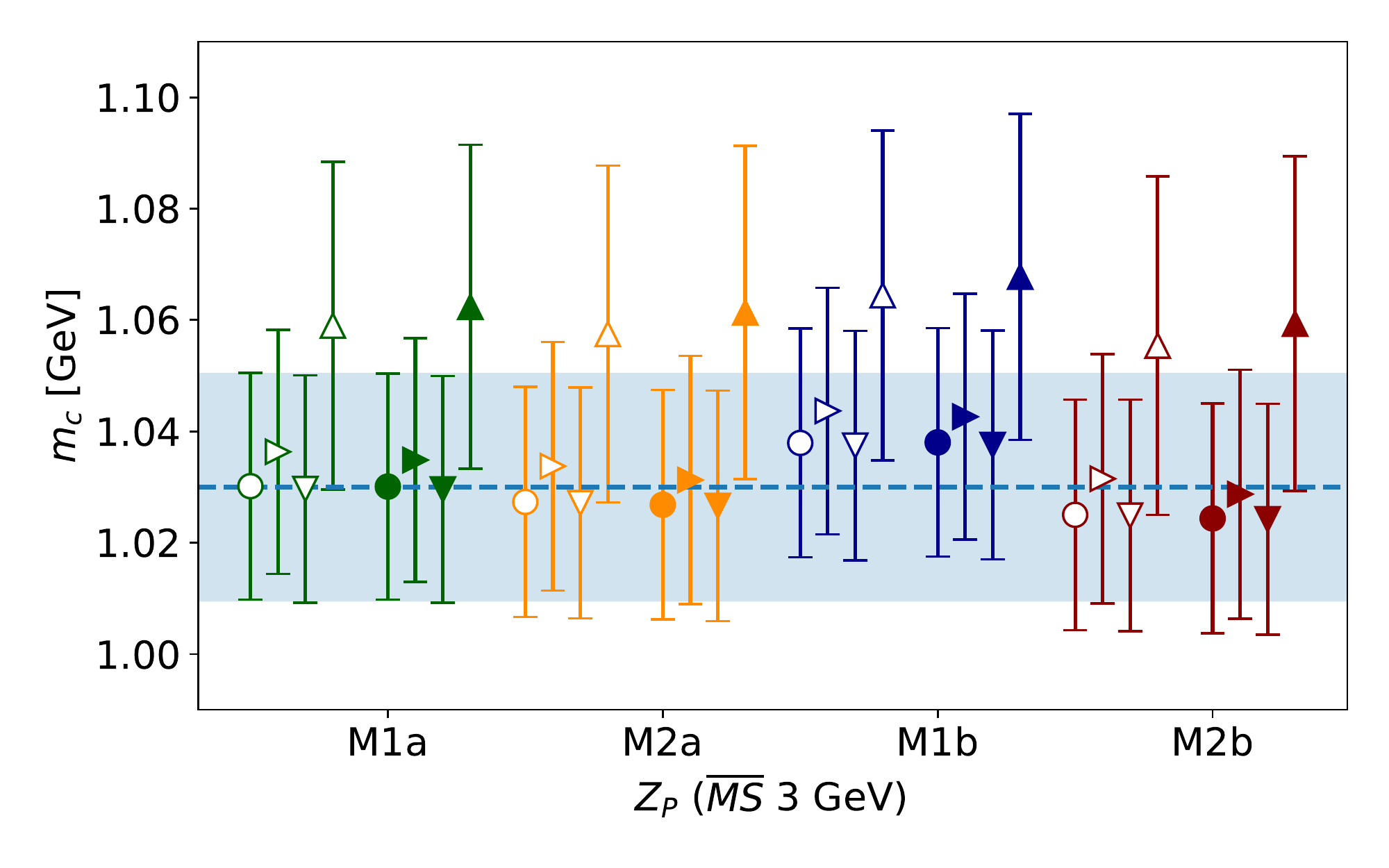}
    \vspace{-0.5cm}
    \caption{We show the values of $m_c$ for different determinations of $Z_P$ and for methods I and II. The notation is the same as that of Fig.~\ref{fig:baryon_mus_summary}. }
    \label{fig:mc_syst}
\end{figure}
We employ the same procedure for the determination of $m_c$ as described in the previous Section for $m_s$ using methods I and II. The mass of $\Lambda_c$ is interpolated linearly in $\mu_c$ using Eq.~(\ref{eq:baryonmass_vs_muc}) within the range spanned by the two $\mu_c$ values. For the chiral extrapolation, we consider a similar expression to that used for $\Omega^-$ as given in Eq.~\eqref{eq:qmass_method2}, namely
\begin{equation}
m_{\Lambda_c} = m_{\Lambda_c}^{(0)} + c_{\Lambda_c}^{(1)} m_{\pi}^2 + d_{\Lambda_c}^{(2)}  a^2\,.
\label{eq:qmass_charm_method2}
\end{equation}
We note that to leading one-loop in ChPT a  $m_\pi^3$-term with an unknown coefficient is present. Including such term in the fit results in a coefficient consistent with zero and a $\chi^2/{\rm d.o.f.}=7.2$.  That such a term is not supported by lattice QCD data  was also found in our previous analysis using a larger set of pion masses~\cite{Alexandrou:2014sha} where this coefficient was found to be consistent with zero. We check that including it does not change the extracted value for $m_c$. Thus, given the larger $\chi^2$ we drop it from our analysis.

Results from method I are reported in Table~\ref{tab:Baryon_mc_res_method1} for all the values of $Z_P$ listed in Table~\ref{tab:ZP_MSb_19}. We also illustrate in Fig.~\ref{fig:baryon_muc_method1} the chiral and continuum extrapolations of $A_{\Lambda_c}$ and $B_{\Lambda_c}$ according to Eq.~\eqref{eq:expansion_m0} and Eq.~\eqref{eq:expansion_Zb} respectively, with $Z_P$ determined using method M1a. The determination of the values of the parameters of Eq.~\ref{eq:qmass_charm_method2} for method II is carried out as for the case of $m_s$ and the results are reported in Table~\ref{tab:Baryon_mc_res_method2} and, for the M1a case, also in Fig.~\ref{fig:baryon_muc_method2}.

The values for $m_c$ from methods I and II as well as how they change by varying the lattice spacings by a standard deviation and by the change in $m_{\Lambda_c}$ by increasing $t_{\rm low}$ by one lattice spacing  in the one-state fit  are presented in Fig.~\ref{fig:baryon_muc_method2}

Again, we observe a very good agreement between the results obtained via method I and method II and among different determinations of the $Z_P$ renormalization constants. We thus average over these values  and compute the systematic error in the same way as for $\mu_s$. We obtain for the charm quark mass $m_c$ in the $\overline{\rm MS}$ scheme at $3$ GeV and the ratio $m_c/m_s$ the following values
\begin{equation}
    m_c = 1030(21)(^{+22}_{-5})~{\rm MeV}\,,\hspace*{0.3cm}   
    m_c/m_s = 12.05(31)(^{+58}_{-15})\,,
\end{equation}
where the errors on the ratio  are combined in quadrature.

%% file: tab_mus_masses.tex
\begin{tabular}{|l|lcc|c|cll|cll|cll|ll|}
 \multicolumn{5}{c}{} & \multicolumn{3}{c}{one-state fit} & 
 \multicolumn{3}{c}{two-state fit} &
 \multicolumn{3}{c}{three-state fit} \\
\hline
Ensemble & $n_{\rm conf}$ & $n_{\rm srcs}$ & $t_{\rm max}/a$ & \multicolumn{1}{c|}{$a\mu_s$} & $t_{\rm low}/a$ & \multicolumn{1}{c}{$\bar{\chi}^2$} & \multicolumn{1}{c|}{$am_{\Omega}$} & $t_{\rm low}/a$ & \multicolumn{1}{c}{$\bar{\chi}^2$} & \multicolumn{1}{c|}{$am_{\Omega}$} & $t_{\rm low}/a$ & \multicolumn{1}{c}{$\bar{\chi}^2$} & \multicolumn{1}{c|}{$am_{\Omega}$} & \multicolumn{1}{c}{$aA_{\Omega}$} & \multicolumn{1}{c|}{$B_{\Omega}/Z_P$} \\ 
\hline
cA211.30.32 & 1260 & 16 & 28 & 0.0182 &                     14 &                    1.0 &             0.7890(22) &                      4 &                    1.1 &             0.7874(19) &                      2 &                    1.1 &             0.7859(30) &  0.645(4) &   7.89(11) \\
            &      &    &    & 0.0227 &                     14 &                    1.3 &             0.8252(18) &                      4 &                    1.2 &             0.8237(17) &                      2 &                    0.8 &             0.8230(22) &       &        \\
            &      &    &    & 0.0273 &                     14 &                    1.2 &             0.8610(15) &                      4 &                    1.1 &             0.8595(15) &                      2 &                    0.9 &             0.8590(19) &       &        \\
\hline
cA211.12.48 & 341  & 64 & 29 & 0.0182 &                     16 &                    0.4 &             0.7797(27) &                      4 &                    0.6 &             0.7796(15) &                      2 &                    0.5 &             0.7783(25) &  0.634(5) &   8.00(13) \\
            &      &    &    & 0.0227 &                     16 &                    0.5 &             0.8164(21) &                      4 &                    0.6 &             0.8164(13) &                      2 &                    0.6 &             0.8155(21) &       &        \\
            &      &    &    & 0.0273 &                     16 &                    0.6 &             0.8527(17) &                      4 &                    0.6 &             0.8527(11) &                      2 &                    0.6 &             0.8521(17) &       &        \\
\hline
cB211.25.32 & 492  & 16 & 28 & 0.0148 &                     16 &                    0.9 &             0.6684(36) &                      8 &                    1.2 &             0.6652(66) &                      2 &                    1.0 &             0.6630(86) &  0.551(7) &   7.94(23) \\
            &      &    &    & 0.0185 &                     16 &                    0.9 &             0.6986(28) &                      8 &                    1.1 &             0.6959(46) &                      2 &                    0.9 &             0.6942(55) &       &        \\
            &      &    &    & 0.0222 &                     16 &                    0.8 &             0.7274(23) &                      8 &                    1.0 &             0.7250(37) &                      2 &                    0.9 &             0.7236(42) &       &        \\
\hline
cB211.25.48 & 651  & 32 & 34 & 0.0148 &                     18 &                    1.6 &             0.6666(21) &                      6 &                    1.0 &             0.6656(17) &                      2 &                    1.1 &             0.6649(24) &  0.551(4) &   7.79(13) \\
            &      &    &    & 0.0185 &                     18 &                    1.4 &             0.6959(16) &                      6 &                    0.9 &             0.6947(14) &                      2 &                    0.9 &             0.6939(22) &       &        \\
            &      &    &    & 0.0222 &                     18 &                    1.2 &             0.7244(13) &                      6 &                    0.8 &             0.7232(13) &                      2 &                    0.9 &             0.7224(20) &       &        \\

\hline
cB211.14.64 & 446  & 16 & 32 & 0.0148 &                     18 &                    0.4 &             0.6603(35) &                      6 &                    0.8 &             0.6583(31) &                      2 &                    0.7 &             0.6528(63) &  0.542(6) &   8.00(21) \\
            &      &    &    & 0.0185 &                     18 &                    0.5 &             0.6905(28) &                      6 &                    0.8 &             0.6883(27) &                      2 &                    0.7 &             0.6838(53) &       &        \\
            &      &    &    & 0.0222 &                     18 &                    0.5 &             0.7197(23) &                      6 &                    0.8 &             0.7174(25) &                      2 &                    0.7 &             0.7136(46) &       &        \\
\hline
cB211.072.64 & 770  & 32 & 34 & 0.0170 &                     18 &                    0.5 &             0.6799(20) &                      7 &                    0.5 &             0.6783(16) &                      2 &                    0.6 &             0.6774(17) &  0.545(4) &   7.90(12) \\
            &      &    &    & 0.0195 &                     18 &                    0.5 &             0.6998(17) &                      7 &                    0.5 &             0.6983(15) &                      2 &                    0.6 &             0.6974(16) &       &        \\
            &      &    &    & 0.0220 &                     18 &                    0.5 &             0.7194(16) &                      7 &                    0.5 &             0.7179(14) &                      2 &                    0.6 &             0.7171(15) &       &        \\
\hline
cC211.20.48 & 205  & 13 & 34 & 0.0150 &                     17 &                    0.7 &             0.5906(30) &                      6 &                    0.9 &             0.5888(30) &                      2 &                    0.7 &             0.5863(32) &  0.474(5) &   7.76(19) \\
            &      &    &    & 0.0170 &                     17 &                    0.8 &             0.6062(28) &                      6 &                    0.9 &             0.6043(28) &                      2 &                    0.7 &             0.6021(29) &       &        \\
\hline
cC211.06.80 & 401  & 16 & 39 & 0.0150 &                     20 &                    0.6 &             0.5766(14) &                      5 &                    0.6 &             0.5759(10) &                      2 &                    0.7 &             0.5756(12) &  0.457(3) &  7.93(1) \\
            &      &    &    & 0.0170 &                     20 &                    0.6 &             0.5926(12) &                      5 &                    0.6 &              0.5919(9) &                      2 &                    0.7 &             0.5917(11) &       &        \\
            &      &    &    & 0.0190 &                     20 &                    0.7 &             0.6083(11) &                      5 &                    0.6 &              0.6077(9) &                      2 &                    0.7 &             0.6075(10) &       &        \\
\hline
\end{tabular}

%% file: tab_muc_masses.tex
\begin{tabular}{|l|lcc|c|cll|cll|cc|}
 \multicolumn{5}{c}{} & \multicolumn{3}{c}{one-state fit} & 
 \multicolumn{3}{c}{two-state fit} & \multicolumn{2}{c}{}  \\
\hline
Ensemble & $n_{\rm conf}$ & $n_{\rm srcs}$ & $t_{\rm max}/a$ & \multicolumn{1}{c|}{$a\mu_c$} & $t_{\rm low}/a$ & \multicolumn{1}{c}{$\bar{\chi}^2$} & \multicolumn{1}{c|}{$am_{\Lambda_c}$} & $t_{\rm low}/a$ & \multicolumn{1}{c}{$\bar{\chi}^2$} & \multicolumn{1}{c|}{$am_{\Lambda_c}$} & \multicolumn{1}{c}{$aA_{\Lambda_c}$} & \multicolumn{1}{c|}{$B_{\Lambda_c}/Z_P$} \\ 
\hline
cA211.30.32 & 287 & 16 & 22 & 0.21476 &                      8 &                    1.0 &             1.0200(26) &                      3 &                    1.2 &             1.0180(32) &  0.632(2) &   1.807(11) \\
            &     &    &    & 0.26786 &                      8 &                    1.0 &             1.1177(29) &                      3 &                    1.1 &             1.1153(39) &       &         \\
            &     &    &    & 0.32214 &                      8 &                    0.9 &             1.2137(34) &                      3 &                    1.0 &             1.2108(48) &       &         \\
\hline
cA211.12.48 & 119 & 16 & 22 & 0.21476 &                      8 &                    1.0 &             1.0038(35) &                      1 &                    0.2 &             1.0011(36) &  0.619(4) &   1.792(23) \\
            &     &    &    & 0.26786 &                      8 &                    0.8 &             1.1009(43) &                      1 &                    0.3 &             1.0980(43) &       &         \\
            &     &    &    & 0.32214 &                      8 &                    0.7 &             1.1957(52) &                      1 &                    0.3 &             1.1925(52) &       &         \\
\hline
cB211.25.32 & 360 & 16 & 26 & 0.17464 &                      8 &                    1.6 &             0.8554(18) &                      1 &                    1.3 &             0.8542(17) &  0.533(2) &  1.8454(80) \\
            &     &    &    & 0.21830 &                      8 &                    1.5 &             0.9375(20) &                      1 &                    1.2 &             0.9362(18) &       &         \\
            &     &    &    & 0.26196 &                      8 &                    1.4 &             1.0163(22) &                      1 &                    1.2 &             1.0148(20) &       &         \\
\hline
cB211.25.48 & 649 & 16 & 31 & 0.17464 &                      9 &                    1.3 &             0.8506(12) &                      1 &                    1.2 &             0.8504(10) &  0.527(1) &  1.8520(58) \\
            &     &    &    & 0.21830 &                      9 &                    1.2 &             0.9331(13) &                      1 &                    1.1 &             0.9327(12) &       &         \\
            &     &    &    & 0.26196 &                      9 &                    1.2 &             1.0121(15) &                      1 &                    1.1 &             1.0115(13) &       &         \\
\hline
cB211.14.64 & 232 & 8  & 27 & 0.20000 &                      9 &                    1.0 &             0.8932(25) &                      2 &                    1.2 &             0.8912(24) &  0.519(2) &   1.870(12) \\
            &     &    &    & 0.22000 &                      9 &                    1.1 &             0.9306(26) &                      2 &                    1.4 &             0.9284(25) &       &         \\
\hline
cB211.072.64 & 400 & 4  & 24 & 0.20000 &                     11 &                    1.2 &             0.8713(47) &                      4 &                    1.1 &             0.8687(57) &  0.501(4) &   1.850(23) \\
            &     &    &    & 0.22000 &                     11 &                    1.2 &             0.9083(50) &                      4 &                    1.1 &             0.9057(60) &       &         \\
\hline

cC211.20.48 & 205 & 4  & 29 & 0.18000 &                     13 &                    0.8 &             0.7869(45) &                      4 &                    0.6 &             0.7835(52) &  0.455(4) &   1.840(24) \\
            &     &    &    & 0.22000 &                     13 &                    0.8 &             0.8605(50) &                      4 &                    0.6 &             0.8571(59) &       &         \\
\hline
cC211.06.80 & 260 & 4  & 28 & 0.18000 &                     12 &                    0.9 &             0.7633(48) &                      3 &                    0.7 &             0.7614(49) &  0.434(4) &   1.825(24) \\
            &     &    &    & 0.24000 &                     12 &                    1.0 &             0.8729(57) &                      3 &                    0.8 &             0.8708(57) &       &         \\
\hline
\end{tabular}

%% file: tab_ms_method1.tex
\begin{tabular}{|l|lccc|lccc|c|}
\hline
\multicolumn{1}{|c}{} & \multicolumn{4}{c}{$A_{\Omega}$} & \multicolumn{4}{c}{$B_{\Omega}$} & \multicolumn{1}{c|}{$\overline{\rm MS}$(2 GeV)} \\
\hline
$Z_P$ & \multicolumn{1}{c}{$\bar{\chi}$} & $c_1$ [GeV] & $c_2$ [GeV$^{-2}$] & $c_3$ [GeV fm$^{-2}$] & $\bar{\chi}'$ &      $c'_1$ & $c'_2$ [GeV$^{-2}$] & $c'_3$ [fm$^{-2}$] & $m_s$ [MeV]\\
\hline
M1a &     3.7 &  1.739(7) &     0.45(7) &       3.3(1.1) &   0.2 &  4.69(11) &    -1.5(1.3) &     -12(19) &        94.6(2.5) \\
M2a &     4.4 &  1.739(6) &     0.44(6) &       5.45(97) &   0.2 &  4.67(12) &    -1.6(1.4) &      23(20) &        94.6(2.3) \\
M1b &     3.2 &  1.739(6) &     0.44(7) &       2.3(1.0) &   0.3 &  4.68(11) &    -1.6(1.3) &     -10(19) &        94.6(2.4) \\
M2b &     4.6 &  1.739(6) &     0.44(6) &       6.19(96) &   0.2 &  4.67(12) &    -1.6(1.4) &      31(20) &        94.5(2.3) \\
\hline
\end{tabular}

%% file: tab_ms_method2.tex
\begin{tabular}{|l|lccc|c|}
\hline
\multicolumn{5}{|c}{} & \multicolumn{1}{c|}{$\overline{\rm MS}$(2 GeV)} \\
\hline
$Z_P$ & \multicolumn{1}{c}{$\bar{\chi}$} & $m_{\Omega}^{(0)}$ [GeV] & $c_{\Omega}^{(1)}$ [GeV$^{-1}$] & $d_{\Omega}^{(2)}$ [GeV fm$^{-2}$] & $m_s$ [MeV]\\
\hline
M1a &     2.4 &     1.6636(16) &  -0.121(21) &     3.7(1.3) &        95.1(2.5) \\
M2a &     2.7 &     1.6637(15) &  -0.120(20) &     5.3(1.2) &        95.2(2.3) \\
M1b &     2.2 &     1.6638(15) &  -0.118(21) &     2.6(1.2) &        95.0(2.4) \\
M2b &     2.7 &     1.6637(15) &  -0.120(20) &     6.0(1.2) &        95.1(2.3) \\
\hline
\end{tabular}

%% file: tab_mc_method1.tex
\begin{tabular}{|l|lccc|lccc|c|}
\hline
\multicolumn{1}{|c}{} & \multicolumn{4}{c}{$A_{\Lambda_c}$} & \multicolumn{4}{c}{$B_{\Lambda_c}$} & \multicolumn{1}{c|}{$\overline{\rm MS}$(3 GeV)} \\
\hline
$Z_P$ & \multicolumn{1}{c}{$\bar{\chi}$} & $c_1$ [GeV] & $c_2$ [GeV$^{-1}$] & $c_3$ [GeV fm$^{-2}$] & $\bar{\chi}'$ &      $c'_1$ & $c'_2$ [GeV$^{-2}$] & $c'_3$ [fm$^{-2}$] & $m_c$ [MeV]\\
\hline
M1a &     2.1 &   2.55(2) &      0.9(2) &      -4.9(2.6) &   1.3 &  1.14(2) &     0.00(15) &  -13.7(2.5) &         1030(21) \\
M2a &     2.1 &   2.55(2) &      0.9(2) &       3.6(2.7) &   1.6 &  1.15(2) &     0.01(16) &   -6.4(2.6) &         1027(20) \\
M1b &     2.2 &   2.54(2) &      0.9(2) &      -3.8(2.6) &   1.0 &  1.14(2) &    -0.03(15) &  -12.8(2.5) &         1037(21) \\
M2b &     2.1 &   2.55(2) &      1.0(2) &       5.5(2.8) &   1.7 &  1.15(2) &     0.01(16) &   -4.8(2.6) &         1025(20) \\
\hline
\end{tabular}

%% file: tab_mc_method2.tex
\begin{tabular}{|l|lccc|c|}
\hline
\multicolumn{5}{|c}{} & \multicolumn{1}{c|}{$\overline{\rm MS}$(3 GeV)} \\
\hline
$Z_P$ & \multicolumn{1}{c}{$\bar{\chi}$} & $m_{\Lambda_c}^{(0)}$ [GeV] & $c_{\Lambda_c}^{(1)}$ [GeV$^{-1}$] & $d_{\Lambda_c}^{(2)}$ [GeV fm$^{-2}$] & $m_c$ [MeV]\\
\hline
M1a &     2.7 &      2.2692(25) &   0.95(14) &    -1.4(2.0) &         1030(21) \\
M2a &     2.7 &      2.2689(26) &   0.96(14) &     5.3(2.2) &         1026(20) \\
M1b &     2.8 &      2.2695(25) &   0.93(14) &    -0.6(2.0) &         1038(21) \\
M2b &     2.7 &      2.2688(26) &   0.97(15) &     6.7(2.2) &         1024(20) \\
\hline
\end{tabular}